\documentstyle[prl,floats,aps]{revtex}

\newcommand{\fracparen}[2]{\left(\frac{#1}{#2}\right)}

\newcommand{\Eqref}[1]{Equation~\ref{#1}} 
\newcommand{\Figref}[1]{Figure~\ref{#1}}

\newcommand{\Secref}[1]{Section~\ref{#1}}

\input{psfig.sty}
\input{epsf.sty}

\renewcommand{\theequation}{\thesection.\arabic{equation}}

\begin{document}

\draft

\title{Time-Symmetric ADI and Causal Reconnection: Stable Numerical
  Techniques for Hyperbolic Systems on Moving Grids.}

\author{Miguel Alcubierre and Bernard F. Schutz}

\address{Department of Physics and Astronomy, University of Wales, \\
College of Cardiff, P.O. Box 913, Cardiff CF1 3YB, UK.}

\maketitle

\begin{abstract}

Moving grids are of interest in the numerical solution of hydrodynamical
problems and in numerical relativity.  We show that conventional integration
methods for the simple wave equation in one and more than one dimension exhibit
a number of instabilities on moving grids.  We introduce two techniques, which
we call {\em causal reconnection} and {\em time-symmetric ADI}, which together
allow integration of the wave equation with absolute local stability in any
number of dimensions on grids that may move very much faster than the wave
speed and that can even accelerate.  These methods allow very long time-steps,
are fully second-order accurate, and offer the computational efficiency of
operator-splitting.

We develop causal reconnection first in the one-dimensional case: we find that
a conventional implicit integration scheme that is unconditionally stable as
long as the speed of the grid is smaller than that of the waves nevertheless
turns unstable whenever the grid speed increases beyond this value.  We
introduce  a notion of {\em local stability} for difference equations with
variable coefficients.    We show that, by ``reconnecting'' the computational
molecule at each time-step  in such a way as to ensure that its members at
different time-steps are within one  another's causal domains, one eliminates
the instability, even if the grid  accelerates. This permits very long
time-steps on rapidly moving grids.  The method extends in a straightforward
and efficient  way to more than one  dimension.

However, in more than one dimension, it is very desirable to use
operator-splitting techniques to reduce the computational demands of implicit
methods, and we find that standard schemes  for integrating the wave equation 
--- Lees' First and Second Alternating Direction Implicit (ADI) methods --- go
unstable for quite small grid velocities.  Lees' first method, which is only
first-order accurate on a shifting grid, has mild but nevertheless significant
instabilities.  Lees' second method, which is second-order accurate, is very
unstable.  

By adopting a systematic approach to the design of ADI schemes, we
develop a new ADI method that cures the instability for all velocities
in any direction up to the wave speed.  This  scheme  is uniquely
defined by a simple physical principle: the ADI difference equations
should be invariant under time-inversion. (The wave equation itself and
the full implicit difference equations satisfy this criterion, but
neither of Lees' methods do.)  This new time-symmetric ADI scheme is,
as a bonus, second-order accurate.  It is thus far more efficient than
a full implicit scheme, just as stable, and just as accurate.

By implementing causal reconnection of the computational molecules, we
extend the time-symmetric ADI scheme to arrive at a scheme that is
second order accurate, computationally efficient and unconditionally
locally stable for all grid speeds and long time-steps.  We have tested
the method by integrating  the wave equation on a rotating grid, where
it remains stable even when the grid speed at the edge is 15 times the
wave speed.  Because  our methods are based on simple physical
principles, they should generalize  in a straightforward way to many
other hyperbolic systems.   We discuss briefly  their application  to
general relativity and their potential generalization to fluid
dynamics.

\end{abstract}


\section{Introduction.}\label{sec:I}

In the numerical study of wave phenomena it is often necessary to use a
reference frame that is moving with respect to the medium in which the
waves propagate.  This could be the case, for example, when studying
the waves generated by a moving source, where it may prove convenient
to use a reference frame attached to this source. In some cases, one
may even need to use a frame that moves faster than the waves
themselves, as in the case of a supersonic flow.  In general
relativity, especially in black-hole problems, one may have to use a
grid that shifts rapidly, even faster than light.  All these problems
arise in more than one spatial dimension, where computational
efficiency may make stringent demands on the algorithm.  It is a common
experience to find that standard algorithms seem to go unstable in
realistic problems.  In this paper, by studying the simple wave
equation, we show that the consistent application of two fundamental
physical  principles --- causality and time-reversal-invariance ---
produces remarkably stable, efficient and accurate integration
methods.  These principles can easily be applied to more complex
physical systems, where we would expect similar benefits.

Our principal motivation for studying these techniques is the
development of suitable algorithms for the numerical simulation of
moving, interacting black holes.  Relativists have long acknowledged
the importance of using shifting grids in some problems, but to our
knowledge there has been no systematic study of the effects of such
shifts on the stability of numerical algorithms.  In the next two
paragraphs we develop this motivation.  Readers not concerned with
numerical relativity may skip these without loss of continuity.

Let us consider the requirements that black-hole problems will make of
our algorithms.  Within the context of the \,$3+1$\, formalism of
General Relativity (\cite{MTW}, \cite{York}), it would seem to be
desirable to develop methods on a quasi-rectangular 3-dimensional grid,
so that no special coordinate features would prevent one from studying
quite general problems.  If we imagine a picture in which a black hole
moves ``through'' such a grid, much the way a star would if it were
interacting with another, then some requirements become clear:
\begin{enumerate} \item Grid points will move from outside to inside
the horizon,  but the grid as a whole should not be sucked in. This may
require an inner  boundary to the grid, say on a marginally trapped
surface, and this  boundary will have to move at faster than the speed
of light.  Grid points may cross this boundary and be forgotten, at
least temporarily, but others will emerge on the other side of the
boundary. \item Grid points that so emerge will then move from inside
to outside the horizon as the hole passes over them; this will require
grids that shift faster than light.  This is inescapable unless one
ties the grid to the hole as it moves.  \item If two black holes begin
in orbit around one another, then it may be desirable to adopt a grid
that rotates with respect to infinity, in which the holes move
relatively slowly at first.  In such a grid one would expect that one
could take long time-steps without losing accuracy, since not much
happens initially.  One therefore would like to be free of the Courant
condition on time-steps, {\em i.e.} one wants to use implicit methods.
\item Integrating the equations of general relativity on a grid with
reasonable resolution will tax the capacity of the best available
computers for some time to come.  Full implicit schemes are very
time-consuming in more than one spatial dimension, because they require
the inversion of huge sparse matrices.  Alternating-direction-implicit
(ADI) schemes reduce this burden enormously by turning the integration
into a succession of one-dimensional implicit integrations, so an ADI
scheme that can cope with grid shifts is very desirable.
\end{enumerate}

In this paper we show that it should be possible to develop stable
methods that satisfy the last three requirements above: ADI schemes
that are absolutely stable and computationally efficient even on grids
that shift at many times the speed of light.  As a bonus, our ADI
methods preserve the second-order accuracy of the full implicit
equations.  The first requirement, that of dealing with an inner
boundary that moves faster than light, is closely related to these
techniques and will be addressed elsewhere.

Having these requirements in mind, we have studied the effects that the
use of a moving reference frame has on the finite difference
approximation to the simple wave equation, centering our attention
particularly in the stability properties.  The wave equation is the
simplest system, so the instabilities we find in the standard ADI
methods should certainly also be present when they are applied to more
realistic physical systems.  Of course, the wave equation is much
simpler than other systems, so it is possible that methods that
stabilize its integration will not extend to other systems.   However,
the principles that we find here are of such a fundamental physical
nature that it seems certain that they should be applied wherever
possible. Other kinds of instabilities may of course arise in complex
systems, especially those directly due to nonlinearity, but we feel
that moving-grid instabilities are likely to be cured by the methods we
describe here.

We shall conclude this introductory section by summarizing the approach
and results of the following sections. In the second section we develop
the mathematical framework of shifting grids.  Then in \Secref{sec:III}
we study the one-dimensional wave equation.  We find simple implicit
finite difference schemes that are locally stable for any speed up to
that of the waves,  even when the grid is accelerating as well as
moving.  When  formulated on a grid that is moving, and even
accelerating, it is not immediately obvious how one defines stability:
solutions of the differential equation do not have simple harmonic
time-dependence in this frame.  We find that  a satisfactory criterion
for  local stability of these simple schemes is  that no solutions of
the difference  equations should grow faster anywhere on the grid than
local solutions of the  differential equation.

However, as soon as the reference frame moves faster than the wave
speed, these schemes become highly unstable.  We trace the origin of
this instability to the fact that the computational molecules no longer
represent in an adequate way the causal relationships between the grid
points.  We find that by modifying the molecules so that they link a
given point on one time-slice with one on the next one that is within
the first point's cone of characteristics (its forward ``light cone''),
one can restore stability.  We discuss one algorithm for doing this in
Appendix~B.

We call this {\em causal reconnection.}  It is important to note that
this has a minimal impact on the integration scheme: for implicit
schemes, the matrix that  must be solved for the solution at a given
time-step is constructed only from the relations between grid-points at
that time-step, while causal reconnection affects only the relations
between points on different time-steps.  Thus, it can be incorporated
into the part of the algorithm that constructs the ``inhomogeneous
terms'' that generate the right-hand-side of the implicit matrix
equation.  For the 1-dimensional wave equation, the extra work involved
in seeking out causally related grid points can be significant, but it
becomes a smaller proportion of the overhead in more than one
dimension, and for complicated systems of equations, such as one has in
general relativity, the overhead will be a negligible fraction of the
total work per time-step.  We have tested causal reconnection and found
it to be stable even on grids moving at many times the speed of the
waves. It is also insensitive to the acceleration of the grid.

We then move our attention in \Secref{sec:IV} to {\em
operator-splitting ADI methods} \cite{Richtmyer}, which are
computationally efficient ways of implementing implicit schemes in more
than one dimension.  We find it helpful to derive ADI schemes from a
more systematic point of view than one usually finds in expositions of
this technique.  The goal is to add extra terms to a set of difference
equations that {\em (i)} do not change its accuracy, but {\em (ii)}
replace the large sparse non-tridiagonal matrix which has to be solved
in implicit schemes with a matrix that is a simple product of
tridiagonal matrices of the 1-D implicit form for each dimension, which
are easy to solve.  The extra terms are related to the ``left-over''
terms that appear as the difference between the true operator and its
factored replacement  acting on the data values on the final
time-step.  These final-time-step terms must be eliminated.  They are
in effect {\em replaced} by similar terms from earlier time-steps,
which replacement makes no difference when $\Delta t \rightarrow 0$,
but which removes them from the matrix inversion and allows them to be
included as part of the inhomogeneous terms in the matrix solution.
Then the new equations will be a valid approximation to the
differential  equation but can be solved by a succession of (rapid) 1-D
tridiagonal  matrix solutions.

When subjected to the same stability analysis as we devised for causal
reconnection,  the standard ADI methods show instabilities even when
the reference frame moves very slowly.  The instability is most marked
in Lees' second method, in which the extra terms added in are of second
order and therefore do not degrade the accuracy of the full implicit
scheme.  The instability is also present, albeit more weakly, in Lees'
first scheme, which is only first-order accurate.

We trace these instabilities to the fact that the extra terms added in
either of the standard methods break the time-reversal invariance
exhibited by the original differential equation and by the full
implicit difference  equations.  {\em Demanding that the extra terms be
time-symmetric uniquely  determines an ADI scheme} that is essentially
a hybrid of Lees' first and  second methods.  This time-symmetric ADI
method turns out to be  fully  stable for all grid shifts up to the
wave speed.  Although not built in as a  requirement, the new method
also turns out to be second-order accurate.

The method can then be extended to grid speeds larger than the wave
speed by a direct generalization of the causal reconnection approach
developed for the one-dimensional case.  We demonstrate this by
performing an integration on a rotating grid whose edge moves faster
than the wave speed.

In Appendix A, we derive the wave equation in the accelerating
coordinate system using the efficient tensorial techniques of
relativity.  In Appendix B, we discuss one method of implementing
causal reconnection.


\section{The wave equation on a moving grid.}\label{sec:II}

The wave equation is a good testing ground for any new algorithms for
hyperbolic systems.  The equations governing many wave  systems can be
reduced to the standard wave equation, and its cone of  characteristics
has the causal structure of space-time.  We shall use it to test
methods for integrating hyperbolic systems on moving grids.

We consider the wave equation in an arbitrary number of spatial
dimensions \,$n$\,,

\begin{equation} \nabla^{2}\phi - \frac{1}{c^{2}} \frac{\partial^{2}
\phi}{\partial t^{2}}\,=\, 0 \, . \end{equation}

written in a standard inertial coordinate system denoted by
\,($t,\xi^i$).

We are interested in finding a finite difference approximation to this
equation using a grid of points that moves with an arbitrary
non-uniform speed.  Moreover, we will assume that the  speed of each
grid point can change with time.  In order to represent  this
situation, we need to introduce a second coordinate system  $(t,x^i)$
that will be comoving with the  grid.  We introduce these coordinates
in the continuous case by  a transformation of the form
\begin{equation} x^i \,=\, x^i(t, \, \xi^k)\, . \end{equation}

We have not changed the time coordinate, so we assume that the
identification of surfaces of constant time does not change.  This is
thus not the usual Lorentz transformation of special relativity, so
there is no reason for the form of the wave equation to remain
invariant.  This  will have the implication that, in finite
differences, the time  interval between $t=\rm const$ slices will be
constant, independent of position.  For  problems in general
relativity, this is somewhat of a restriction,  but we do not feel it
is a serious one.  If the causal relations are properly taken into
account, then a spatial dependence in the lapse function ought not to
change our physical conclusions.

In Appendix~A we show that the wave equation takes the following form
in the new coordinates $(t,x^i)$:

\begin{equation} (g^{i\,k} - \beta^i \beta^k) \, \frac{\partial^2
\phi}{\partial x^i \, \partial x^k} + \frac{2 \beta^i}{c} \,
\frac{\partial^2 \phi}{\partial x^i \, \partial t} - \Gamma^i \,
\frac{\partial \phi}{\partial x^i} - \frac{1}{c^2} \, \frac{\partial^2
\phi}{\partial t^2}  \,= \, 0 \, .  \label{eq:wave2} \end{equation}

The following quantities derived from the coordinate transformation
appear  in the last equation:

\begin{eqnarray}  \beta^i &:=&\, - \frac{1}{c} \, \frac{\partial
x^i}{\partial t} \, ,  \label{eq:shift1} \\  g_{\,i\,j} &:=&
\sum_{l=1}^{n} \, \frac{\partial \xi^l}{\partial  x^i} \,
\frac{\partial \xi^l} {\partial x^j} \rule[0mm]{0mm}{8mm} \, ,
\label{eq:3metric} \\   g^{\,i\,j} &:=& \sum_{l=1}^{n} \,
\frac{\partial x^i}{\partial \xi^l}  \, \frac{\partial x^j} {\partial
\xi^l} \, ,  \label{eq:3metricinv} \\  g &:=& \det \left(g_{ij}\right)
\rule[0mm]{0mm}{6mm} \, , \label{eq:3metricdet} \\ \Gamma^i &:=&
-\frac{1}{\sqrt{g}}\left\{\frac{\partial}{\partial t}
\left(\sqrt{g}\beta^i\right)+\frac{\partial}{\partial x^j}\left[
\sqrt{g}\left(g^{ij}-\beta^i\beta^j\right)\right]\right\}
\rule[0mm]{0mm}{9mm} \, . \label{eq:gamma}  \end{eqnarray}

Each of these quantities has a physical interpretation, which we now
explain.  Readers familiar with these ideas may skip to the next
section.

The {\em shift vector} \,$\beta^i$\, gives the relationship between
the  two coordinate systems on nearby surfaces of constant
time.\footnote{This is just the standard definition of the  shift
vector in the 3+1 formalism of Numerical Relativity \cite{MTW}.} Let
the line of constant \,$\{\xi^i\}$\, have coordinates  \,$\{x^i_t\}$\,
at the lower hypersurface and  \,$\{x^i_{t+dt}\}$\ at the upper
hypersurface.  From the definition of the shift vector in
\Eqref{eq:shift1}, it is   clear that

\begin{equation} x^i(\xi^j,t+dt) \,\approx \, x^i(\xi^j,t) \,-\, c \,
\beta^i dt \, .  \label{shift:def}  \end{equation}

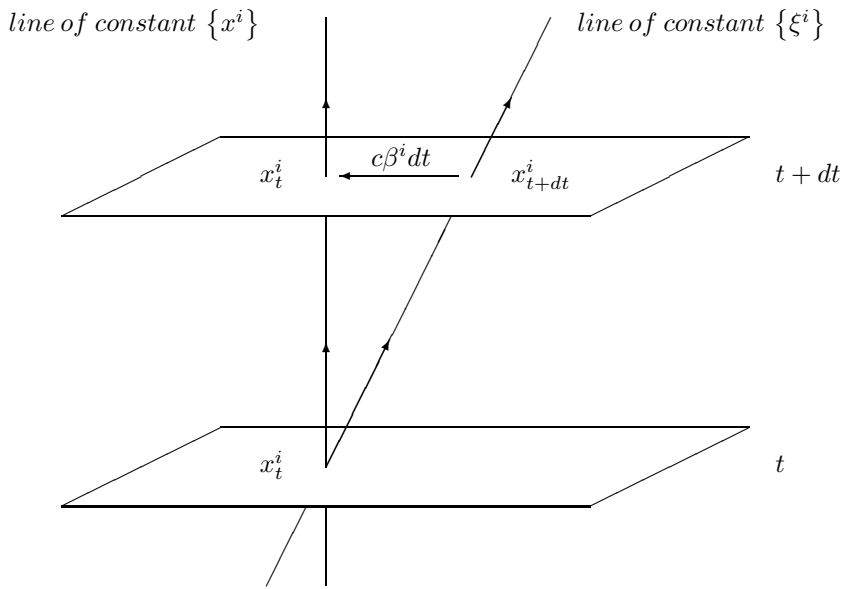
\begin{figure} \begin{center} \begin{picture}(400,200)
\multiput(70,30)(60,30){2}{\line(1,0){200}}
\multiput(70,30)(200,0){2}{\line(2,1){60}}
\multiput(70,140)(60,30){2}{\line(1,0){200}}
\multiput(70,140)(200,0){2}{\line(2,1){60}} \put(170,0){\line(0,1){30}}
\put(170,45){\line(0,1){95}} \put(170,155){\line(0,1){60}}
\put(170,45){\vector(0,1){47.5}} \put(170,155){\vector(0,1){30}}
\put(147.5,0){\line(1,2){15}} \put(170,45){\line(1,2){47.5}}
\put(225,155){\line(1,2){30}} \put(170,45){\vector(1,2){24}}
\put(225,155){\vector(1,2){15}} \put(220,155){\vector(-1,0){45}}
\put(340,43){$t$} \put(340,153){$t+dt$} \put(145,43){$x^i_{t}$}
\put(145,153){$x^i_t$} \put(240,153){$x^i_{t+dt}$} \put(187,159){$c
\beta^i dt$} \put(50,210){$line \: of \: constant \, \left\{ x^i
\right\}$} \put(265,210){$line \: of \: constant \, \left\{ \xi^i
\right\}$} \end{picture} \end{center}  \caption{Shift vector
$\beta^i$.}\label{fig:shift} \end{figure}

As we illustrate in \Figref{fig:shift}, if one starts at  any given
point at time $t$,  then by time $t+dt$ the $\{x^i\}$  coordinates will
have shifted by an amount  equal to the shift vector  times $c\,dt$
relative to the $\{\xi^i\}$ (inertial)  coordinates.  The  shift vector
\,$\beta^i$\, will in general be a function  of both \,$\{x^i_t\}$\,
and t.

We now introduce the {\em spatial  metric tensor} \,$g_{\,i\,j}$, which
we have defined in \Eqref{eq:3metric}. Its name comes from the fact
that the distance \,$dl$\, between two points whose coordinates differ
by $d\xi^l$ in the original coordinates and by  \,$dx^i$\, in the
shifting coordinates is  given by the Pythagorean Theorem:

\begin{equation} dl^{\,2} \,=\, \sum_{l=1}^n d\xi^l\,d\xi^l =
\sum_{i,j = 1}^n g_{\,i\,j} \, dx^i dx^j
\label{eq:pythag}\end{equation}

That this gives \Eqref{eq:3metric} for $g_{ij}$ is readily seen by
substituting the following transformation from $d\xi^l$ to $dx^i$ into
the first version of the Pythagorean Theorem: \begin{equation} d\xi^l =
\sum_{i=1}^n \frac{\partial \xi^l}{\partial x^i} dx^i. \end{equation}
\hspace{2em} The next tensor that appears in  the general form of the
wave equation  is the  {\em inverse metric tensor} given by
\Eqref{eq:3metricinv}.  This   is the matrix inverse of the metric
tensor,  \begin{equation} \sum_{j=1}^n g^{ij}g_{jk} = \delta^i_k,
\end{equation}  as can easily be seen by substituting
Equations~\ref{eq:3metric} and  \ref{eq:3metricinv} into the above.

The final quantity we need is $\Gamma^i$, a measure of the
acceleration  of the shifting coordinates with respect to the old ones,
given by  \Eqref{eq:gamma}.  We will leave the full derivation of
$\Gamma^i$  to Appendix A, but to illustrate our interpretation of it
as an  acceleration term, we shall explicitly evaluate it in the case
where  the new coordinates are obtained from the inertial ones by a
simple  shift independent of position.  Then  the shift vector
\,$\beta^i$\, is only a function of time, and the spatial metric
\,$g_{\,i\,j}$\, is just the  unit matrix:

\begin{equation} \frac{\partial \beta^i}{\partial x^j} \,=\, 0 \, ,
\hspace{10mm}  g_{\,i\,j} \,=\, \delta_{\,i\,j} \, . \end{equation}

It is not difficult to see that in this case the coefficients
\,$\Gamma^i$\, reduce to: \begin{equation} \Gamma^i \,=\, -\,
\frac{1}{c} \, \frac{d \beta^i}{dt} \, . \end{equation} Since the shift
vector gives the speed of the $\{x^i\}$ coordinates,  the last
expression implies that the \,$\Gamma^i$\, coefficients are
essentially the acceleration.

Notice that if there is no acceleration, the only essential difference
from the normal wave equation is the {\em transport} term
$\vec{\beta}\cdot\nabla\dot{\phi}$, which arises as well in
hydrodynamical problems.  We will see that the local stability
properties of  the algorithms we study are determined mainly by
$\beta^i$, not  $\Gamma^i$, which is one reason we expect our analysis
to have much  wider applicability than just to problems involving the
wave equation.

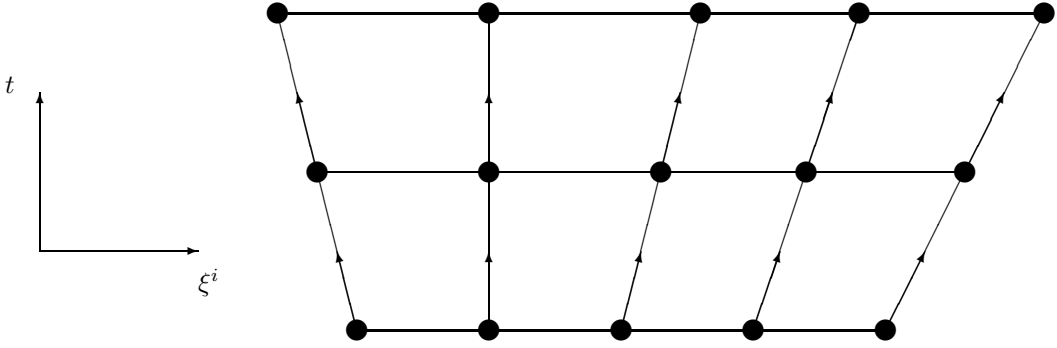
\begin{figure} \begin{center}  \begin{picture}(400,140)
\put(140,20){\line(1,0){200}}  \put(125,80){\line(1,0){245}}
\put(110,140){\line(1,0){290}}  \put(140,20){\line(-1,4){30}}
\multiput(140,20)(-15,60){2}{\vector(-1,4){7.5}}
\multiput(140,20)(-15,60){3}{\circle*{8}} \put(190,20){\line(0,1){120}}
\multiput(190,20)(0,60){2}{\vector(0,1){30}}
\multiput(190,20)(0,60){3}{\circle*{8}} \put(240,20){\line(1,4){30}}
\multiput(240,20)(15,60){2}{\vector(1,4){7.5}}
\multiput(240,20)(15,60){3}{\circle*{8}} \put(290,20){\line(1,3){40}}
\multiput(290,20)(20,60){2}{\vector(1,3){10}}
\multiput(290,20)(20,60){3}{\circle*{8}} \put(340,20){\line(1,2){60}}
\multiput(340,20)(30,60){2}{\vector(1,2){15}}
\multiput(340,20)(30,60){3}{\circle*{8}} \put(20,50){\vector(1,0){60}}
\put(20,50){\vector(0,1){60}} \put(7,110){$t$} \put(80,35){$\xi^i$}
\end{picture} \end{center} \caption{Grid in original coordinates,
showing true distances.}\label{fig:gridoriginal} \end{figure}

\begin{figure}[t] \begin{center} \begin{picture}(400,160)
\multiput(140,20)(50,0){5}{\line(0,1){120}}
\multiput(140,20)(0,60){3}{\line(1,0){200}}
\multiput(140,20)(50,0){5}{\vector(0,1){30}}
\multiput(140,80)(50,0){5}{\vector(0,1){30}}
\multiput(140,20)(50,0){5}{\circle*{8}}
\multiput(140,800)(50,0){5}{\circle*{8}}
\multiput(140,140)(50,0){5}{\circle*{8}} \put(20,50){\vector(1,0){60}}
\put(20,50){\vector(0,1){60}} \put(7,110){$t$} \put(80,35){$x^i$}
\put(353,106){$\Delta \, t$} \put(307,150){$\Delta \, x$} \end{picture}
\end{center} \caption{Grid in new coordinates.}\label{fig:gridnew}
\end{figure}
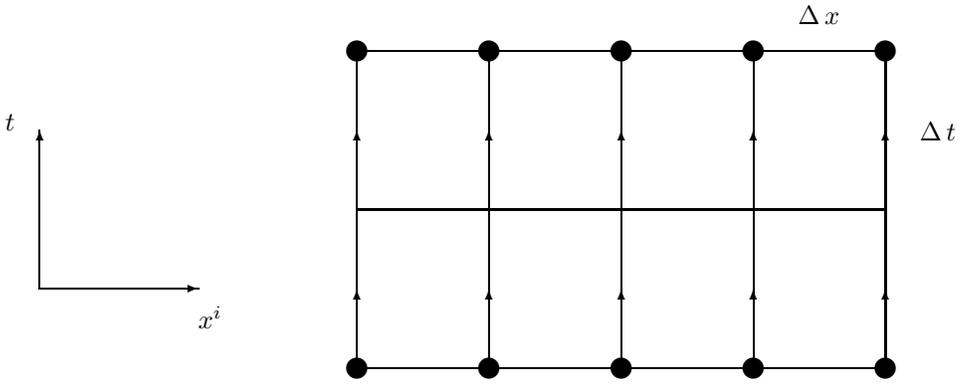

Having derived the form of the wave equation in our new coordinates,
we  now establish a grid for formulating difference equations in these
coordinates.  By assumption, we take the time-interval $\Delta  t$
between successive surfaces of constant time to be uniform
(independent of position) and constant (the same for any pair of
surfaces).  We take each grid point to have a fixed spatial  coordinate
position $x^i$, and for convenience we take the spacing between grid
points $\Delta x^i$ to be uniform in each coordinate  direction.  As
seen in the inertial frame, the grid deforms itself as  in
\Figref{fig:gridoriginal}. The corresponding picture in the
$x^i$-coordinate frame looks much more regular (\Figref{fig:gridnew}).

\setcounter{equation}{0}

\section{The one dimensional case.}\label{sec:III}

\subsection{Finite difference approximation.}\label{sec:3sub1}

The one-dimensional wave equation allows us to study shifting grids in
a relatively simple fashion.  The added complication of extra
dimensions will be treated in the next section.

In one spatial dimension, the metric, shift, and acceleration
coefficients reduce to scalar functions: \begin{equation} \left.
\begin{array}{lll} g_{\,1\,1}(x,t) \,=\,  g(x,t) \, ,
\\ \\ \beta^1(x,t) \,=\, \beta(x,t) \, , \\ \\  \Gamma^1(x,t) \,=\,
\Gamma(x,t) \, . \end{array} \right\}   \end{equation}

Because the metric scales the {\em squares} of the coordinate distances
(\Eqref{eq:pythag}),  it is convenient to define the linear {\it scale
function} \,$s(x,t)$\,  by \begin{equation} s(x,t) \,:=\, \sqrt{g(x,t)}
= \frac{\partial\xi}{\partial x} \, , \end{equation}  so that the
spatial proper distance is given by \begin{equation} d\xi \,=\, s(x,t)
\, dx \, .  \end{equation} Using this expression, \Eqref{eq:wave2}
becomes:

\begin{equation} (\frac{1}{s^2} - \beta^2) \, \frac{\partial^2
\phi}{\partial x^2} + \frac{2 \beta}{c} \, \frac {\partial^2
\phi}{\partial x \, \partial t} - \Gamma \, \frac{\partial
\phi}{\partial x} - \frac{1}{c^2} \, \frac{\partial^2 \phi}{\partial
t^2} \,=\, 0 \label{eq:wave1D} \, . \end{equation}

For the finite difference approximation to this equation we employ the
usual notation: \begin{equation} \phi^j_i \,:=\, \phi \, (i \, \Delta
x,j \, \Delta t) \, . \end{equation} We define the first and second
centered spatial differences as: \begin{equation} \left.
\begin{array}{ll} \delta_x \, \phi^j_i \,:=\, \phi^j_{i+1} -
\phi^j_{i-1} \, , \\ \\  \delta^2_x \, \phi^j_i \,:=\, \phi^j_{i+1} -2
\phi^j_i + \phi^j_{i-1} \, . \end{array} \right\} \label{differences}
\end{equation} It is important to note that with the last definitions
$\left( \delta_x \right)^2 \,\neq\, \delta^2_x$.  We can also define
analogous differences for the time direction.

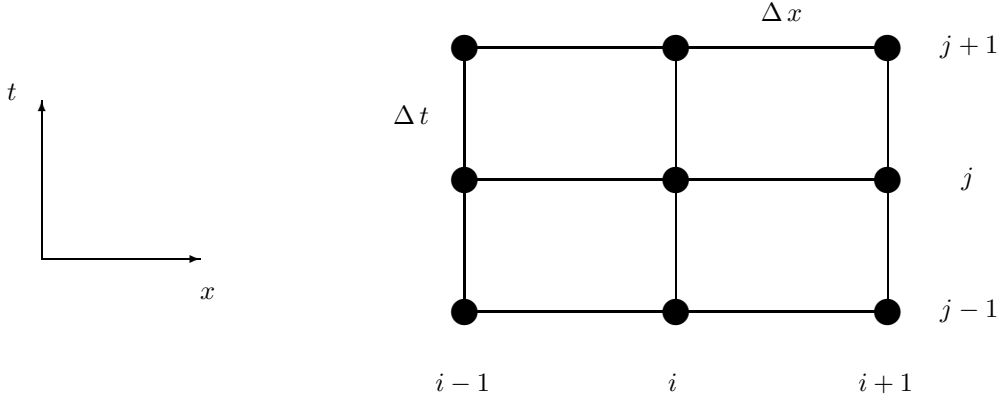
\begin{figure} \begin{center} \begin{picture}(400,140)
\multiput(180,40)(80,0){3}{\line(0,1){100}}
\multiput(180,40)(0,50){3}{\line(1,0){160}}
\multiput(180,40)(80,0){3}{\circle*{10}}
\multiput(180,90)(80,0){3}{\circle*{10}}
\multiput(180,140)(80,0){3}{\circle*{10}} \put(20,60){\vector(1,0){60}}
\put(20,60){\vector(0,1){60}} \put(7,120){$t$} \put(80,45){$x$}
\put(292,150){$\Delta \, x$} \put(153,111){$\Delta \, t$}
\put(169,10){$i-1$} \put(257,10){$i$} \put(329,10){$i+1$}
\put(360,38){$j-1$} \put(368,88){$j$} \put(360,138){$j+1$}
\end{picture} \end{center} \caption{Computational molecule.}
\label{fig:compmol}\end{figure}

We now write the finite difference approximation to the differential
operators that appear in \Eqref{eq:wave1D} using the computational
molecule shown in \Figref{fig:compmol}.  We have: \begin{equation}
\left( \partial^2_t \, \phi \right)^j_i \,=\,  \frac{\phi^{j+1}_i - 2
\phi^j_i + \phi^{j-1}_i}{\left( \Delta t  \right)^2} + E_{tt} \, ,
\end{equation} where \,$E_{tt}$\, is the truncation error whose
principal part is: \begin{equation} E_{tt} \,=\, - \frac{\left( \Delta
t \right)^2}{12} \,  \left( \partial^4_t \, \phi \right)^j_i \, +
\ldots \, . \end{equation} Similarly, for the mixed derivative in space
and time we find:  \begin{equation} \left( \partial_x \partial_t \,
\phi \right)^j_i \,=\,  \frac{\delta_x \,  \phi^{j+1}_i - \delta_x \,
\phi^{j-1}_i}{4 \, \Delta  x \, \Delta t} + E_{xt} \, ,
\end{equation}   with:  \begin{equation} E_{xt} \,=\, - \frac{1}{6}
\left[ \left( \Delta x  \right)^2 \, \left( \partial^3_x \partial^{}_t
\, \phi \right)^j_i +  \left( \Delta t \right)^2 \, \left(
\partial^{}_x \partial^3_t \, \phi  \right)^j_i \right] \, + \, \ldots
\, . \end{equation}

For the second derivative in the \,$x$\, direction we use an {\it
implicit} approximation of the following form:
\begin{equation}\left(\partial^2_x\phi\right)^j_i= \frac{\theta_1}{2}
\, \left[ \frac{\delta^2_x \, \phi^{j+1}_i}{\left(  \Delta x \right)^2}
+ \frac{\delta^2_x \, \phi^{j-1}_i}{\left( \Delta x  \right)^2} \right]
+ \left( 1 - \theta_1 \right) \, \left[  \frac{\delta^2_x \,
\phi^j_i}{\left( \Delta x \right)^2} \right] +  E_{xx} \, ,
\end{equation} where \,$\theta_1$\, is an arbitrary parameter that
gives the  weight of the implicit terms.  If \,\mbox{$\theta_1 \,=\,
0$}\, the  approximation is explicit, while if \,\mbox{$\theta_1 \,=\,
1$}\, all  the weight is given to the initial and final time-steps of
the molecule  in the figure.  Note that the last equation is symmetric
in time.  The  error for this \,$x$\, derivative is: \begin{equation}
E_{xx} \,=\, - \frac{\left( \Delta x \right)^2}{12} \,  \left(
\partial^4_x \, \phi \right)^j_i - \theta_1 \, \frac{\left(  \Delta t
\right)^2}{2} \, \left( \partial^2_x \partial^2_t \, \phi  \right)^j_i
\, + \, \ldots \, .  \end{equation}

Finally, for the first derivative in \,$x$\, we take: \begin{equation}
\left( \partial_x \, \phi \right)^j_i \, = \,  \frac{\theta_2}{2}
\left[ \frac{\delta_x \, \phi^{j+1}_i}{2 \, \Delta  x} \,+\,
\frac{\delta_x \, \phi^{j-1}_i}{2 \, \Delta x } \right] \,+\,  \left( 1
- \theta_2 \right) \, \left[ \frac{\delta_x \, \phi^j_i}{2 \,  \Delta x
} \right]  \,+\, E_x \, , \end{equation} where we have used again an
implicit approximation with a different parameter \,$\theta_2$\,.  The
truncation error \,$E_x$\, is:  \begin{equation}  E_x \,=\, - \left[
\frac{\left( \Delta x \right)^2}{6}  \, \left( \partial^3_x \, \phi
\right)^j_i \,+\, \theta_2 \,  \frac{\left( \Delta t \right)^2}{2} \,
\left( \partial_x \partial_t^2  \, \phi \right)^j_i \right] \,+\,
\ldots \, .  \end{equation}

We can now write down a second order finite difference approximation to
\Eqref{eq:wave1D}: \begin{eqnarray}  \rho^2 \, \left( \frac{1}{s^2} -
\beta^2\right) \,  \left\{ \frac{\theta_1}{2} \, \left[ \delta^2_x \,
\phi^{j+1}_i +  \delta^2_x \, \phi^{j-1}_i \right] + \left( 1 -
\theta_1 \right) \,  \left[ \delta^2_x \, \phi^j_i \right] \right\} \,
&&\nonumber \\ {}+ \frac{\rho \beta}{2} \, \left[ \delta_x \,
\phi^{j+1}_i \,-\,  \delta_x \, \phi^{j-1}_i \right] \,-\, \left[
\phi^{j+1}_i - 2 \phi^j_i  + \phi^{j-1}_i \right] \,&&\nonumber
\\ {}-\frac{\rho \, \left( c \, \Delta t \right)}{2} \:\:  \Gamma \,
\left\{  \frac{\theta_2}{2} \left[ \delta_x \, \phi^{j+1}_i \,+\,
\delta_x \,  \phi^{j-1}_i \right] \,+\, \left( 1 - \theta_2 \right) \,
\left[ \delta_x \, \phi^j_i \right] \right\} \,&=&\, 0 \, ,
\label{eq:finite1D} \end{eqnarray} where \,$\rho$\, is the `Courant
parameter' \cite{recepies} given by:

\begin{equation} \rho \,:=\, \frac{c \, \Delta t}{\Delta x} \, .
\end{equation} The coefficients \{$s$, $\beta$, $\Gamma$\} appearing
in  \Eqref{eq:finite1D} should be evaluated at the point \,$\left( i,j
\right)$\, that corresponds to the center of the molecule.

To arrive at the final form of the difference equation we multiplied
it through by \mbox{\,$\left( \Delta t \right)^2$\,}.  This means  that
the overall truncation error is now

\begin{equation} E_{\mbox{\scriptsize \ref{eq:finite1D}}} = {\cal O}
\left[ \left( \Delta x \right)^2 \, \left( \Delta t \right)^2 \right]
\,+\,  {\cal O} \left[ \left( \Delta t \right)^4 \right] \, .
\end{equation}

\Eqref{eq:finite1D} is well studied in the particular case  when
\mbox{\,$\beta \,=\, \Gamma \,=\, 0$\,} and \mbox{\,$s \,=\, 1$\,}
\cite{Richtmyer}.  It is important to note that, because we use
centered differences in the transport term,  the above finite
difference approximation will be implicit whenever the shift vector is
different from zero, even when \,$\theta_1\,=\,\theta_2 \,=\, 0$\,.
Therefore the use of implicit approximations for the spatial
derivatives does not add any extra numerical difficulty.

We shall need to know how much numerical work is involved in using the
implicit scheme.  Suppose there are $N$ spatial grid points.  Then
\Eqref{eq:finite1D} is to be solved for the $N$ values
$\{\phi^{j+1}_i,\, i=1,\ldots,N\}$ at the final time-step.  The
equation for index $i$ relates three such values, at points
$\{i-1,\,i,\,i+1\}$.  The system of equations therefore has the matrix
form

\begin{equation}\label{eqn:1Dmatrix} \hat{\cal Q}_x\,\phi^{j+1}=
f(\phi^j,\,\phi^{j-1}), \end{equation} where $\hat{\cal Q}_x$ is a
tridiagonal $N\times N$ matrix, and the  inhomogeneous term $f$ is
constructed from field values at the first  two time-steps.   Solving a
tridiagonal matrix involves \mbox{\,${\cal O}\left( N \right)$\,}
operations.  Since we also need \mbox{\,${\cal  O}\left( N \right)$\,}
operations for the solution of an explicit scheme,  we see that the use
of an implicit method in one dimension will increase  the number of
operations per time-step by at most a multiplier, independent  of the
number of grid points.  Against this, the implicit scheme for certain
choices of $\theta_1$ and $\theta_2$ can, on a fixed grid, take much
larger  time-steps, limited only by accuracy considerations.  In the
next section,  we shall show that this property of the implicit scheme
in one dimension  can, with suitable modifications,  be extended to
grids that shift essentially  arbitrarily fast.

\subsection{Local stability: \hspace{1mm} definition and  analysis of
the implicit scheme on a shifting grid.}\label{sec:3sub2}

It is well known \cite{Richtmyer} that the implicit approximation to
the wave equation can be made unconditionally stable in the case when
\mbox{\,$\beta \,=\, \Gamma \,=\, 0$\,} and \mbox{\,$s \,=\, 1$\,} by
using an implicit parameter \mbox{\,$\theta_1 \, \geq \, 1/2$\,}.  We
are interested in studying under what conditions this property is
preserved in the case of a shifting grid.  The shifting grid introduces
a major difficulty: the coefficients in the equation generally  depend
on both position and time.  This complicates the definition of
stability.

This difficulty means that an analytic stability analysis must be  {\em
local}:  we will actually only consider the stability of the
difference equation obtained from \Eqref{eq:finite1D} by, at  each
point $(x,t)$, taking the coefficients to be constant, with the  values
corresponding to that point.  We feel that this is not a very
restrictive assumption, since in practice instabilities usually  appear
as local phenomena \cite{Richtmyer}, with the fastest growing  modes
having wavelengths comparable to the grid spacing.  Moreover, if  the
coefficients in the difference equation are not practically constant
over a few grid points, then we are probably not  approximating the
original differential equation adequately anyway.

We will start then by considering the nature of the solutions of the
{\it differential} equation in a very small region around the point
\,$\left( x,t \right)$\,.  As usual, we look for a solution of the
form: \begin{equation} \phi(x,t) \,=\, e^{\imath\alpha t} \, e^{\imath
kx} \,. \end{equation} Substituting this in \Eqref{eq:wave1D} gives the
following ``dispersion relation'' for $\alpha$: \begin{equation}
\alpha_\pm \,=\, k \beta c \pm c \, \left[  \frac{k^2}{s^2} + i k
\Gamma \right]^{1/2} \, .  \label{eq:dispersion1D}\end{equation} The
general solution for a wavenumber $k$ is \begin{equation} \phi(x,t) =
e^{\imath kx}\left[ Z_+e^{\imath\alpha_+t}
+Z_-e^{\imath\alpha_-t}\right]. \end{equation}

Clearly, if $\Gamma \neq 0$, then one of the independent solutions
will  grow with time, but the other one will decay because of what we
shall  call the {\em analytic boundedness condition}: \begin{equation}
\left| e^{\imath\alpha_+t}e^{\imath\alpha_-t}  \right| \,=\, 1 \, .
\label{eq:bounded}\end{equation}  This does not mean that the system is
physically unstable, but only that in  an accelerating coordinate
system \mbox{\,$ \left( \Gamma \neq 0 \right) $\,}  the wave equation
does not have purely sinusoidal solutions.  One can understand this
intuitively in the following way:  Consider a sinusoidal solution in a
static coordinate system.
From the point of view of an accelerating observer, the frequency of
this
solution will be changing with time (he will be seeing more and more
crests per unit time).  This change in frequency will have important
local effects. As a crest approaches our accelerated observer, he will
see the wave function rising faster than an observer moving at the same
speed, but not accelerating, would.  Hence the appearance of locally
growing modes in our analysis. Similarly, after a crest is reached, the
accelerated observer will see the value of the wave function falling
faster than a uniformly moving observer would.  It is not difficult to
see that the difference between the growth rate in the first case and
the decay rate in the other, as seen by our two observers, will be the
same.  This is the origin of the analytic boundedness condition given
above.

The presence of the growing modes is crucial for our local stability
analysis.  Since the solutions  of the  differential equation can grow
with time, we can not ask the solutions of the  finite difference
approximation not to do so.  What we  are entitled to ask is  for the
numerical solutions not to grow faster than the corresponding normal
modes of the differential equation.   Our stability criterion is,
therefore: {\em a difference equation is  locally stable if every
solution for a given wavenumber $k$ is  bounded in time by a solution
of the differential equation for the same  $k$.}

Bearing this in mind, we now proceed to an analogous analysis of the
solutions of the finite difference scheme. We look for a local
stability condition around the point \,$\left( n,m \right)$\, by
making  the substitution:  \begin{equation} \phi^m_n \,=\, \left( \psi
\right)^{m} \,  e^{i \, k \, n \, \Delta x} \, . \end{equation}
Substituting the last expression in the finite difference
approximation (\Eqref{eq:finite1D}) we find a quadratic equation in
\,$\psi $\, of the form: \begin{equation}  A \, \left( \psi \right)^2 +
B \, \psi + C \,=\, 0 \, , \end{equation} with coefficients given by:
\begin{eqnarray}  A \,&=&\, \left\{ \theta_1 \, \rho^2 \left(
\frac{1}{s^2} - \beta^2 \right) \left[ \cos \left( k \, \Delta x
\right) \,-\, 1 \rule[0mm]{0mm}{5mm} \right] \,-\, 1 \right\} \nonumber
\\  &&{}+\, i \rho \:  \sin \left( k \, \Delta x \right) \, \left[
\beta  \,-\, \frac{\theta_2}{2} \left( c \, \Delta t \right) \Gamma
\right] \,  ,\\  \nonumber \\  B \,&=&\, \left\{ 2 \left( 1 - \theta_1
\right) \, \rho^2 \left(  \frac{1}{s^2} - \beta^2 \right) \left[ \cos
\left( k \, \Delta x  \right) \,-\, 1 \rule[0mm]{0mm}{5mm} \right]
\,+\, 2 \right\}  \nonumber \\  &&{}-\, i \rho \left( 1 - \theta_2
\right)  (c \, \Delta t)  \Gamma \: \sin \left( k \, \Delta x \right)
\rule[0mm]{0mm}{6mm} \, , \\  \nonumber \\  C \,&=&\, \left\{ \theta_1
\, \rho^2 \left( \frac{1}{s^2} - \beta^2  \right) \left[ \cos \left( k
\, \Delta x \right) \,-\, 1 \rule[0mm]{0mm}{5mm} \right] \,-\,  1
\right\} \nonumber \\ &&{}-\, i\rho \: \sin \left( k \, \Delta x
\right) \, \left[ \beta  \,+\, \frac{\theta_2}{2} \left( c \, \Delta t
\right) \Gamma \right] \,  .  \end{eqnarray} The two roots of this
equation are:  \begin{equation} \psi_\pm \,=\, \frac{-B \pm \left( B^2
- 4 A C \right)^{1/2}}{2A} \, , \end{equation} and the general solution
of the difference equation is \begin{equation} \phi^m_n= e^{\imath
kn\Delta x}\left[Z_+\left(\psi_+\right)^m +
Z_-\left(\psi_-\right)^m\right]. \end{equation}

It is not difficult to see that the coefficients \,$A$\, and \,$C$\,
have  the property: \begin{equation} \left| A \right|^2 \,=\, \left| C
\right|^2 \,-\, 2 \, \theta_2 \, \rho^2 \beta \, \left( c \, \Delta t
\right) \, \Gamma \, \sin \left( k \Delta x \right) \, ,
\label{eq:AandC}\end{equation}  which implies:  \begin{equation} \left|
\psi_+\psi_-\right| = \left| \frac{C}{A}  \right| \, \neq \, 1 \, .
\end{equation}

This contrasts with the differential case, \Eqref{eq:bounded}, where
the product  of the magnitudes of the two fundamental solutions was 1.
Since the ratio $|C/A|$ depends on the value of $k$ in
\Eqref{eq:AandC}, there will always exist wavenumbers for which the
product $\left|\psi_+\psi_-  \right|$ exceeds 1.  This would seem to be
undesirable from the point  of view of stability, but we can eliminate
it as a potential problem by  setting from now on \begin{equation}
\theta_2 \,=\, 0 \, . \end{equation} This means that we will use an
implicit approximation only for the  second spatial derivatives
\mbox{\,$\left( \theta_1 \, \neq \, 0  \right)$\,}, and not for the
first spatial derivatives \mbox{\,$\left(  \theta_2 \,=\, 0
\right)$\,}. Since from now on we will have only one  $\theta$
parameter, we will change notation now and define   \mbox{\,$\theta
\,:=\, \theta_1$\,}.   The solutions of the difference equation now
satisfy: \begin{equation} \left| \psi_+\psi_-\right| \,=\, 1 \, .
\end{equation} \hspace{2em} Next we introduce the {\em amplification
measure} \,$M$\,: \begin{equation} M \,:=\, \max_k \, \left( \left|
\psi_+  \right|^2, \left| \psi_- \right|^2 \right) \, , \end{equation}
and analogously for the solutions of the differential  equation. The
amplification measure bounds the growth in the magnitude of any  normal
mode in one time-step. Our local stability  condition is then
equivalent to \begin{equation} M_{\rm Num} \, \leq \, M_{\rm Ana} \, ,
\label{eq:stabilityID} \end{equation} where \,$M_{\rm Num}$\, and
\,$M_{\rm Anal}$\, are the  amplification measures for the finite
difference approximation and  the differential equation respectively.

When all the parameters are free to take any value,
\Eqref{eq:stabilityID} is very complicated, and it is then difficult to
find its consequences analytically.  We shall therefore study this
equation numerically, in order to find regions of the parameter space
in which the finite difference scheme is stable.

\begin{figure}
\psfig{file=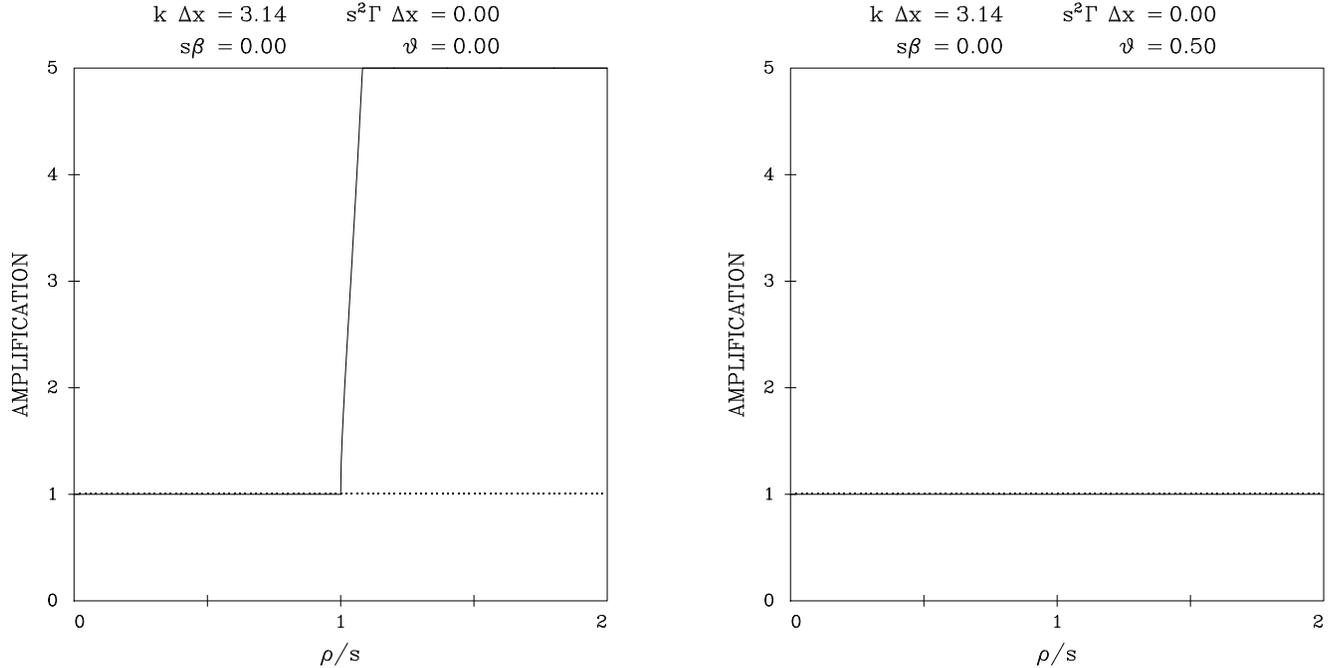,bbllx=0cm,bblly=0cm,bburx=20cm,bbury=12cm,height=9cm,width=20cm}
\caption{Stability on a static grid.  In the left-hand  figure, we
treat the explicit scheme, where we find, as expected, that instability
sets in for Courant parameter $\rho/s>1$.  On the right, we see that a
fully implicit scheme ($\theta=0.5$) is stable for all time-steps,
again as expected.} \label{fig:stability:simple}\end{figure}

First let us consider the case of a static grid, $\beta \,=\, 0 \, ,
\Gamma \,=\, 0$.  This case has, of course, been studied analytically
\cite{Richtmyer}, and it is  known that if $\theta < 1/2$ the
generalized Courant stability condition is: \begin{equation}
\fracparen{\rho}{s}^2 \, \leq \, \frac{1}{\left( 1 - 2 \theta \right)}
\, , \end{equation} while if $\theta\ge 1/2$ the scheme is absolutely
stable. In \Figref{fig:stability:simple} we show graphs of both the
numerical amplification measure (solid line) and the one corresponding
to the differential equation (dotted line).  We have only plotted the
functions for \mbox{\,$k \, \Delta x \,=\, \pi$\,} because this turns
out to be the worst case.   The first graph shows how for
\mbox{\,$\theta \,=\, 0$\,} the scheme is stable  for values of the
Courant parameter \,$\rho/s$\,  smaller than one.  However,  when this
parameter takes values slightly larger than one, the numerical
amplification measure begins to grow very fast.  In the second case we
see  that for\mbox{$\,\theta \,=\, 1/2$\,} the scheme is locally stable
for all values  of the Courant parameter, in agreement with the known
stability condition  given above.

\begin{figure}
\psfig{file=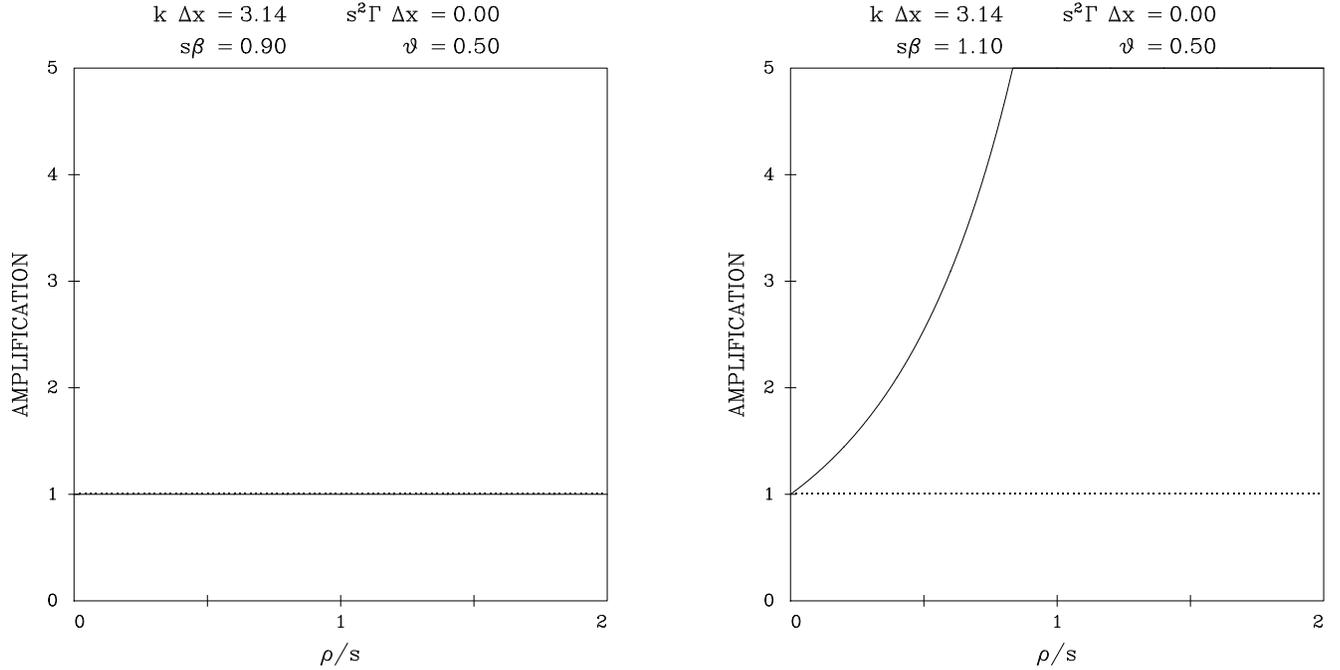,bbllx=0cm,bblly=0cm,bburx=20cm,bbury=12cm,height=9cm,width=20cm}
\caption{Stability on a uniformly shifting grid.  The figure on the
left has a grid speed $0.9$ times the wave speed.  On the right the
grid moves at $1.1$ times the wave speed.  In both cases we have set
$\theta=1/2$ and $\Gamma=0$ (no acceleration).}
\label{fig:stability:velocity} \end{figure}

The next group of graphs (\Figref{fig:stability:velocity}) shows the
effect of a uniform shift.  In both graphs we have assumed that there
is no acceleration \mbox{\,$\left( \Gamma \,=\, 0 \right)$\,}, and we
have taken \mbox{\,$\theta \,=\, 1/2$\,} in order to avoid any
instability of the type seen in \Figref{fig:stability:simple}.  The
first of these shows  that the scheme remains locally stable for all
values of the Courant parameter, even when the grid shift speed \,$s
\beta$\, is  very close to 1. However, in the second graph we see that,
as soon as \,$s \beta$\,  becomes larger than 1, the scheme turns
unstable for {\em all} values of \,$\rho$\,.   In this last case there
is no stable choice of time-step.  This is a very important  property:
{\it The finite difference scheme becomes unconditionally unstable
whenever the shift is faster than the speed of the waves.}

\begin{figure}
\psfig{file=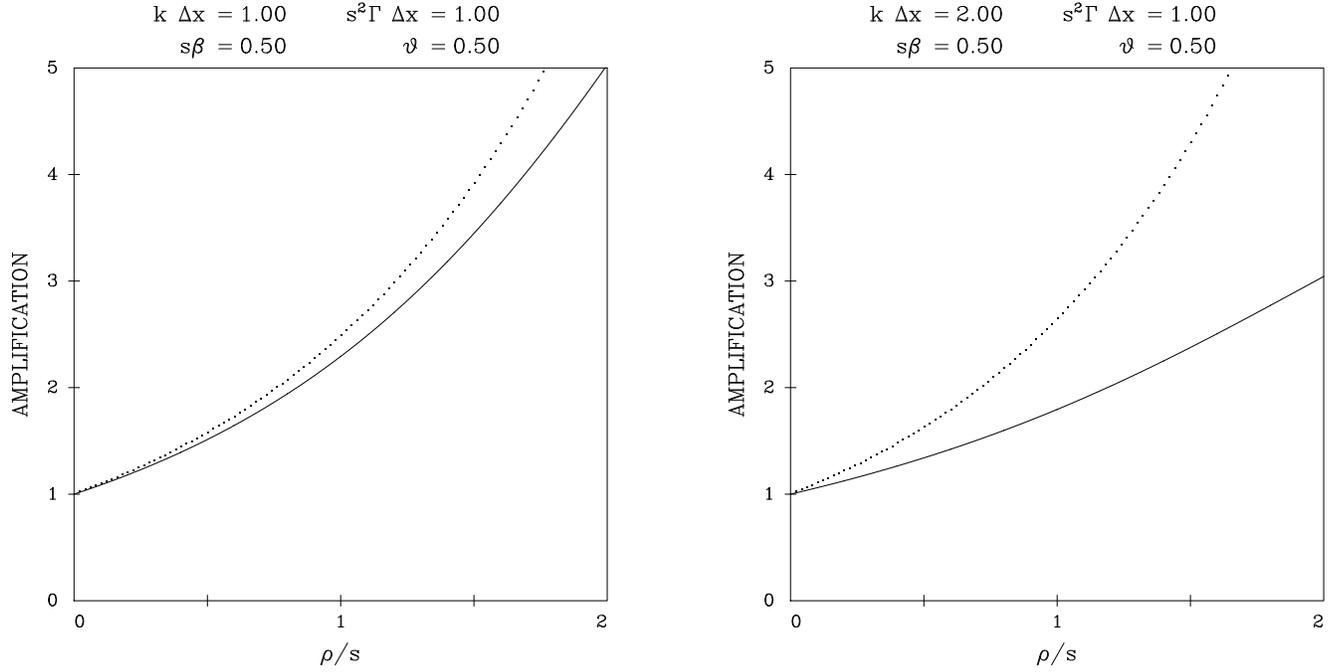,bbllx=0cm,bblly=0cm,bburx=20cm,bbury=12cm,height=9cm,width=20cm}
\caption{Stability on an accelerating grid.  For two different modes,
the finite-difference amplification measure (solid curve) lies below
that of the differential equation (dotted curve).  This means the
finite-difference scheme is stable, at least for these modes.}
\label{fig:stability:accel1}\end{figure}

\begin{figure}
\psfig{file=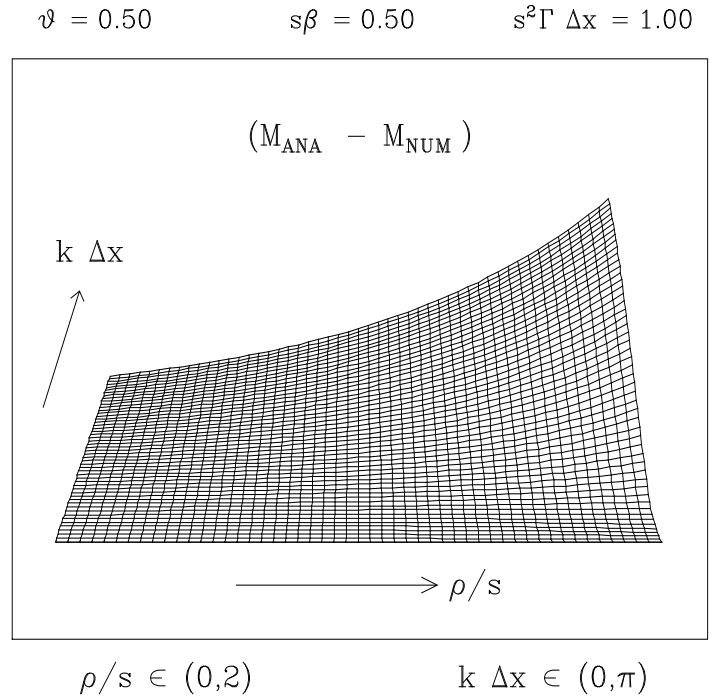,bbllx=0cm,bblly=15cm,bburx=23cm,bbury=24.5cm,height=9cm,width=20cm}
\caption{Stability on an accelerating grid. Here we show a surface plot
of the difference between the analytical and numerical amplification
measures.  This difference is always positive, which means that the
finite-difference scheme is stable in the whole region.}
\label{fig:stability:accel2}\end{figure}

Finally, in Figures \ref{fig:stability:accel1} and
\ref{fig:stability:accel2} we consider the effects of an accelerating
grid for the particular case when: \mbox{\,$\theta \,=\, 1/2$\,},
\mbox{\,$s \,\beta \,=\, 1/2 $\,}, and \mbox{\,$s^2 \, \Gamma \, \Delta
x \,=\, 1$\,}. In \Figref{fig:stability:accel1} we show the behavior of
the amplification measure for the finite-difference equation and the
differential equation for two different normal modes (two values of
$k$).   As we expect, the amplification measure corresponding to the
differential equation, $M_{\rm Ana}$ is no longer  \,$1$\,.  For the
first graph we have \mbox{\,$k \, \Delta x \,=\, 1$\,} and for the
second  {\,$k \, \Delta x \,=\, 2$\,}.   For the smaller wave number
(larger wavelength) the amplification measures  for the differential
and finite-difference cases are relatively close to each other.   As
the wavenumber increases, the finite-difference amplification measure
falls  further below that of the differential equation, so that the
finite-difference  scheme remains stable (although less accurate).
\Figref{fig:stability:accel2} shows a surface plot of \mbox{\,$(M_{\rm
Ana} \,-\, M_{\rm Num})$\,} in the region: \[ \rho/s \,\in\, (0,2)
\hspace{20mm} k \, \Delta x \,\in\, (0,\pi) \,, \]  We clearly see how
\mbox{$(M_{\rm Ana} \,-\, M_{\rm Num}) \,\geq\, 0$} in the whole
region.  Since \mbox{$k \, \Delta x \,=\, \pi$\,} corresponds to the
smallest wavelength that can be represented on the grid
\mbox{\,$(\lambda \,=\, 2 \, \Delta x)$\,}, we find that the
finite-difference scheme will be stable for all modes.

We have searched through other values of  \,$\Gamma$\,, and we have
found that, although the details of the graphs change, the qualitative
behavior is preserved.  {\it The acceleration parameter \,$\Gamma$\,
thus seems to have no important effect on the stability of the scheme.}

In summary, our stability analysis shows that the finite difference
scheme given by \Eqref{eq:finite1D} will be locally stable for all
values of the Courant  parameter \,$\rho$\, if the  following
conditions are satisfied:

\begin{equation} \left. \begin{array}{lll}  \bullet \hspace{5mm}
\theta_1 \,\geq\, 1/2 \, , \\ \\  \bullet \hspace{5mm} \theta_2 \,=\, 0
\, , \\ \\ \bullet \hspace{5mm} \left| \, s \beta \, \right| \,<\, 1 \,
, \\ \\ \bullet \hspace{5mm} \Gamma  \hspace{5mm} irrelevant \, .
\end{array} \right\} \label{eq:stability1D:causal} \end{equation} The
limit on $\beta$ is inconvenient in many problems, where it is
desirable to have grids shifting faster than the wave speed.  We turn
now to a method for removing this restriction.

\subsection{Causal reconnection of the computational molecules.}
\label{sec:3sub3}

\subsubsection{Causality problem.}

\begin{figure}
\psfig{file=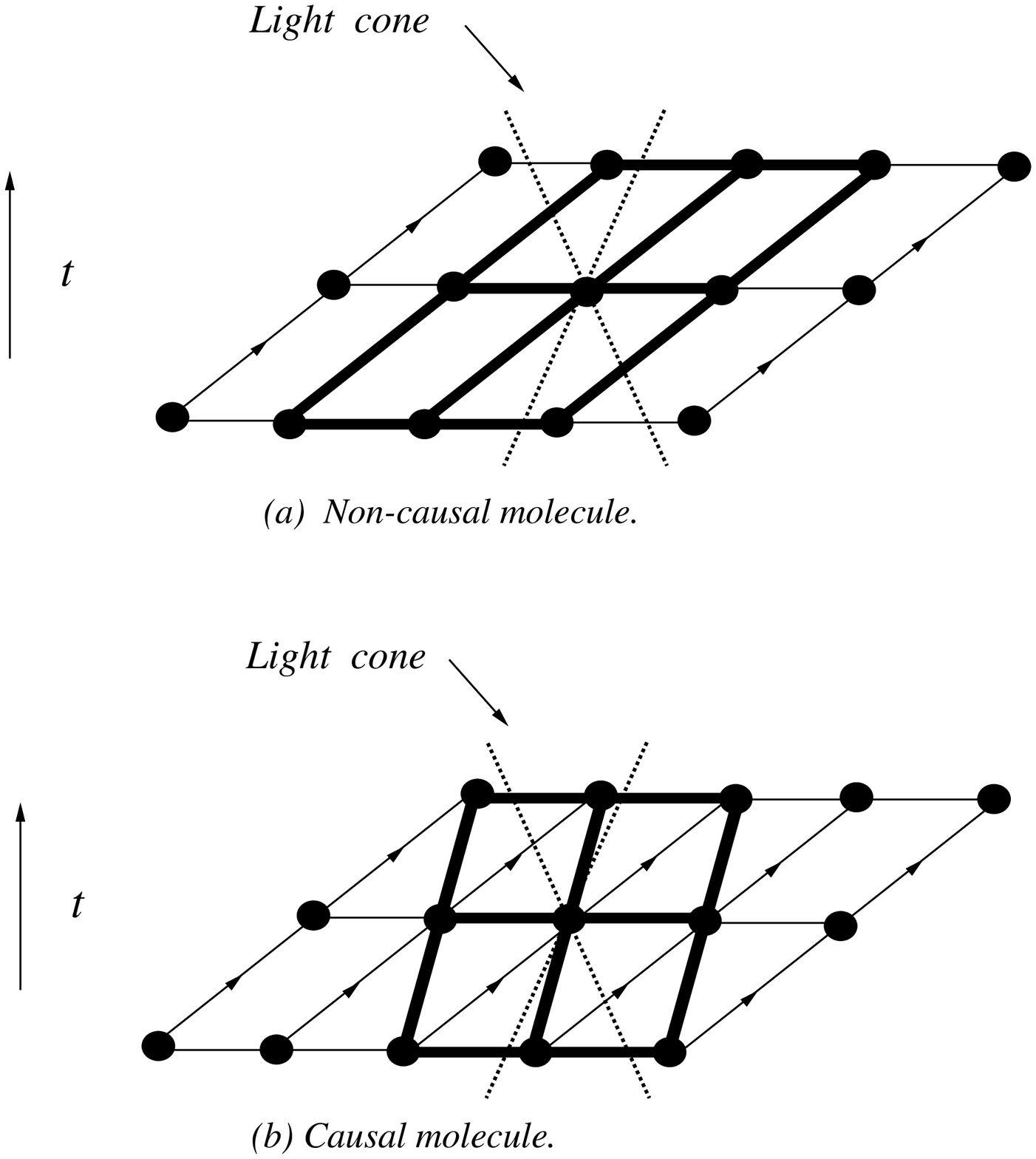,bbllx=0cm,bblly=0cm,bburx=23cm,bbury=24cm,height=20cm,width=20cm}
\caption{Causal computational molecule. In both figures, the dashed lines
represent the light-cone, and the thick solid lines show the  computational
molecule.  Figure ({\em a}) shows the usual molecule that follows the motion of
the grid points.  Figure ({\em b}) shows the reconnected molecule, where we pay
attention to the causal structure instead.} \label{fig:reconnection} 
\end{figure}

The causal structure of a grid shifting faster than the wave speed is
particularly clearly illustrated in the original \,$\left( \xi,t \right)$\,
coordinates.  In \Figref{fig:reconnection}{\em a} we see how, for a very large
shift, the individual grid points move faster than the waves, that is, they
move outside the light-cone.\footnote{From here on we will adopt the language
of relativity and refer to the characteristic cone of the hyperbolic equation
as the `light-cone'.}  Since the differential equation propagates data along
this cone, it seems plausible that the instability found in the previous
section arises in the fact that the difference scheme attempts to determine the 
solution at points on the final time-step using data that are outside the past 
light-cone of these points.

This suggests that we should not build the computational molecules from grid 
points with fixed index labels, but instead use those points that have the
closest {\em causal} relationship (\Figref{fig:reconnection}{\em b}).  We shall
now proceed to show analytically how such a reconnection can stabilize the
scheme.

In order to build this causal molecule let us consider then an individual grid
point at the last time level.  We look for that grid point in the previous time
level that is closest to it in the causal sense. Having found this point, we
repeat the procedure to find the closest causally connected point in the first
time level.  In Appendix B we give a simple algorithm for finding these points
in an integration of the wave equation.  The algorithm adapts easily  to other
linear hyperbolic systems.  We shall return in a later paper to its 
generalization to nonlinear equations, and in particular to the case of shocks 
in hydrodynamics.  First we consider the general constraints on the  time-step
that causal reconnection imposes, and then address the issue of how much extra
computational effort causal reconnection may involve.

\subsubsection{The causal reconnection condition.}

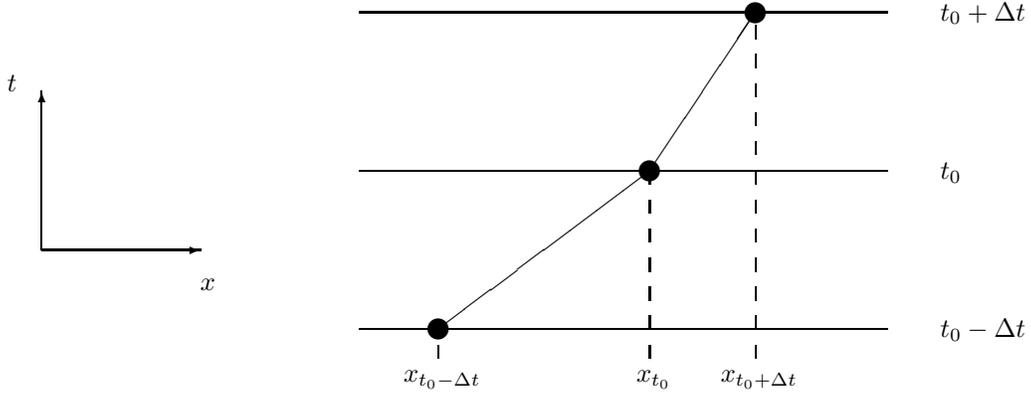
\begin{figure} \begin{center} \begin{picture}(400,160)
\multiput(140,20)(0,60){3}{\line(1,0){200}} \put(170,20){\circle*{8}}
\put(250,80){\circle*{8}} \put(290,140){\circle*{8}}
\put(170,20){\line(4,3){80}} \put(250,80){\line(2,3){40}}
\put(170,9){\line(0,1){5}}  \multiput(250,9)(0,11){7}{\line(0,1){5}}
\multiput(290,9)(0,11){12}{\line(0,1){5}}  \put(360,17){$t_0-\Delta t$}
\put(360,77){$t_0$}  \put(360,137){$t_0+\Delta t$}  \put(157,0){$x_{t_0
- \Delta t}$} \put(245,0){$x_{t_0}$} \put(277,0){$x_{t_0 + \Delta t}$}
\put(20,50){\vector(1,0){60}} \put(20,50){\vector(0,1){60}}
\put(7,110){$t$} \put(80,35){$x$} \end{picture} \end{center}
\caption{Causally connected grid points.}
\label{fig:moving}\end{figure}

Since we permit the grid to move with an arbitrary non-uniform speed,
there is no reason that these causally connected points should be in a
straight line in either the original inertial reference frame \,$(\xi,
t)$\, or in the moving reference frame \,$(x, t)$.  In the moving
coordinate system \,$\left( x,t \right)$\, the relationship among these
three points may generically look something like that shown in
\Figref{fig:moving}.

Accordingly, we introduce a new local coordinate system \,$\left( x',t'
\right)$\, adapted to the three given points.  In order to do this,  it
is convenient to introduce the interpolating second order polynomial
that  can be obtained from these three points: \begin{equation} P
\left( t \right) \,=\, A \, \frac{\left( t - t_0 \right)^2}{2} +  B \,
\left( t - t_0 \right) + x_0 \, , \end{equation} where \,$t_0$\, is the
time at the central point of the molecule,  \,$x_{t_0}$\, the position
of that point, and: \begin{equation} A \,:=\, \left( \frac{x_{t_0 +
\Delta t} - 2 x_{t_0} + x_{t_0 - \Delta t}} {\left( \Delta t \right)^2}
\right)  \, , \hspace{20mm}  B \,:= \, \left(\frac{x_{t_0 + \Delta t} -
x_{t_0 - \Delta t}}{2 \, \Delta t} \right) \,
. \end{equation}

We define the new local coordinate system adapted to the causal
molecule by:  \begin{equation}  \left.  \begin{array}{ll} x' \,:=\,
\left( x - x_{t_0} \right) - P \left( t \right) \, , \\ \\  t' \,\,:=\,
t \, .  \end{array} \right\} \end{equation}

It can easily be seen that this new coordinate system \,$\left( x',t'
\right)$\, moves with respect to the old one \,$\left( x,t \right)$\,
with a speed \,$B$\, at time \,$t \,=\, t_0$\,, and with a constant
acceleration \,$A$\,.  Since in general the value of the coefficients
\,$A$\, and \,$B$\, will change from molecule to molecule, the above
change of variables must be repeated for each molecule.  We assume that
this can be done in a smooth manner; this may not be possible in a
nonlinear system, if the characteristics depend on the solution.

In the primed coordinate system, the differential equation has exactly
the same form as before (see \Eqref{eq:wave1D}), except for the
following substitutions:  \begin{equation} \beta \,\longrightarrow\,
\beta + \frac{A \, \left( t - t_0 \right) + B}{c} \, , \hspace{20mm}
\Gamma \,\longrightarrow\, \Gamma - \frac{A}{c^2} \, . \end{equation}
In the same way, the finite difference approximation will have the same
form as before (\Eqref{eq:finite1D}), except for the substitutions:
\begin{equation} \beta \,\longrightarrow\, \beta + \frac{B}{c} \, ,
\hspace{20mm} \Gamma \,\longrightarrow\, \Gamma - \frac{A}{c^2} \, ,
\end{equation}  where the term with \,$\left( t - t_0 \right)$\, has
disappeared because in this case the coefficients should be evaluated
at the center of the molecule where \mbox{\,$t \,=\, t_0$\,}.

Since the original finite difference approximation could be made stable
as long as \Eqref{eq:stability1D:causal} was satisfied (and \,$\theta
\, \geq \, 1/2 $), the analogous condition for a finite-difference
scheme adapted to the new local coordinates takes the form:
\begin{equation} \frac{1}{s^2} \,-\, \left( \beta + \frac{B}{c}
\right)^2 \, \geq \, 0 \, .  \label{proper_causal_molecule}
\end{equation}

We will say that the three given points form a {\em proper causal
molecule} when the last condition is satisfied. In order to find when
this happens, we will  start by defining the {\em effective numerical
light-cone} of the point \,$x_{t_0}$\,  as the region between the
lines: \begin{equation} x_{\pm} \left( t \right) \,:=\, x_{t_0} \,+\,
\left( -\beta \,\pm\, \frac{1}{s} \right) c \, \left( t - t_0 \right)
\,+\, \frac{1}{2} \, \Gamma c^2 \left( t - t_0 \right)^2 \, .
\end{equation} This numerical light-cone will coincide with the exact
light-cone when: \begin{equation} \frac{\partial \beta}{\partial x}
\,=\, 0 \hspace{20mm} \frac{\partial \Gamma} {\partial x} \,=\,
\frac{\partial \Gamma}{\partial t} \,=\, 0 \, . \end{equation}

We will also define the axis of the numerical light-cone as the line:
\begin{equation} x_{a} \left( t \right) \,:=\, x_{t_0} \,-\, \beta c \,
\left( t - t_0 \right) \,+\, \frac{1}{2} \,  \Gamma c^2 \left( t - t_0
\right)^2 \, .  \end{equation} We will now show that if \,$x_{t_0 +
\Delta t}$\, and \,$x_{t_0 - \Delta t}$\, are inside the numerical
light-cone of \,$x_{t_0}$\, then the three points  will form a proper
causal molecule.  From the definition of the numerical  light-cone we
see that if \,$x_{t_0 + \Delta t}$\, and \,$x_{t_0 - \Delta t}$\,  are
inside it then \begin{equation} x_{t_0 + \Delta t} \,=\, x_{a} \left(
\Delta t \right) \,+\, D_{+} \hspace{20mm} x_{t_0 - \Delta t} \,=\,
x_{a} \left( -\Delta t \right) \,+\, D_{-} \, , \end{equation} with
\begin{equation} D_{+}, D_{-} \,\in\, \left[ -\frac{c \, \Delta
t}{s}\,,\, \frac{c \, \Delta t}{s} \, \right] \, . \end{equation}

The coefficients \,$A$\, and \,$B$\, will then be given by
\begin{equation} B \,=\, -\beta \, c \,+\, \frac{D_{+} - D_{-}}{2
\left( \Delta t \right)} \, , \hspace{20mm} A \,=\, \Gamma \, c^2 \,+\,
\frac{D_{+} + D_{-}}{\left( \Delta t \right)^2} \, , \end{equation}
which in turn means \begin{equation} \left( \beta + \frac{B}{c} \right)
\,\in\, \left[ -\frac{1}{s} \,,\, \frac{1}{s} \,\, \right] \, ,
\hspace{20mm}\left( \Gamma - \frac{A}{c^2} \right) \,\in\, \left[
-\frac{2}{s \, \left( c \Delta t \right)}\,,\,\frac{2}{s \, \left( c
\Delta t \right)} \, \right] \, . \end{equation}

From this it is easy to see that condition~\ref{proper_causal_molecule}
is
indeed satisfied, that is, the three points do form a proper causal
molecule.  Moreover, the absolute value of the acceleration in the new
local coordinates will be bounded, and even though this doesn't affect
the stability of the finite difference scheme, it does improve its
accuracy.

In order to be able to form proper causal molecules everywhere, we must
guarantee that two logically distinct conditions hold.  If we call the
central point at $t_0$ as the ``parent'' and the points at
$t_0\pm\Delta t$ the ``children'', then every parent must have two
children and every child must have a parent:

\begin{enumerate} \item  {\em Every parent must have two children.}
There must always be at least one grid point in the upper and lower
time levels inside the numerical light-cone of any given point in the
middle time level.  This can be guaranteed if we ask for the distance
between grid points to be smaller than the spread of the smallest
light-cone, that is, \begin{equation} \Delta x \,\leq\, \frac{2 c \,
\Delta t}{\max \left( s \right)} \, , \end{equation} which implies
\begin{equation} 2 \, \rho \, \min \left( \frac{1}{s} \right) \,\geq\,
1 \, . \label{reconnection1_1D} \end{equation} (Without loss of
generality, we assume in this section that $\Delta t$ and hence $\rho$
are positive.)

\item  {\em Every child must have a parent.}  All the grid points in
the upper and lower time levels must be inside the  numerical
light-cone of at least one point in the middle time level.  This
requires that the distance between the axes of the numerical
light-cones of two  consecutive grid points must be smaller than the
spread of the minimum light-cone.

Let us therefore consider two consecutive grid points \,$x_1$\, and
\mbox{\,$x_2 \,=\, x_1 + \Delta x$\,}.  The distance between the axis
of their light-cones  at the next time level is given by:
\begin{equation} d_{+} \,=\, \left| \, \left( {x_a} \right)_2 \left(
\Delta t \right) - \left( {x_a} \right)_1  \left( \Delta t \right) \,
\rule[0mm]{0mm}{5mm} \right| \, .  \end{equation} Using the definition
of \,$x_a$\, we find that: \begin{equation} d_{+} \,=\, \left| \,
\Delta x \,-\, c \, \Delta t \, \left[ \beta \left( x_2 \right) -
\beta \left( x_1 \right) \rule[0mm]{0mm}{5mm} \right] \,+\,
\frac{1}{2}  \left( c \, \Delta t \right)^2 \, \left[ \Gamma \left( x_2
\right) \,-\,  \Gamma \left( x_1 \right)  \rule[0mm]{0mm}{5mm} \right]
\, \right| \, .  \end{equation}

In the same way we find that the distance between the axis of the
light-cones at the previous time level is: \begin{equation} d_{-} \,=\,
\left| \, \Delta x \,+\, c \, \Delta t \, \left[ \beta \left( x_2
\right) -  \beta \left( x_1 \right) \rule[0mm]{0mm}{5mm} \right] \,+\,
\frac{1}{2}  \left( c \, \Delta t \right)^2 \, \left[ \Gamma \left( x_2
\right) \,-\,  \Gamma \left( x_1 \right) \rule[0mm]{0mm}{5mm} \right]
\, \right| \, . \end{equation}

The maximum of these two is: \begin{equation} d \,=\, \left| \, \Delta
x \,+\, c \, \Delta t \, \left| \beta \left( x_2 \right) -  \beta
\left( x_1 \right) \rule[0mm]{0mm}{5mm} \right| \,+\, \frac{1}{2}
\left( c \, \Delta t \right)^2 \, \left[ \Gamma \left( x_2 \right)
\,-\,  \Gamma \left( x_1 \right) \rule[0mm]{0mm}{5mm} \right] \,
\right| \, . \end{equation}

Let us assume  now that both \,$\beta$\, and \,$\Gamma$\, are
continuous functions.  Then  we can expand them in a Taylor series
around the point:  \begin{equation} \bar{x} \,:=\, \frac{x_1 \,+\,
x_2}{2} \, . \end{equation}

We then find that, to second order in \,$\Delta x$\,: \begin{equation}
d \,=\, \left| \, \Delta x \,+\, c \, \Delta t \, \left| \,
\frac{\partial \beta}{\partial x} \,  \Delta x \, \right| \,+\,
\frac{1}{2} \left( c \, \Delta t \right)^2 \, \left[ \,  \frac{\partial
\Gamma}{\partial x} \, \Delta x \, \right] \, \right| \, ,
\end{equation} where the derivatives are evaluated at the point
\,$\bar{x}$\,.

The condition that the maximum value of this distance should be
smaller than the spread of the minimum light-cone is now:
\begin{equation} \max \left( d \right) \,\leq\, \frac{2 c \, \Delta
t}{\max \left( s \right)} \, , \end{equation} Using now the expression
for \,$d$\,, we can rewrite this as:  \begin{equation} 2 \, \rho \,
\min \left( \frac{1}{s} \right) \,\geq\, \left| \, 1 \,+\, \rho \,
\Delta x \, \max \left| \, \frac{\partial \beta}{\partial x} \,
\right|  \,+\,  \frac{1}{2} \, \rho^2 \, \left( \Delta x \right)^2 \,
\max \left[ \,  \frac{\partial \Gamma}{\partial x} \, \right] \,
\right| \, .  \label{reconnection2_1D} \end{equation} This is the ``no
orphans'' condition, that every child point should have a  parent.
Since we want this to be true for all grid points,   it must hold for
all \,$x$\,. \end{enumerate}

We call equations~\ref{reconnection1_1D} and \ref{reconnection2_1D}
the  {\em causal reconnection conditions:} when they are satisfied, one
is guaranteed that causal molecules can be formed everywhere.

It is clear that if the derivatives of \,$\beta$\,  and \,$\Gamma$\,
are too large, it will be impossible to satisfy the second of the
causal reconnection conditions, \Eqref{reconnection2_1D} (no
orphans).   This can be avoided if we require  that \,$\beta$\, and
\,$\Gamma$\, change very little from one grid point  to another:
\begin{equation} \Delta x \, \max \left| \, \frac{\partial
\beta}{\partial x} \, \right| \,\ll\, 1 \hspace{20mm}  \Delta x \, \max
\left| \, \frac{\partial \Gamma}{\partial x} \, \right| \,\ll\, 1
\label{bound} \end{equation}

As we mentioned before, if this is not the case our finite difference
approximation  is unlikely to be good anyway, and a more refined grid
spacing should be used.

Another interesting feature of condition~\ref{reconnection2_1D} is the
fact that,  whenever \,$\Gamma$\, is not uniform,  there will always be
a value of \,$\rho$\,  large enough for the condition to be violated.
These sets an {\em upper bound}  on the  time-step, which can be
understood if we examine the effects of a non-uniform  acceleration on
two adjacent numerical light-cones.  If the acceleration increases
with  \,$x$\,, the numerical light-cones will eventually converge and
pass through each other  at a large enough time, even if they were
diverging initially.  Similarly, if the acceleration decreases with
\,$x$\,,  the numerical light-cones will  eventually diverge, even if
they intersect each other for a while.  Clearly these  situations do
not arise in the exact (differential) case because they would break
the causal structure of the solutions.  We must therefore conclude that
the  numerical light-cones will not approximate the real light-cones
properly when we have  a time-step large enough for these effects to
occur.   However, if \,$\Gamma$\, is such  that \ref{bound} holds, then
the upper limit on \,$\Delta t$\, will be very large indeed, much
larger than the Courant limit, and so large that the accuracy of the
integration must be breaking down anyway.  Moreover, for any given
time-step condition~\ref{reconnection2_1D}  can always be satisfied for
a small enough grid spacing \,$\Delta x$\,.

\subsubsection{Numerical overheads of causal reconnection.}

Notice that causal reconnection does not change the fundamental
structure of the difference equation, since it does not affect the
relations between points at the  final time-step.  Therefore, even with
causal reconnection, the equation will  have the form \begin{equation}
\hat{\cal Q}_x\,\phi^{j+1}= f'(\phi^j,\,\phi^{j-1}) \, ,
\label{eqn:causaleqn} \end{equation} where $f'$ is a different
function, which reflects the fact that causal reconnection identifies
different points at time-steps $j$ and $j-1$ to use to generate the
points  at the final time-step.  Therefore, any algorithms that are
used without causal  reconnection for the solution of this tridiagonal
system of equations can be  used equally well with causal
reconnection.

There will, of course, be an overhead associated with the search for
causally related grid points.  In a one-dimensional problem with $N$
grid points, this will  require only ${\cal O}(N)$ operations, since
once the causal molecule of one  grid point has been constructed, the
causal molecule of its neighbor will, by  continuity, usually differ by
at most one spatial shift at any time level.  In more  than one
dimension, the search should still scale linearly with the number of
grid points, since again by continuity the causal molecule of any point
can  usually be guessed from that of any of its neighbors.

We have found  that for the one-dimensional wave equation,  the
implementation of causal reconnection given in Appendix~B can  multiply
the computation time by something like a factor of two.   But for a
more realistic problem, such as general relativity, where there are
many dependent variables per  grid point, the overhead of searching for
the causal structure will be no  different than for the simple wave
equation, so it will represent a small  percentage of the overall
computing time.

\subsection{Numerical examples of causal reconnection.}\label{sec:3sub4}

As an example of the methods that we have developed in the last
sections, we will consider a grid that is oscillating in the original
coordinates \,$\left( \xi,t \right)$\,.  The scale and shift functions
are given by: \begin{equation} s \left( x,t \right) \,=\, 1  \, ,
\hspace{20mm}  \beta \left( x,t \right) \,=\, A \: \cos \left( \omega t
\right) \, ,  \end{equation}  from which we deduce  \begin{equation}
\Gamma \,=\, A \, \omega \: \sin \left( \omega t \right) \, .
\end{equation}

This grid turns out to give a very good illustration of all the
properties  we have mentioned so far.  In the calculations we have
taken \mbox{\,$c \,=\, 1$\,}.

\begin{figure}
\psfig{file=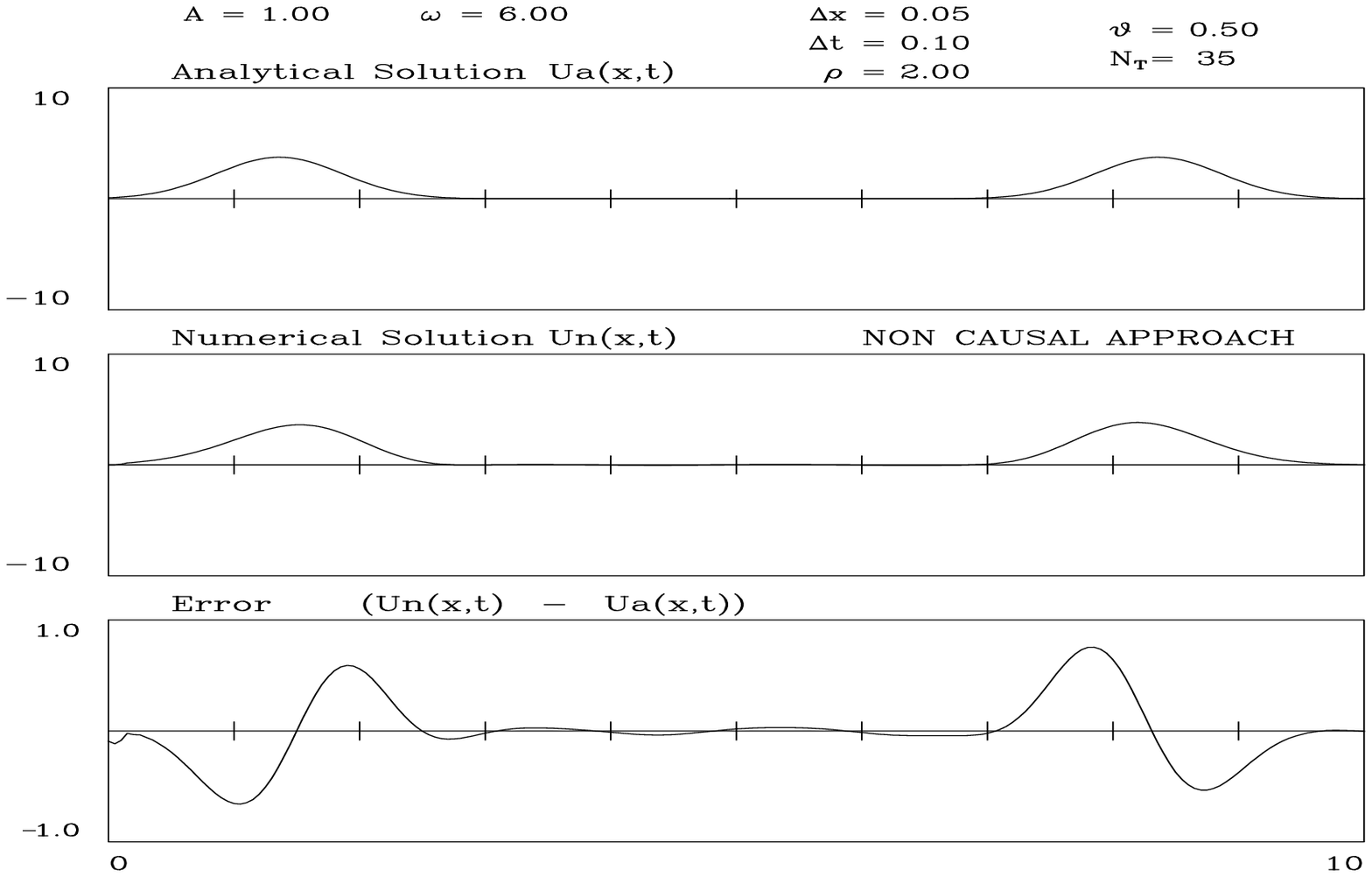,bbllx=0cm,bblly=-12.5cm,bburx=18cm,bbury=12cm,height=22cm,width=15cm}
\caption{Oscillating grid, non-causal approach.}
\label{fig:example:noncausal}\end{figure}

In \Figref{fig:example:noncausal} we show two calculations using the
finite difference approximation given by \Eqref{eq:finite1D}.  In these
examples we have not taken into consideration the causal structure. The
figures show  the evolution of a Gaussian wave packet that was
originally at rest at the center of the grid.  In both cases we have
taken $\omega \,=\, 6$, and we show the situation after 35 time-steps
(3.5 periods of oscillation of the grid).

In the first graph $ A \,=\, 1$: the maximum shift equals the speed of
the waves, but for all the rest of the time the shift is less than 1.
At the end of the calculation there is no evidence of any
instabilities.  In fact, we have integrated this for a very large
number of  time-steps with the same result: no instabilities appear.

For the second graph we have taken$ A \,=\, 1.1$: the maximum shift is
now slightly larger than the speed of the waves, although even here the
grid spends most of its time at speeds less than 1.  By the end of the
calculation an instability has started to form.  It exhibits the
characteristic  feature of finite-difference instabilities, that the
shortest wavelengths are  the most unstable.

\begin{figure}
\psfig{file=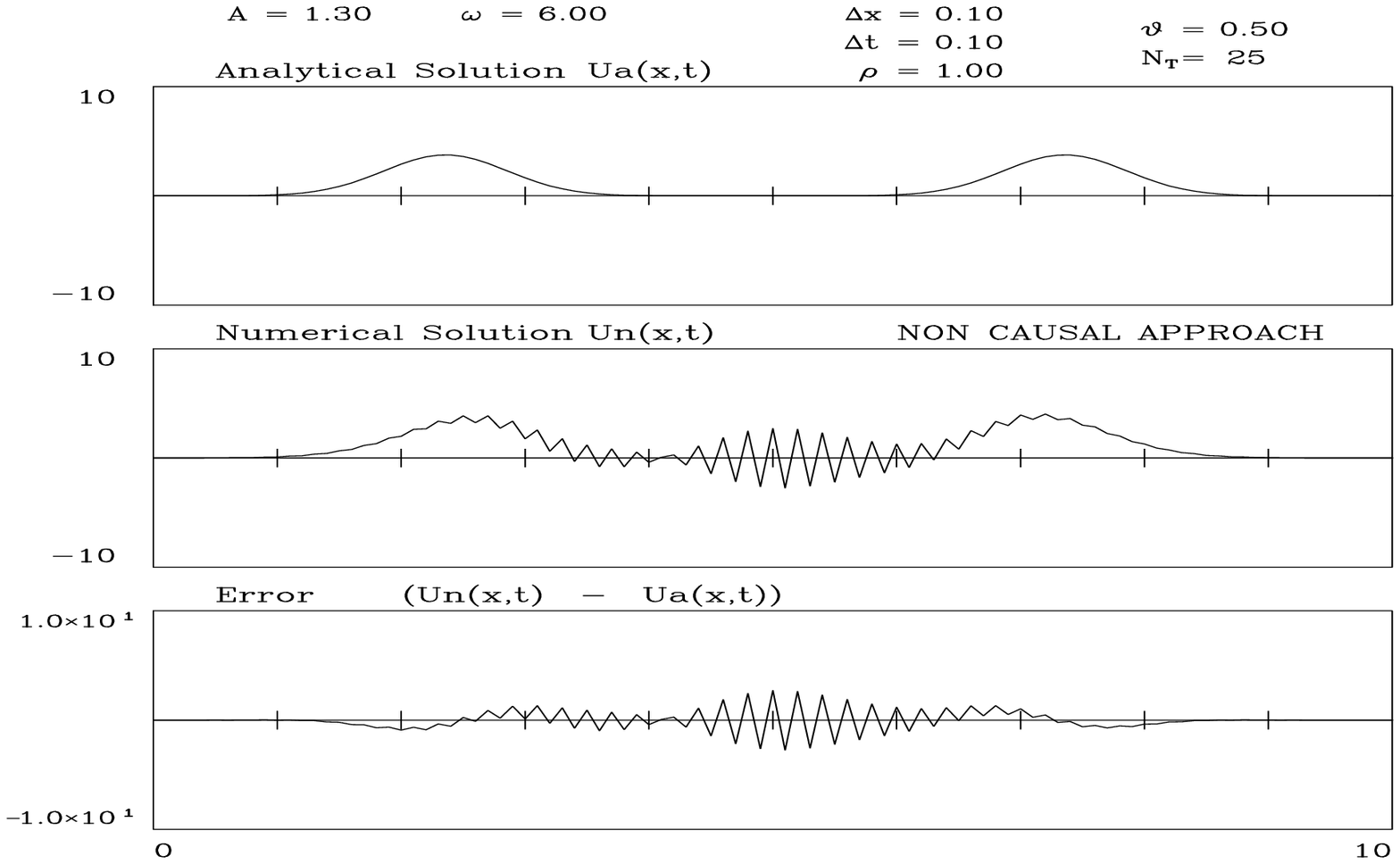,bbllx=0cm,bblly=-12.5cm,bburx=18cm,bbury=12cm,height=22cm,width=15cm}
\caption{Oscillating grid, causal reconnection.}
\label{fig:example:causal} \end{figure}

In \Figref{fig:example:causal} we compare the direct, non-causal,
approach  and the causal approach for a larger shift amplitude.  We use
the same initial Gaussian wave packet as before and take $ A \,=\,
1.3$, $\omega \,=\, 6$.  Here the instability appears very fast in the
direct approach (after only 25 time-steps).   With causal reconnection,
however, the calculation remains stable.  We have  carried out the same
calculation for many more time-steps, and also for larger  values of
\,$A$\, (up to \mbox{\,$A \,=\, 15$\,}), and the results are the same:
no  instabilities develop in the causal approach.

Therefore causal reconnection of the computational molecules seems to
cure  all the local instabilities on grids that move faster than the
waves.  We will see that in more than one dimension it will also
guarantee local stability  for rapid shifts, but only after we cure a
further instability that arises in operator-splitting methods for small
velocities.


\setcounter{equation}{0}

\section{The multi-dimensional case.}\label{sec:IV}

\subsection{How to design an ADI scheme for a hyperbolic
system.}\label{sec:4sub1}

\subsubsection{Fully implicit scheme.}\label{sec:fullimplicit}

We shall begin our discussion of stable integration schemes in more
than  one dimension by introducing ADI schemes in a way that makes our
time-symmetric ADI method emerge naturally, and which makes it clear
how to generalize it to other hyperbolic systems in  a straightforward
manner.  ADI is basically a device for implementing an implicit
integration scheme in many dimensions without the enormous
computational  overheads that the direct implicit scheme would
involve.  We begin our discussion,  therefore, with an examination of
the direct implicit scheme and its computational  demands.  We shall
concentrate on two dimensions, but the generalization to more  is
straightforward.

The general wave equation (\Eqref{eq:wave2}) in two dimensions is
\begin{eqnarray}  \left[ g^{x\,x} - \left( \beta^x \right)^2 \right] \,
\frac{\partial^2 \phi}{\partial x^2} + 2 \left( g^{x\,y} - \beta^x
\beta^y \right) \, \frac{\partial^2 \phi}{\partial x \, \partial y} +
\left[ g^{y\,y} - \left( \beta^y \right)^2 \right] \, \frac{\partial^2
\phi}{\partial y^2} & & \nonumber \\  {}+\frac{2 \beta^x}{c} \,
\frac{\partial^2 \phi}{\partial x \, \partial t} + \frac{2 \beta^y}{c}
\, \frac{\partial^2 \phi}{\partial y \, \partial t} - \Gamma^x \,
\frac{\partial \phi}{\partial x} - \Gamma^y \, \frac{\partial
\phi}{\partial y} - \frac{1}{c^2} \, \frac{\partial^2 \phi}{\partial
t^2} &=& 0 \, . \label{eq:wave2D}  \end{eqnarray}

In the finite difference approximations to this equation, we will use
the notation \begin{equation} \phi^j \,:=\, \phi^j_{i_x,i_y} \, ;
\end{equation} that is, we will suppress the spatial indices and write
them only when the possibility of confusion arises. The finite
difference approximations to all the differential operators that appear
in \Eqref{eq:wave2D} have the same form as in the one dimensional case,
except for a new term that  did not exist before: \begin{equation}
\left( \partial_x \partial_y \, \phi \right)^j \,=\, \frac{\delta_x
\delta_y \, \phi}{4 \, \left( \Delta x \right)^2} + E_{xy} \, ,
\end{equation} where the spatial differences are defined in the same
way as before, and the truncation error is: \begin{equation}  E_{xy}
\,=\, - \frac{1}{6} \, \left( \Delta x \right)^2 \, \left[ \left(
\partial^3_x \partial_y \, \phi \right)^j + \left( \partial_x
\partial^3_y \, \phi \right)^j \right] \, +\, \ldots\, .
\end{equation}

As we learned to do in the one-dimensional case, we will use only
explicit approximations for the first spatial derivatives.  We can then
write our second-order implicit finite difference approximation to
\Eqref{eq:wave2D} in the form: \begin{eqnarray}  \rho^2 \left[ g^{x\,x}
- \left( \beta^x \right)^2 \right] \left[ \frac{\theta}{2} \, \left(
\delta^2_x \, \phi^{j+1} + \delta^2_x \, \phi^{j-1} \right) + \left( 1
- \theta \right) \, \left( \delta^2_x \, \phi^j \right) \right] \, & &
\nonumber \\  {}+\rho^2 \left[ g^{y\,y} - \left( \beta^y \right)^2
\right] \left[ \frac{\theta}{2} \, \left( \delta^2_y \, \phi^{j+1} +
\delta^2_y \, \phi^{j-1} \right) + \left( 1 - \theta \right) \, \left(
\delta^2_y \, \phi^j \right) \right]  \,& & \nonumber \\
{}+\frac{\rho^2}{2} \left( g^{x\,y} - \beta^x \beta^y \right) \left(
\delta_x \delta_y \, \phi^j \right) \,+\, \frac{\rho \beta^x}{2} \,
\left( \delta_x \, \phi^{j+1} - \delta_x \, \phi^{j-1} \right) \,& &
\nonumber \\   {}+ \frac{\rho \beta^y}{2} \, \left( \delta_y \,
\phi^{j+1} - \delta_y \, \phi^{j-1} \right) \,-  \frac{\rho \, \left( c
\, \Delta t \right)}{2} \: \Gamma^x \, \left( \delta_x \, \phi^j
\right) & & \nonumber \\  {}-\, \frac{\rho \, \left( c \, \Delta t
\right)}{2} \: \Gamma^y \, \left( \delta_y \, \phi^j \right) \,-\,
\left( \phi^{j+1} - 2 \phi^j + \phi^{j-1} \right) &=& 0 \, .
\label{eq:finite2D} \end{eqnarray}

In this equation we have assumed for convenience that the spatial
increment is the same in both directions, and we have defined the
Courant parameter in the same way as before.  As in the one dimensional
case, to arrive at the last expression we have multiplied through by
\mbox{\,$\left( \Delta t \right)^2$\,}. The truncation error is
therefore of order: \begin{equation}
E_{\mbox{\scriptsize\ref{eq:finite2D}}} = {\cal O}  \left[ \left(
\Delta x \right)^2 \, \left( \Delta t \right)^2 \right] \,+\, {\cal O}
\left[ \left( \Delta t \right)^4 \right] \, .
\label{eq:finite2Derr}\end{equation}

\Eqref{eq:finite2D} is the most direct finite difference approximation
to the original differential equation in $2$ dimensions. We call it the
``fully implicit''  scheme.  As in the one-dimensional case, it takes
the form of a matrix equation  \begin{equation}\label{eqn:nDmatrix}
\hat{\cal Q}_2\phi^{j+1}=f(\phi^j,\,\phi^{j-1}). \end{equation}
However, as is well-known, the numerical solution of this equation is
considerably more time-consuming than in the one dimensional case, and
the  computational demands increase very rapidly with the number of
dimensions.   This is due to the fact that, if we have \,$N$\, grid
points in each of $n$ spatial  directions, the matrix $\hat{\cal Q}_n$
will have $ N^n$ rows and columns.   Most importantly, this matrix will
{\em not} be tridiagonal: it may be possible  to arrange that the
nearest neighbors in, say, the $x$-direction of any point  should
occupy adjacent columns, but those in other directions will be far
away  in another part of the matrix.  The matrix  will still be sparse,
but the number of operations involved in solving it may be very large
indeed, in the worst case involving of order $N^{3n}$ operations at
each time-step.  Even if a well-designed relaxation method is used, the
number of operations will in general increase faster than $N^{n}$.

ADI schemes offer a systematic way around this problem, usually
affording considerable savings  in computational effort with, as we
will show, no sacrifice in accuracy.  However, while the fully implicit
scheme may be expected to be as stable against grid shifts in $n$
dimensions as in one,  this is not true of ADI schemes, and we shall
have to be careful to design  a stable one.

\subsubsection{Designing ADI schemes: how to make the operator
factorizable.}

Alternating Direction Implicit (ADI) methods \cite{Richtmyer} reduce
the numerical work involved in an $n$-dimensional problem by modifying
the finite-difference scheme in such a way as to replace the original
large sparse matrix $\hat{\cal Q}_n$ by one that can be factored into a
product of tridiagonal matrices related to $\hat{\cal Q}_x$ for each
spatial direction.  If we assume that we have the same number \,$N$\,
of grid points in all directions, we will have to invert a series of
\,$N^{n-1}$\, tridiagonal matrices  of size \,$N \times N$\, for each
spatial dimension. This means that we will  need only \mbox{\,${\cal O}
( n N^n )$\,} operations to solve the system.  We see then that the
number of operations for the ADI scheme will scale  with the number of
grid points in the same way as it does for an explicit method.

The reason that one can contemplate replacing the original operator
$\hat{\cal Q}_n$ with a different one is that the fully implicit
finite-difference equation is only an approximation to the differential
equation, so if we modify it by adding extra high-order terms that are
of the same order as those neglected in the original approximation, the
accuracy of the scheme will  not be affected.  If we can then choose
these extra terms to change the  operator acting on the function at the
last time level \,$\phi^{j+1}$\, into a factorizable one, we  will have
speeded up the solution by a huge amount.

For our two-dimensional wave equation, the operator acting on
\,$\phi^{j+1}$\, is (see \Eqref{eq:finite2D}): \begin{eqnarray}
\hat{\cal Q}_2 &:=& -\, 1 \,+\, \frac{\rho}{2} \, \left( \beta^x \,
\delta_x + \beta^y \, \delta_y \right) \nonumber \\  & & {}+\rho^2 \,
\frac{\theta}{2} \, \left\{ \, \left[ g^{x\,x} - \left( \beta^x
\right)^2 \right] \,  \delta^2_x +  \left[ g^{y\,y} - \left( \beta^y
\right)^2 \right] \, \delta^2_y \, \right\} \, .  \label{operator}
\end{eqnarray} We want to add high-order terms to this expression to
transform it into a product of one-dimensional operators of the form of
the similar term we had in the one-dimensional case
(\Eqref{eq:finite1D} with  \mbox{\,$\theta_2 \,=\, 0$\,})
\footnote{This form of the factorized operator is not unique, there are
many different operators that one could choose instead of
\mbox{\,$\hat{\cal Q}_x \: \hat{\cal Q}_y$\,}. See for example
\cite{Strang(1968)} and \cite{Marchuck(1982)}.}:  \begin{eqnarray}
\hat{\cal Q}'_2&=& \hat{\cal Q}_x \:  \hat{\cal Q}_y \nonumber \\  &:=&
- \left\{ \, 1 \,-\, \frac{\rho \, \beta^x}{2} \, \delta_x \,-\, \rho^2
\, \frac{\theta}{2} \, \left[ g^{x\,x} - \left( \beta^x \right)^2
\right] \, \delta^2_x \, \right\} \nonumber \\  & & \quad{}\times
\left\{ \, 1 \,-\, \frac{\rho \, \beta^y}{2} \, \delta_y \,-\, \rho^2
\,  \frac{\theta}{2} \, \left[ g^{y\,y} - \left( \beta^y \right)^2
\right] \, \delta^2_y \, \right\} \, .
\label{eq:factorizable}\end{eqnarray}

Let us define $\hat{\cal S}$ to be the difference between these
operators:  \begin{equation}\label{eqn:sdef} \hat{\cal S} := \hat{\cal
Q}'_2 \,-\, \hat{\cal Q}_2. \end{equation} Then we can rearrange the
fully implicit equation (\Eqref{eqn:nDmatrix}) to read
\begin{equation}\label{eqn:rearranged} \hat{\cal
Q}'_2\phi^{j+1}=f(\phi^j,\,\phi^{j-1})+\hat{\cal S}\phi^{j+1}.
\end{equation} Now, this is not directly any help, since although we
have the factorizable  operator $\hat{\cal Q}'_2$ on the
left-hand-side, we have unknown terms in  $\phi^{j+1}$ on the right.
However, let us consider the following related equation:
\begin{equation}\label{eqn:related} \hat{\cal
Q}'_2\phi^{j+1}=f(\phi^j,\,\phi^{j-1})+\hat{\cal S}\phi^{j}.
\end{equation} This equation {\em is} in a form that can be solved
easily, since the unknown $\phi^{j+1}$ appears only with the operator
$\hat{\cal Q}'$.  Moreover,  since in the limit $\Delta t \rightarrow
0$ we have $\phi^j\rightarrow\phi^{j+1}$, in that limit
\Eqref{eqn:related} approaches \Eqref{eqn:rearranged}, which  is the
original fully implicit equation.  This fully implicit approximation is
itself only a valid approximation to the original differential equation
in the same limit, so it follows that \Eqref{eqn:related} also
approximates the differential equation in that limit.  (It need not be
as good an approximation, of course: the error terms of the original
\Eqref{eqn:rearranged} may be smaller than those introduced by the
change to \Eqref{eqn:related}.  We will address this point below.)

There is nothing unique about changing $\hat{\cal S}\phi^{j+1}$ to
$\hat{\cal S}\phi^{j}$ to make the equation factorizable.  One could
change $\hat{\cal S}\phi^{j}$ to any combination of terms that limits
to  $\hat{\cal S}\phi^{j+1}$ as $\Delta t\rightarrow 0$.  The different
ADI  methods make different choices of these terms.  We shall see that
the  standard choices produce equations that are very unstable when
the  grid shifts, but that by imposing the simple physical requirement
of  time-reversibility one gets a uniquely defined ADI scheme  that is
stable and just as accurate as the original fully implicit method.

It will be helpful to write out explicitly what the operator $\hat{\cal
S}$ defined in \Eqref{eqn:sdef} is: \begin{eqnarray}   \hat{\cal S} &=&
- \left[ \, \frac{\rho^2}{4} \, \beta^x \, \delta_x \left( \beta^y
\delta_y \right) \,+\, \frac{\rho^3 \theta}{4} \, \left[ \beta^x \,
\delta_x \left\{ \left[ g^{y\,y} - \left( \beta^y \right)^2 \right] \,
\delta_y^2 \right\} \,+\,  \left[ g^{x\,x} - \left( \beta^x \right)^2
\right] \, \delta_x^2  \left( \beta^y \delta_y \right) \right] \right.
\nonumber \\  & & \qquad{}+\,\left.\frac{\rho^4 \theta^2}{4} \, \left[
g^{x\,x} - \left( \beta^x \right)^2 \right] \, \delta_x^2  \left\{
\left[ g^{y\,y} - \left( \beta^y \right)^2 \right] \, \delta_y^2
\right\} \, \right] \, .  \label{eq:defS}\end{eqnarray}

It is important to note here that the expression for the operator
\,$\hat{\cal S}$\, includes terms in which the difference operators in
the \,$x$\, direction act on the functions \,$g^{yy}$\, and
\,$\beta^y$\,,  because these functions in general depend on both
\,$x$\, and \,$y$\,.   Had we defined \,$\hat{\cal Q}'_2$\, as
\,$\hat{\cal Q}_y \: \hat{\cal Q}_x$\,  we  would have had a different
expression for \,$\hat{\cal S}$\, , because,  whenever the coefficients
in the finite difference approximation change  with position, the
operators \,$\hat{\cal Q}_x$\, and \,$\hat{\cal Q}_y$\,  do not
commute.

\subsubsection{ADI schemes old and new.}

We first cast the two original, and still standard, ADI methods for
the  wave equation, Lees' first and second methods, into the notation
we  have used above.  They serve to illustrate how our approach to ADI
methods works on a familiar method, and we will subsequently analyze
the stability of these schemes.  Then we will introduce the scheme
that  will turn out to be stable for shifting grids, the time-symmetric
scheme.

\paragraph{Lees' first method.} The most straightforward approach is
that introduced by Lees in 1962  (\cite{Lees}, \cite{Fairweather}) for
the case of the ordinary wave equation on a fixed grid.  It is
convenient to describe it in terms of the extra terms that one adds to
the left-hand-side of \Eqref{eq:finite2D} to produce a factorizable
equation:  \begin{equation}\label{term:leesfirst} \hat{\cal S} \,
\left( \phi^{j+1} - \phi^{j-1} \right) \, . \end{equation} This
effectively produces the equation \begin{equation}\label{eqn:leesfirst}
\hat{\cal Q}'_2\phi^{j+1}=f(\phi^j,\,\phi^{j-1})+\hat{\cal
S}\phi^{j-1}. \end{equation} This is a simple change from
\Eqref{eqn:related}.

It is clear that as $\Delta t \rightarrow 0$ the extra term
(\ref{term:leesfirst}) will vanish,  and we will recover the original
differential equation. However, we will  see below that the extra terms
do not vanish as fast as the errors in  \Eqref{eq:finite2D}, which are
given in \Eqref{eq:finite2Derr}.   Lees' first method  is only
first-order accurate on a shifting grid.  It is known that Lees' first
method is  absolutely stable on a static grid.  However, it will turn
out to be subject  to weak but significant instabilities when used on
shifting grids.

\paragraph{Lees' second method.} Another way to modify the equation is
to add instead a second-time-difference term.  This is known as Lees'
second method:  \begin{equation} \label{term:leessecond} \hat{\cal S}
\, \left( \phi^{j+1} - 2\phi^j + \phi^{j-1} \right) \, .
\end{equation}

Here again we recover the original differential equation in the limit
$\Delta t \rightarrow 0$.  The result is in fact a linear combination
of Equations~(\ref{eqn:related}) and~(\ref{eqn:leesfirst}).  We will
see below that this method does not sacrifice  accuracy: the introduced
terms are of the same order as the original truncation  error on
shifting grids. Moreover, it is absolutely stable on a static grid.
However,  our stability analysis will reveal that this  method shows
strong instabilities when used on shifting grids.

\paragraph{The time-symmetric ADI method.} A more general  approach
would be to separate the operator \,$\hat{\cal S}$\, into different
pieces and use a first time difference with some of them, and a second
time difference with the rest.  We will call this a mixed ADI method.
One can try many different mixed methods, but there is one that is
natural from the point of view of the original  differential equation.
This is the one we call the time-symmetric ADI method.

The original differential equation, \Eqref{eq:wave2D}, has, in common
with all fundamental physics differential equations, the property of
time-reversal invariance.  In this case, the equation is invariant if
we make the replacements  \begin{equation}  t \longrightarrow \, -t
\qquad \mbox{and} \qquad \beta^i\longrightarrow\, -\beta^i.
\label{eq:timereverse} \end{equation} The fully implicit difference
equation, \Eqref{eq:finite2D}, also has this property, since the
approximations used for the time-derivatives are centered  differences:
they do not bias the direction of time.  In our case, time-reversal is
implemented by the exchange of the time-step indices $j+1$ and $j-1$.
However,   replacing the fully implicit operator $\hat{\cal Q}_2$ with
$\hat{\cal Q}'_2$ breaks this invariance, because this change modifies
the way that $\phi^{j+1}$ enters  the equation without automatically
modifying the $\phi^{j-1}$ terms in a symmetrical way.

If we look at the definition of the operator $\hat{\cal S}$ in
\Eqref{eq:defS}, we  see that it is not itself invariant: it contains
terms both linear and quadratic in  $\beta^i$.  Therefore, since Lees'
first and second schemes both add terms  in which $\hat{\cal S}$
operates on an expression with a definite time-symmetry,  neither
scheme is time-reversal invariant.  What we need to do is to separate
$\hat{\cal S}$ into parts $\hat{\cal S}_e$  and $\hat{\cal S}_o$ that
are even  and odd with respect to $\beta^i$ and then to allow them to
act, respectively, on even and odd extra terms.  That is, we add to
\Eqref{eq:finite2D}  the term \begin{equation}  \hat{\cal S}_e\,\left(
\phi^{j+1} -  2\phi^j + \phi^{j-1}\right) \, + \, \hat{\cal
S}_o\,\left( \phi^{j+1} - \phi^{j-1} \right)\, ,
\label{eq:tsADIdef}\end{equation} where the even and odd parts of the
operator $\hat{\cal S}$ are defined by \begin{eqnarray}  \hat{\cal S}_e
&:=&  - \, \frac{\rho^2}{4} \, \beta^x  \, \delta_x \left( \beta^y
\delta_y \right) \,-\, \frac{\rho^4 \theta^2}{4} \, \left[ g^{x\,x} -
\left( \beta^x \right)^2 \right] \,  \delta_x^2 \left\{ \left[ g^{y\,y}
- \left( \beta^y \right)^2 \right] \delta_y^2 \right\} \, ,\\ \hat{\cal
S}_o &:=& -  \frac{\rho^3 \theta}{4} \, \left[ \beta^x \, \delta_x
\left\{ \left[ g^{y\,y} - \left( \beta^y \right)^2 \right] \delta_y^2
\right\} \,+\, \left[ g^{x\,x} - \left( \beta^x \right)^2 \right] \,
\delta_x^2 \left( \beta^y \delta_y \right) \right]  \, .
\label{eq:defSeo}\end{eqnarray}

This effectively ensures that we apply the same modification to the
$\phi^{j-1}$  terms as to the $\phi^{j+1}$ terms in producing a
factorizable equation that limits  to the fully implicit equation as
the time-step goes to zero.  We will see that, by so  preserving the
time-symmetry of the original equation, we have also produced a method
that is just as accurate as the fully implicit  method and, perhaps
more importantly, is unconditionally locally stable  on grids shifting
at any speed up to the wave speed.

\subsubsection{Intermediate values and the implementation of ADI
schemes.}

Whichever ADI method we choose to use, we will always produce an
equation of the form: \begin{equation}  \hat{\cal Q}'_2 \, \phi^{j+1}
\,=\, \hat{\cal A} \, \phi^j \,+\,\hat{\cal B} \, \phi^{j-1} \, ,
\label{ADI:unsplit} \end{equation} where \,$\hat{\cal A}$\, and
\,$\hat{\cal B}$\, are spatial finite difference operators whose
specific form will depend on the method chosen.  Looking at the
definition of \,$\hat{\cal Q}'_2$\, in \Eqref{eq:factorizable},  we see
that the last equation can be decomposed into a system of two coupled
equations in the following way: \begin{eqnarray}   \left\{ \, 1 \,-\,
\frac{\rho \, \beta^y}{2} \, \delta_y \,-\, \rho^2 \, \frac{\theta}{2}
\, \left[ g^{y\,y} - \left( \beta^y \right)^2 \right] \, \delta^2_y \,
\right\} \, \phi^{j+1} \, &:=& \left. \phi^* \right.^{j+1}  ,
\\ \left\{ \, 1 \,-\, \frac{\rho \, \beta^x}{2} \, \delta_x \,-\,
\rho^2 \, \frac{\theta}{2} \, \left[ g^{x\,x} - \left( \beta^x
\right)^2 \right] \, \delta^2_x \, \right\} \, \left. \phi^*
\right.^{j+1} &=& \hat{\cal A} \, \phi^j \,+\, \hat{\cal B} \,
\phi^{j-1} , \end{eqnarray} where the first equation defines the
so-called {\em intermediate value}   ${\phi^*}^{j+1}$.

These two equations give the simplest ADI split of the finite
difference approximation.  In order to solve the system, one first
solves the second equation for \mbox{\,$\left. \phi^* \right.^{j+1}$\,}
using values of \,$\phi$\, in the previous time levels.  This operation
involves solving a tridiagonal system of equations for each fixed value
of the $y$-index.  One then solves for  \,$\phi^{j+1}$\, using the
first equation, again solving only tridiagonal equations.   In the
general case of an \,$n$\, dimensional problem, this procedure will
take  us to a system of \,$n\,$ equations and $n-1$ intermediate
values.  Each  equation employs an operator acting only in one of the
spatial directions.

It is important to realize that the splitting of
Equation~\ref{ADI:unsplit} given  above is by no means unique.  One may
find many different splittings of the same  equation, and some may
prove to be more computationally efficient than the one  we have given
above.  However, the differences will only be in the algebra (and  in
roundoff errors): different  splittings are only different ways of
writing the {\em same} ADI scheme.

\subsubsection{Accuracy of ADI methods.}

The different methods of forming \,$\hat{\cal A}$\, and \,$\hat{\cal
B}$\, will differ in general in their accuracy and stability.  To find
the accuracy of the different ADI schemes on shifting grids, we start
by considering Lees' first method.  In this case we must add to the
left hand side of \Eqref{eq:finite2D} the following term:
\begin{eqnarray} \hat{\cal S} \left( \phi^{j+1} - \phi^{j-1} \right)
&=& {}- \frac{\rho^2}{4} \, \beta^x \, \delta_x \left[ \beta^y \,
\delta_y  \, \left( \phi^{j+1} - \phi^{j-1} \right) \right] \nonumber
\\  & & {}- \frac{\rho^3 \theta}{4} \, \left\{ \beta^x  \, \delta_x
\left[ \left( g^{y\,y} - \left( \beta^y \right)^2 \right) \delta_y^2 \,
\left( \phi^{j+1} - \phi^{j-1} \right) \right] \right. \nonumber \\ & &
{} + \left.  \left( g^{x\,x} - \left( \beta^x \right)^2 \right) \,
\delta_x^2 \left[ \beta^y \, \delta_y \left( \phi^{j+1} - \phi^{j-1}
\right) \right] \right\} \nonumber \\  & & {}- \frac{\rho^4
\theta^2}{4} \, \left( g^{x\,x} - \left( \beta^x \right)^2 \right)
\delta_x^2 \left[ \left( g^{y\,y} - \left( \beta^y \right)^2 \right)
\delta_y^2  \left( \phi^{j+1} - \phi^{j-1} \right) \right] \, .
\end{eqnarray}

The order of this term is found by replacing differences with
derivatives:  \begin{eqnarray} \hat{\cal S} \, \left( \phi^{j+1} -
\phi^{j-1} \right) &\approx& {}-\,  2 c^2 \left( \Delta t \right)^3 \,
\beta^x \frac{\partial}{\partial x} \left[ \beta^y \, \frac{\partial^2
\phi}{\partial y \partial t} \right] \nonumber \\  & & {}-\, \theta c^3
\left( \Delta t \right)^4 \, \left\{ \, \beta^x \,
\frac{\partial}{\partial x} \left[ \left( g^{y\,y} - \left( \beta^y
\right)^2 \right) \, \frac{\partial^3 \phi}{\partial y^2 \partial t}
\right] \right. \nonumber \\  & & \left. {}+\, \left( g^{x\,x} - \left(
\beta^x \right)^2 \right) \frac{\partial^2}{\partial x^2} \left[
\beta^y \, \frac{\partial^2 \phi}{\partial y \partial t} \right] \,
\right\} \nonumber \\  & & {}-\, \frac{\theta^2}{2} c^4 \left( \Delta t
\right)^5 \, \left( g^{x\,x} - \left( \beta^x \right)^2 \right)
\frac{\partial^2}{\partial x^2} \left[ \left( g^{y\,y} - \left( \beta^y
\right)^2 \right) \, \frac{\partial^3 \phi}{\partial y^2 \partial t}
\right] \nonumber \\  & = & {\cal O} \left( \left( \Delta t \right)^3
\right) \,+\, {\cal O} \left( \left( \Delta t \right)^4 \right) \,+\,
{\cal O} \left( \left( \Delta t \right)^5 \right) \, . \end{eqnarray}

From this we can see that the principal part of the truncation error
introduced
by the new terms is of order \,$\left( \Delta t \right)^3$\,,  which is
in fact one order less than the original accuracy of
\Eqref{eq:finite2D}.  Using Lees' first ADI decomposition reduces the
accuracy of the original scheme.  This  is only true, however, when we
consider a moving grid.  From the last  expression it is clear that for
a fixed grid \mbox{\,$ ( \beta^x \,=\, \beta^y \,=\, 0 ) $\,},  the
truncation error introduced by this method will only be of  order
\,$\left( \Delta t \right)^5$\,, as is well known.

When we do the same analysis for the case of the ADI scheme based on
{\em Lees' second method}, we find that the principal part of the
truncation error introduced by the new terms is of order \,$\left(
\Delta t \right)^4$\, for a shifting  grid.  The accuracy of the
original equation is therefore preserved.  In principle, one would
therefore prefer Lees' second method.  However, we will see below that
the second method is far more unstable than the first when the grid
shifts, so its higher accuracy is of limited usefulness.

For the {\em time-symmetric ADI scheme}, the principal part of the
truncation error introduced by the extra terms is again of order
\,$\left( \Delta t \right)^4$\,.  This method is thus as accurate as
the fully implicit one.  We will see that it is also stable.

\subsection{Local stability analysis.}\label{sec:4sub2}

We turn now to the all-important question of the stability of the ADI
schemes that we have described in the last section.  In the same way as
in the one dimensional case, we will start by studying the nature of
the solutions of the differential equation (\Eqref{eq:wave2D}), and we
will again consider the solutions in a very small region around the
point \,$\left( x^i,t \right)$\,, assuming that the coefficients remain
constant in this region.

Moreover, for simplicity we will assume that we can take the functions
\,$g^{yy}$\, and \,$\beta^y$\, out of the difference operators in the
expression for \,$\hat{\cal S}$\, (\Eqref{eq:defSeo}).  Again, if these
functions change rapidly from one grid point to the next,  the accuracy
of the finite-difference scheme on this grid is probably poor anyway.

\subsubsection{Solutions of the differential equation.}

Following the same procedure as in the one-dimensional case, we will
look for solutions of the differential equation \Eqref{eq:wave2D} that
take the form:  \begin{equation}  \phi \, \left( x,t \right) \,=\,
e^{\imath \, \alpha (\vec{k}) \, t} \, e^{\imath \,  \vec{k} \cdot
\vec{x}} \, ,  \end{equation} where $\vec{k}$ is the 2-dimensional wave
vector.  We denote its  components by $k^i$, and define the associated
covector (one-form)  components $k_i$ by \begin{equation} k_i =g_{ij}
\, k^j.  \end{equation} Substituting into \Eqref{eq:wave2D} and solving
for $\alpha$, we find the  dispersion relation:  \begin{equation}
\alpha_\pm \,=\, c \, \left( k_j \beta^j \right) \,\pm\, c \left[
\left( k_j k^j \right) + i \left( k_j \Gamma^j \right) \right]^{1/2} \,
.  \label{eq:dispersion2D}\end{equation}  This is a simple
generalization of
\Eqref{eq:dispersion1D}.  Again we have  the analytic boundedness
condition \begin{equation} \left| e^{\imath\alpha_+t}
e^{\imath\alpha_-t}\right|^2 \,=\, 1 \, . \label{eq:normal}
\end{equation} Therefore, if one of the solutions is growing, the other
is dying at the same rate.

\subsubsection{Local stability of the difference equations.}

We will now proceed to the local stability analysis of the different
numerical approximations. We look for numerical solutions to the finite
difference approximations of the following form: \begin{equation}
\phi^m_{n_x n_y} \,=\, \psi^{m} \, e^{i \, \left( k_x n_x +k_y n_y
\right) \, \Delta x} \, , \end{equation} where we have used our
simplifying assumption that $\Delta y = \Delta x$.   By substituting
this equation into any of the finite difference approximations, we
shall always obtain a quadratic equation for \,$\psi$\, of the form:
\begin{equation} A \, \phi^2 + B \, \psi + C \,=\, 0 \, .
\label{eq:quadratic}\end{equation} The coefficients in this equation
will depend on the particular approximation used.  We call the two
solutions of this equation $\psi_\pm$.

As in the one-dimensional case, we define the numerical amplification
measure:  \begin{equation} M_{\rm Num} \,:=\, \max_{\vec{k}} \, \left(
\left| \psi_+ \right|^2, \left| \psi_- \right|^2 \right) \,,
\end{equation} and we take our local stability condition to be:
\begin{equation}  M_{\rm Num}\, \leq \, M_{\rm Ana} \, .
\label{eq:stability2D} \end{equation}

One general consideration applies to all difference schemes.     It is
not difficult to see that the analytic boundedness condition
\Eqref{eq:normal} will also hold in the finite difference case when the
following condition on the coefficients in  \Eqref{eq:quadratic} is
satisfied: \begin{equation}  |A |\,=\, |C| .  \label{complex}
\end{equation}

\paragraph{Lees' first scheme.} Lees showed that his methods are stable
for all time-steps if $\theta\ge1/2$, for a static grid.  In the
shifting case,   the coefficients of the quadratic equation for  Lees'
first method are:  \begin{eqnarray}  A &=& \left\{ 1 \,-\, i \rho
\beta^x \: \sin \left( k_x \Delta x \right) \,-\, \rho^2 \theta \,
\left[ g^{xx} - \left( \beta^x \right)^2 \right] \: \left[ \cos \left(
k_x \Delta x \right) - 1 \right] \right\} \nonumber \\  & &{}\times
\left\{ 1 \,-\, i \rho \beta^y \: \sin \left( k_y \Delta x \right)
\,-\, \rho^2 \theta \, \left[ g^{yy} - \left( \beta^y \right)^2 \right]
\: \left[ \cos \left( k_y \Delta x \right) - 1 \right] \right\} \, ,\\
\nonumber \\  B &=& -\left( 2 \rho^2 \left( 1 - \theta \right) \left\{
\left[ g^{xx} - \left( \beta^x \right)^2 \right] \left[ \cos \left( k_x
\Delta x \right) - 1 \right]   + \left[ g^{yy} - \left(
\beta^y\right)^2 \right] \left[ \cos \left( k_y \Delta x  \right) - 1
\right] \right\}\right. \nonumber \\  & & \quad{} \,-\, 2\rho^2 \left(
g^{xy} - \beta^x \beta^y \right) \: \sin  \left( k_x \Delta x \right)
\: \sin \left( k_y \Delta x \right)  \nonumber \\  & & \quad{}\,-\,
\left.  i \rho^2 \Delta x \, \left[\Gamma^x \: \sin  \left( k_x \Delta
x \right) \, +\, \Gamma^y \: \sin \left( k_y \Delta x\right) \right]
\,+\, 2 \right) \, ,\\  \nonumber \\  C &=& - \left( \left\{ 1\,-\, i
\rho \beta^x \: \sin \left( k_x \Delta x \right) \, -\, \rho^2 \theta
\, \left[g^{xx} - \left( \beta^x \right)^2 \right] \:  \left[ \cos
\left( k_x \Delta x \right) -1 \right] \right\} \right. \nonumber \\  &
& \qquad{}\times\left\{ 1 \,-\, i \rho\beta^y \: \sin \left( k_y \Delta
x \right) \, -\, \rho^2 \theta \, \left[ g^{yy} -\left( \beta^y
\right)^2 \right] \:  \left[ \cos \left( k_y \Delta x \right) -
1\right] \right\} \nonumber \\  & & \quad{}+ 2 \rho^2 \theta \, \left\{
\left[ g^{xx} \,-\, \left( \beta^x \right)^2 \right] \left[ \cos \left(
k_x \Delta x \right) \,-\, 1 \right] + \left[ g^{yy} \,-\,
\left(\beta^y \right)^2 \right] \left[ \cos \left( k_y \Delta x \right)
\,-\, 1 \right]\right\} \nonumber \\  & & \quad{} \,-\, \left.
\rule[0mm]{0mm}{4mm} \, 2 \right) \, .  \end{eqnarray} It is clear that
these coefficients do not satisfy the boundedness condition
\Eqref{complex}.

We shall test the stability condition given by \Eqref{eq:stability2D}
on Lees' first method by numerically calculating the amplification
measure from the roots of \Eqref{eq:quadratic}.  For simplicity, we
consider the case where:  \begin{equation}  \left. \begin{array}{rcccl}
g^{xx} &=& g^{yy}&=& 1 \, ,\\ g^{xy}&=&0 \, .&& \\  \Gamma^i & = & 0 \,
. && \end{array} \right\} \end{equation} We do not believe this
restricts the generality of our conclusions: from our analysis of the
one-dimensional case, the restriction on $\Gamma^i$ should not cause a
problem, and the particular values of the metric tensor are  unlikely
to have a determining effect on stability.

\begin{figure}
\psfig{file=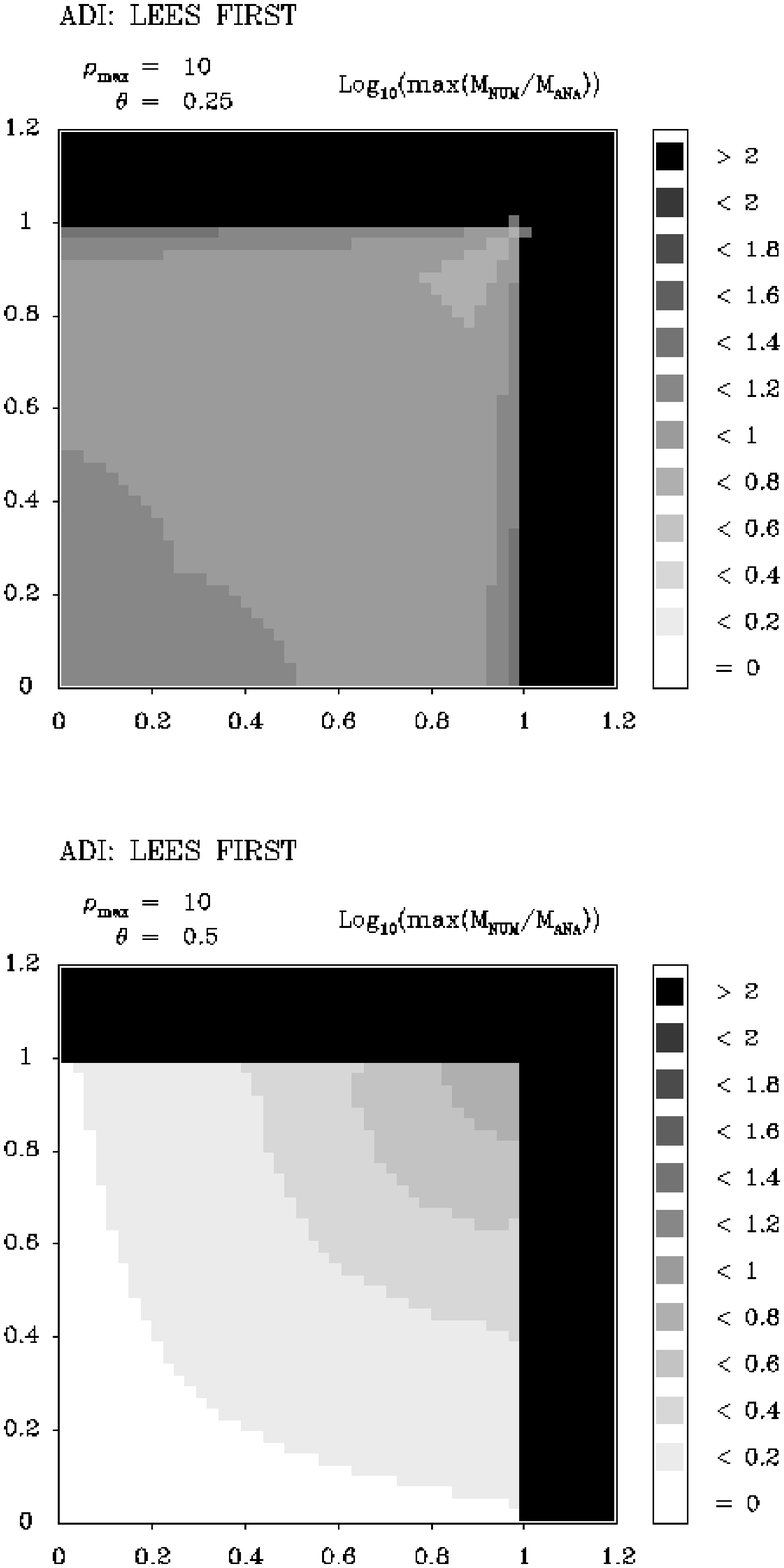,bbllx=-7cm,bblly=-4cm,bburx=15cm,bbury=30cm,height=22cm,width=15cm}
\caption{Stability for a method of Lees' first type.}
\label{fig:stability:lees1} \end{figure}

The results of our local stability analysis appear in
\Figref{fig:stability:lees1},  where we show the following region of
the shift vector space: \[ \beta^x, \,\beta^y \,\in\, \left( 0, 1.2
\right) \, . \] Since $\left| \vec{\beta} \right| \,=\, 1$ corresponds
to a grid shifting with a speed \,$c$\,, the region considered in the
graphs will include grids that shift faster than the waves. We have
considered \mbox{\,$50 \times 50 $\,} uniformly spaced values of the
shift vector  inside this region.  For a given point in the shift
vector space, we find the maximum value of the quantity: \[ R \,:=\,
\frac{M_{\rm Num}}{M_{\rm Ana}} \, , \] using \mbox{\,$10 \times 10$\,}
different values of the wave vector \,$\vec{k}$\,: \[ k_x,\,k_y \,\in\,
\left( 0, 2 \pi \right) \, , \] and, for each wave vector, \,$100$\,
different values of the Courant parameter \,$\rho$\, in the interval
\,$(0,10)$\,.

Having found \,$R_{\max}$\, we plot its value on a logarithmic scale.
In the graphs, values of \mbox{\,$\log_{10}(R_{\max})$\,} smaller than
or equal to \,$0$\, (\mbox{$R_{\max} \,\leq\, 1$}) are represented by
clear regions,  and values larger than \,$2$\, (\mbox{$R_{\max}
\,\geq\, 100$}) by the  darkest regions.  It is not difficult to see
that the clear regions will correspond  to values of the shift vector
for which the finite difference scheme is locally stable (at least for
\mbox{\,$\rho \,\in\, \left( 0, 10 \right) $\,}), and dark regions to
values of the shift that give rise to instabilities.  The darker the
region, the more violent the instabilities.  Finally, the solid arc in
the graphs marks the end of the light-cone.

It is important to note that the presence of a dark region does not
mean that for the given value of \,$!p vec{\beta}$\, the scheme will be
unstable for all \mbox{\,$\rho \,\in\, (0, 10)$\,}, but only that we
must expect instabilities for at least some values of the \,$\rho$\, in
that interval.

In the upper graph in \Figref{fig:stability:lees1}, we show the case
where \mbox{\,$\theta \,=\, 1/4$\,}.  The finite difference scheme is
unstable for at least some value of \,$\rho$\, at all values of the
shift vector.  This instability becomes much stronger whenever one of
the components of the shift vector is greater than the speed of the
waves. We recall that even in the one-dimensional case, the implicit
scheme for $\theta = 1/4$ is only conditionally stable, so the behavior
here is no surprise.  This figure also shows how the introduction of an
operator splitting has broken the rotational symmetry of our problem:
it is no longer the light-cone which is the important feature, but the
rectangular region in which the light-cone is inscribed.

The lower graph corresponds to the case \mbox{\,$\theta \,=\, 1/2$\,},
which is unconditionally stable in the one-dimensional case.  In two
dimensions, the scheme is still locally stable for values of the shift
vector along the direction  of the coordinate axis, just as the 1-D
scheme was.  However, instabilities  appear for speeds in other
directions. These instabilities are weak compared  to those for speeds
faster than the wave speed, but their presence will  nevertheless be
significant, as we will show in the examples of numerical  integrations
that we give  below.

We have looked at larger values of the parameter \,$\theta$\,, but the
situation doesn't improve beyond $\theta=1/2$.  Lees' first scheme is
therefore not very useful for grid speeds that are not aligned with the
coordinate axis.

\paragraph{Lees' second scheme.} We next turn  to Lees' second method,
for which the coefficients of the quadratic equation are:

\begin{eqnarray}  A &=& \left\{ 1 \,-\, i \rho \beta^x \: \sin \left(
k_x \Delta x \right) \,-\, \rho^2 \theta \, \left[ g^{xx} - \left(
\beta^x \right)^2 \right] \: \left[ \cos \left( k_x \Delta x \right) -
1 \right] \right\} \nonumber \\  & &{}\times\left\{ 1 \,-\, i \rho
\beta^y \: \sin \left( k_y \Delta x \right) \,-\, \rho^2 \theta \,
\left[ g^{yy} - \left( \beta^y \right)^2 \right] \: \left[ \cos \left(
k_y \Delta x \right) - 1 \right] \right\} , \\  \nonumber \\  B &=& -\,
\left( 2\rho^2 \left( 1 - \theta \right) \left\{ \left[ g^{xx} - \left(
\beta^x \right)^2 \right] \left[ \cos \left( k_x \Delta x \right) - 1
\right] + \,  \left[ g^{yy} - \left( \beta^y \right)^2 \right] \left[
\cos \left( k_y \Delta x \right) - 1 \right] \right\} \right.
\nonumber \\  & & \quad{}\,-\, 2 \rho^2 \, g^{xy} \: \sin \left( k_x
\Delta x \right) \: \sin \left( k_y \Delta x \right) -\, i \rho^2
\Delta x \, \left[ \Gamma^x \: \sin \left( k_x \Delta x \right) \,+\,
\Gamma^y \: \sin \left( k_y \Delta x \right) \right] \nonumber \\  & &
\quad{}+\, 2 i \rho^3 \theta \, \left\{ \beta^x \, \left[ g^{yy} -
\left( \beta^y \right)^2 \right] \: \sin \left( k_x \Delta x \right) \:
\left[ \cos \left( k_y \Delta x \right) - 1 \right] \right. \nonumber
\\  & &\quad \qquad\left. {}+\, \beta^y \, \left[ g^{xx} - \left(
\beta^x \right)^2 \right] \:  \sin \left( k_y \Delta x \right) \:
\left[ \cos  \left( k_x \Delta x \right) - 1 \right] \right\} \nonumber
\\ & & \quad{}+\, 2 \rho^4 \theta^2\left[ g^{xx} - \left( \beta^x
\right)^2 \right] \left[ g^{yy} - \left( \beta^y\right)^2 \right]
\left[ \cos \left( k_x \Delta x \right) - 1 \right] \left[ \cos \left(
k_{y} \Delta x \right) - 1 \right]\nonumber \\  & & \quad{}\left.
\,+\, 2 \rule[0mm]{0mm}{4mm} \, \right) , \\ \nonumber \\  C &=& -\,
\left( \rho^2 \theta \left\{ \left[ g^{xx} - \left( \beta^x \right)^2
\right] \left[ \cos \left( k_x \Delta x \right) - 1 \right] \,+\,
\left[ g^{yy} - \left( \beta^y \right)^2 \right] \left[ \cos \left( k_y
\Delta x \right) - 1 \right] \right\} \right. \nonumber \\  &
&\quad{}-\, i \rho \left[ \beta^x \: \sin \left( k_x \Delta x \right) +
\beta^y \: \sin \left( k_y \Delta x \right) \right] \,+\, \rho^2
\beta^x \beta^y \:  \sin \left( k_x \Delta x \right) \: \sin \left( k_y
\Delta x \right) \nonumber \\  & & \quad{}-\, i \rho^3 \theta\left\{
\beta^x \, \left[ g^{yy} - \left( \beta^y \right)^2 \right] \:
\sin\left( k_x \Delta x \right) \: \left[ \cos \left( k_y \Delta x
\right) - 1 \right] \right. \nonumber \\  & &\quad \qquad{}\left. +\,
\beta^y \, \left[ g^{xx} - \left( \beta^x\right)^2 \right] \:  \sin
\left( k_y \Delta x \right) \: \left[ \cos \left( k_x \Delta x\right) -
1 \right] \right\} \nonumber \\  & &\quad{} -\, \rho^4 \theta^2 \left[
g^{xx} - \left( \beta^x \right)^2 \right] \left[ g^{yy} - \left(
\beta^y \right)^2 \right] \left[ \cos \left( k_x \Delta x \right) - 1
\right] \left[ \cos \left( k_y \Delta x \right) - 1 \right]\nonumber
\\  & &\quad{} \left.  \,-\, 1 \, \rule[0mm]{0mm}{4mm} \right) .
\end{eqnarray}

\begin{figure}
\psfig{file=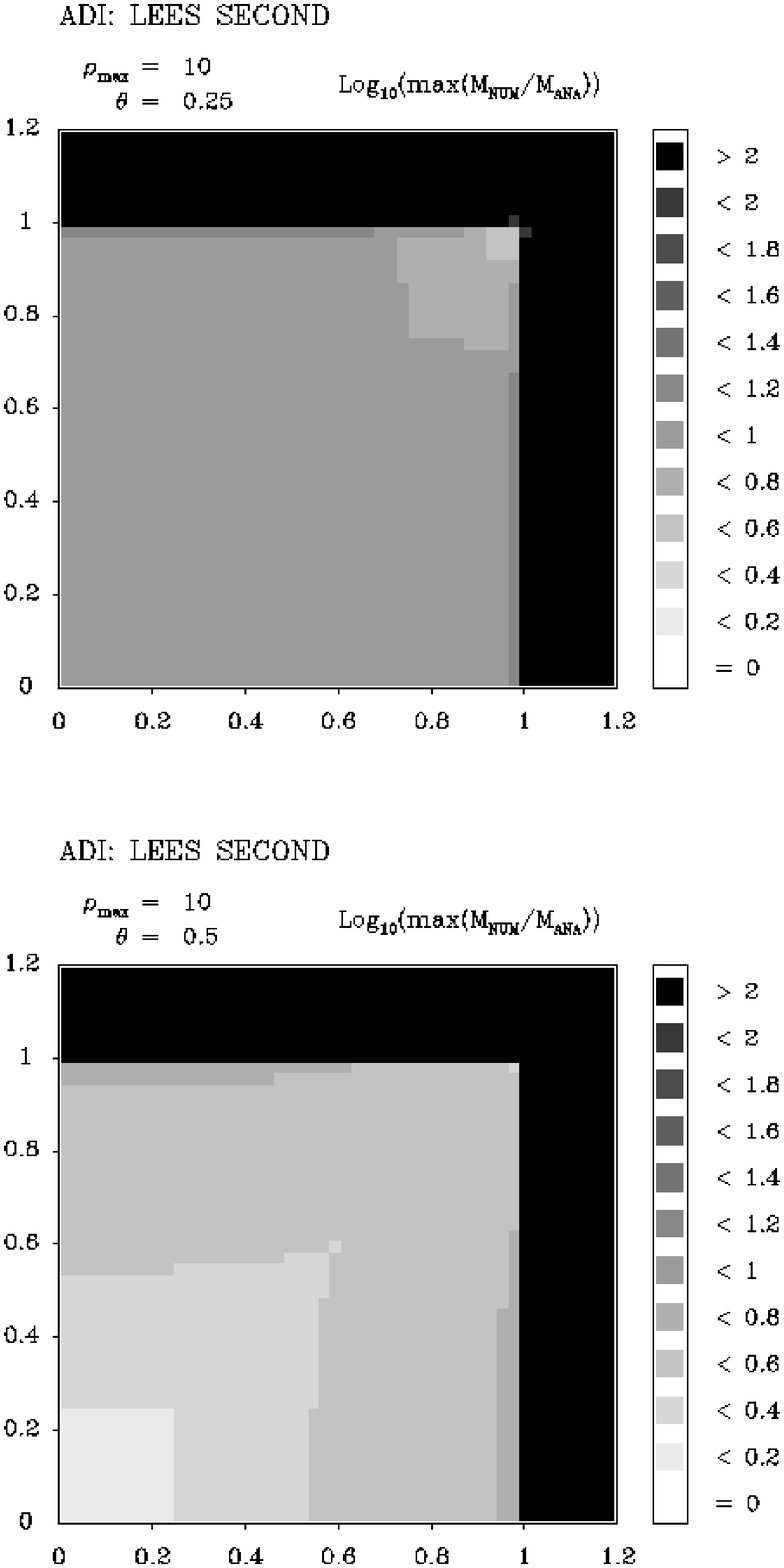,bbllx=-7cm,bblly=-4cm,bburx=15cm,bbury=30cm,height=22cm,width=15cm}
\caption{Stability for a method of Lees' second type.}
\label{fig:stability:lees2} \end{figure}

Again, these do not satisfy the boundedness condition \Eqref{complex}.
In \Figref{fig:stability:lees2} we again portray the cases for
\mbox{\,$\theta \,=\, 1/4$\,} and \mbox{\,$\theta \,=\, 1/2$\,} .   The
situation is even worse than before: the instabilities in Lees' second
scheme grow faster than for the first scheme, and even for
\,$\theta>1/2$\, there is only a very  small region of stability just
around the origin.  Clearly, this  scheme will not be practical for
any moving grid.

\paragraph{The time-symmetric scheme.} We have found that both standard
ADI methods become unstable  when the reference frame is moving.
Neither satisfies the symmetry condition \Eqref{complex}.  Now we look
in the same way at  the time-symmetric scheme.  The fact that this
scheme does indeed  satisfy \Eqref{complex} can readily be  seen from
the form that the coefficients of the quadratic equation take  in this
case:

\begin{eqnarray}  A &=& \left\{ 1 \,-\, i \rho \beta^x \: \sin \left(
k_x \Delta x \right) \,-\, \rho^2 \theta \, \left[ g^{xx} - \left(
\beta^x \right)^2 \right] \: \left[ \cos \left( k_x \Delta x \right) -
1 \right] \right\} \nonumber \\  & &\quad{}\times\left\{ 1 \,-\, i \rho
\beta^y \: \sin \left( k_y \Delta x \right) \,-\, \rho^2 \theta \,
\left[ g^{yy} - \left( \beta^y \right)^2 \right] \: \left([\cos \left(
k_y \Delta x \right) - 1 \right] \right\} \, , \\  \nonumber
\\ \nonumber \\ B &=& - \, \left(2 \rho^2 \left( 1 - \theta \right)
\left\{ \left[ g^{xx} -  \left( \beta^x \right)^2 \right] \left[ \cos
\left( k_x \Delta x \right) - 1 \right] + \left[g^{yy} - \left( \beta^y
\right)^2 \right] \left[ \cos \left( k_y \Delta x \right) - 1 \right]
\right\}\right. \nonumber \\  & & \quad {} \,-\, 2 \rho^2 \, g^{xy} \:
\sin  \left( k_x \Delta x \right) \: \sin \left( k_y \Delta x \right)
-\, i \rho^2  \Delta x \, \left[ \Gamma^x \: \sin \left( k_x \Delta x
\right) \,+\, \Gamma^y \: \sin \left( k_y \Delta x \right) \right]
\nonumber \\  & & \quad{} +\, 2 \rho^4\theta^2 \left[ g^{xx} - \left(
\beta^x \right)^2 \right]  \left[ g^{yy} - \left( \beta^y \right)^2
\right] \left[ \cos\left( k_x \Delta x \right) - 1 \right] \left[ \cos
\left( k_y \Delta x \right) - 1\right]\nonumber \\  & & \quad\left. {}
\,+\, 2 \rule[0mm]{0mm}{4mm} \, \right) \, , \\  \nonumber \\  C &=&
\left\{ 1\,+\, i \rho \beta^x \: \sin \left( k_x \Delta x \right) \,-\,
\rho^2 \theta \,  \left[ g^{xx} - \left( \beta^x \right)^2 \right] \:
\left[ \cos \left( k_x \Delta x \right) - 1 \right] \right\} \nonumber
\\  & &\quad {}\times \left\{ 1 \,+\, i \rho \beta^y \: \sin \left( k_y
\Delta x \right) \,-\, \rho^2 \theta \, \left[ g^{yy} - \left( \beta^y
\right)^2 \right] \: \left[ \cos \left( k_y \Delta x \right) - 1
\right] \right\} \, . \end{eqnarray}

\begin{figure}
\psfig{file=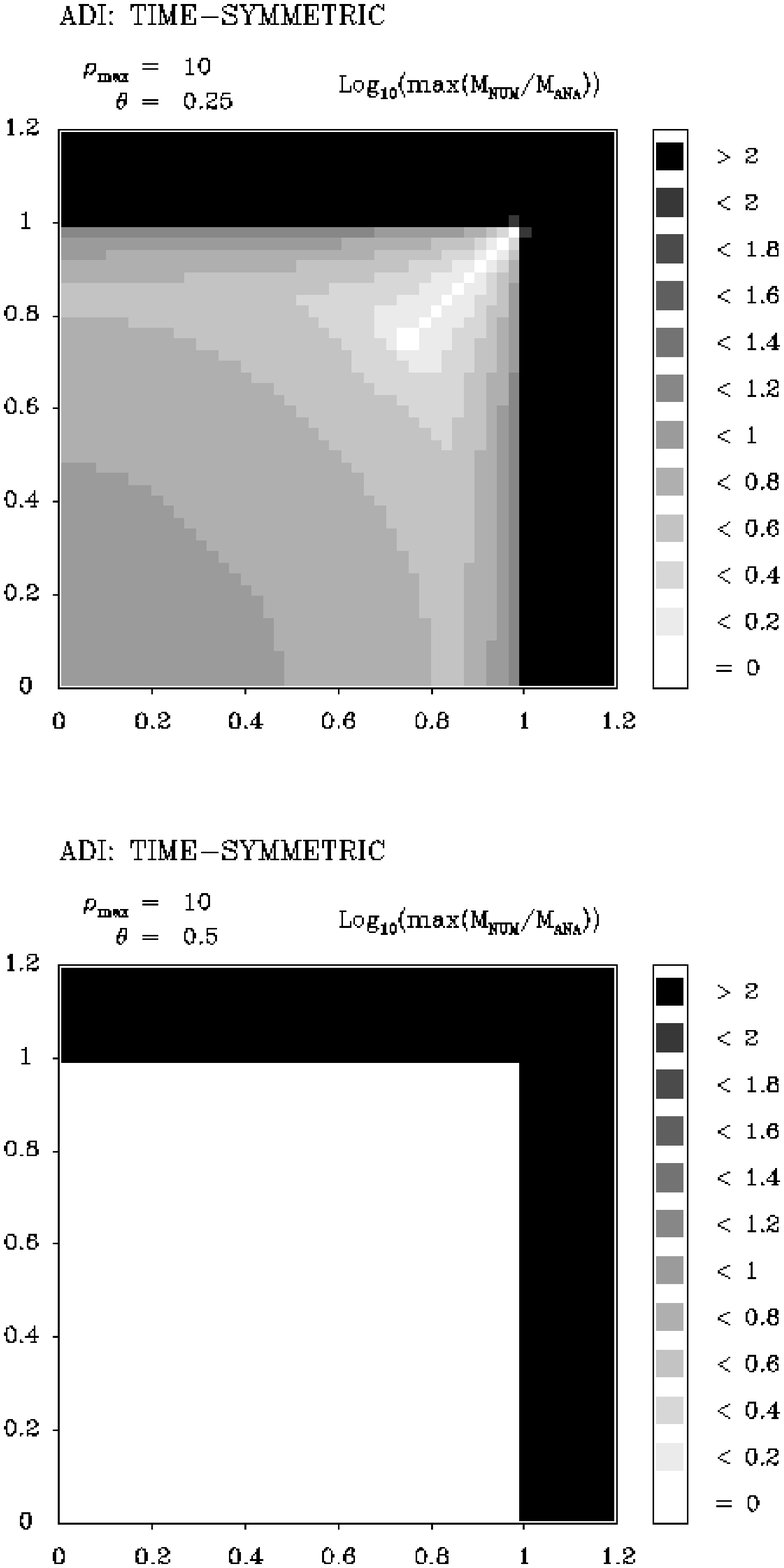,bbllx=-7cm,bblly=-4cm,bburx=15cm,bbury=30cm,height=22cm,width=15cm}
\caption{Stability for the time-symmetric scheme.}
\label{fig:stability:timesymm} \end{figure}

\Figref{fig:stability:timesymm} shows the local stability analysis for
this scheme, where again we show what happens for \mbox{\,$\theta \,=\,
1/4$\,} and \mbox{\,$\theta \,=\, 1/2$\,}.

For the first case, the situation is no better than before: the scheme
is unstable for practically every value of the shift vector.  However,
when we set $\theta = 1/2$, the value that gave absolute stability in
the  one-dimensional case, the scheme becomes locally stable {\em for
every value of the shift vector inside the rectangular region that
inscribes the light-cone}.  We find that this stability is maintained
for larger values of \,$\theta$\, .

What we see here is effectively a `light-cone stability condition',
except for the fact that instead of a cone we now have a rectangle, as
a  consequence of the fact that the ADI splitting breaks the
rotational  symmetry of the problem.  In the general case, this local
stability condition  can be expressed in the following way:
\begin{equation} g^{i\,i} \,-\, \left( \beta^i \right)^2 \, \geq \, 0
\qquad   (\rm no \:\: sum) \, , \label{eq:stability2D:causal}
\end{equation} where \,$i $ may refer to any spatial direction.

The time-symmetric scheme has also the important property that in the
stable region the numerical solutions will always be non-dissipative
(at least in the non-accelerating case), that is:  \[ max \left(
\frac{M_{\rm Num}}{M_{\rm Ana}} \right) \,=\, min \left( \frac{M_{\rm
Num}}{M_{\rm Ana}} \right) \,=\, 1 \]

This can be easily proved from the fact that this scheme satisfies
condition~\ref{complex}:  If the scheme has a dissipative solution with
\mbox{\,$M_{\rm Num} \,<\, M_{\rm Ana}$\,}, then
condition~\ref{complex} together with the analytic boundedness
condition (\Eqref{eq:normal}) implies that it must also have an
unstable solution with \mbox{\,$M_{\rm Num} \,>\, M_{\rm Ana}$\,}.

With the time-symmetric scheme we have then found what we are looking
for:  an absolutely locally stable, second-order accurate ADI
decomposition for the finite difference approximation to the wave
equation in a reference  frame moving at any speed up to the wave speed
in any direction.  This  scheme can be easily generalized to any number
of spatial dimensions, as can be seen from its definition in the last
section.

The restriction to frames moving slower than the wave speed is
expected, of course.  To remove it, we now  define  causal reconnection
for the 2-dimensional case, using the time-symmetric ADI scheme as our
starting point.

\subsection{Causal reconnection in 2 dimensions.} \label{sec:4sub3}

In the  last section we found that the time-symmetric ADI scheme had
stability properties superior to both the schemes of Less' first and
Lees'  second types.  However, even for the time-symmetric scheme,
instabilities   appear as soon as one of the components of the grid
speed becomes  larger than the  speed of the waves.  In the one
dimensional case, we saw that this instability could be avoided if we
used a computational molecule  based on the causal structure of the
wave equation, and not in the motion  of the individual grid points.
We now want to generalize this approach to the two dimensional case.

We will again look for a computational molecule that guarantees that
the light-cone is properly represented in the immediate vicinity of the
central point.  For the moment we will assume that we have already
found the points that form such a molecule.  We then introduce the
local coordinate system \,$\{\left. x^i \right.',t'\}$\, adapted to the
causal molecule as a direct generalization of the one dimensional
case:  \begin{equation}  \left.  \begin{array}{ll} \left. x^i \right.'
\,:=\, \left( x^i - x^i_{t_0} \right) - P^i \left( t \right) \, , \\
\\ t'\:\:\:\,:=\, t \, ,  \end{array} \right\} \end{equation}  where
\,$\{x^i_{t_0},t_0\}$\, are the coordinates of the central point of
the  molecule, and where \begin{eqnarray}   P^i \left( t \right) &=&
A^i \, \frac{\left( t - t_0\right)^2}{2} +  B^i \, \left( t - t_0
\right) +  x^i_0 \: , \\  A^i &:=& \left(\frac{x^i_{t_0 + \Delta t} - 2
x^i_{t_0} + x^i_{t_0 - \Delta t}} {\left( \Delta t \right)^2} \right)
\rule[0mm]{0mm}{8mm} \, , \\  B^i &:=& \left( \frac{x^i_{t_0 + \Delta
t} - x^i_{t_0 - \Delta t}}{2 \, \Delta t} \right) \, .  \end{eqnarray}

In the new coordinate system, the wave equation has the same form as
before, except for the substitutions: \begin{equation}  \beta^i
\,\longrightarrow\, \beta^i + \frac{A^i \, \left( t - t_0 \right) +
B^i}{c} \, , \hspace{20mm} \Gamma^i \,\longrightarrow\, \Gamma^i -
\frac{A^i}{c^2} \, .  \end{equation} Since in the finite difference
approximation the coefficients should be evaluated at the center of the
molecule, the above expressions will reduce to:  \begin{equation}
\beta^i \,\longrightarrow\, \beta^i + \frac{B^i}{c} \, , \hspace{20mm}
\Gamma^i \,\longrightarrow\, \Gamma^i - \frac{A^i}{c^2} \, .
\end{equation}

We know that the original finite difference approximation was locally
stable as long as \Eqref{eq:stability2D:causal} was satisfied.  This
implies that the approach based on the reconnected molecule will be
stable if \begin{equation} g^{i\,i} - \left( \beta^i + \frac{B^i}{c}
\right)^2 \, \geq \,0  \hspace{20mm} \forall i \, .  \end{equation}

As in the one-dimensional case, we will say that the given three points
form a proper causal molecule if the last condition is satisfied.

We will now define a generalization to two dimensions of the concept
of effective numerical light-cone.  We do this by defining first the
axis of  this numerical light-cone as the line: \begin{equation} x^i_a
\left( t \right) \,:=\, x^i_{t_0} \,-\, \beta^i c \, \left( t - t_0
\right) \,+\, \frac{1}{2} \, \Gamma^i c^2 \left( t - t_0 \right)^2 \, .
\end{equation} The numerical light-cone will then be defined by taking
at each time the  region covered by a rectangle that is centered at the
axis and has sides: \begin{equation} 2 c \, \Delta t \, \left( g^{i\,i}
\right)^{1/2} \, . \end{equation}

With this definition, the numerical ``light-cone'' is really a prism
and not a cone.  It is not difficult to prove now that if the points
\,$x^i_{t_0 - \Delta t}$\,  and \,$x^i_{t_0 + \Delta t}$\, are inside
the numerical light-cone of \,$x^i_{t_0}$\,, then we will have:
\begin{eqnarray} \left( \beta^i + \frac{B^i}{c} \right) &\in& \left[
-\left( g^{i\,i} \right)^{1/2} \,,\,  \left( g^{i\,i} \right)^{1/2}
\,\, \rule[-1mm]{0mm}{7mm} \right] \, , \\ \left( \Gamma^i -
\frac{A^i}{c^2} \right) &\in& \left[ -\frac{2 \left( g^{i\,i}
\right)^{1/2}} {\left( c \Delta t \right)}\,,\, \frac{2 \left( g^{i\,i}
\right)^{1/2}}{\left( c \Delta t \right)} \, \right] \, .
\end{eqnarray} This means that the three points do form a proper causal
molecule, and  also that the acceleration in the new local coordinates
will be bounded.

As in the one-dimensional case, we can guarantee that proper causal
molecules can be formed everywhere if we ask for two conditions:

\begin{enumerate}

\item {\em Every parent has at least two children.}  There must always
be at least one grid point in the upper and lower time levels inside
the numerical light-cones of all points in the middle time level.   It
is easy to see that this will require: \begin{equation} 2 \, \rho \,
\min \left( g^{i\,i} \right)^{1/2} \,\geq\, 1 \hspace{20mm} \forall i
\, . \label{reconnection1_2D} \end{equation}

\item {\em Every child has a parent.}   All the grid points in the
upper and lower time levels must be inside the  numerical light-cone of
at least one point in the middle time level.  We  guarantee this by
asking that the light-cones of the points in the middle time level
should cover completely the upper and lower time levels, in other words
that the union of the intersections of these light-cones with both the
upper and lower  levels should be the entire grid.

Let us consider a square of nearest neighbours in the middle time
level.  We want their numerical light-cones to cover the whole
quadrilateral area  defined by the points where the axis of those
light-cones intersect the adjacent  time levels.  A sufficient
condition for this to happen is to ask for the sides  of this
quadrilateral to be smaller than the spread of the smallest light-cone
divided by \,$\sqrt{2}$\, (this factor arises from the fact that the
diagonals of a  square are \,$\sqrt{2}$\, times larger than its
sides).

Following now the same procedure as before, we can show that this
condition takes the form: \begin{eqnarray} \frac{2}{\sqrt{2}} \,\, \rho
\, \min \left( g^{i\,i} \right)^{1/2} &\geq& \left\{ 1 \,+\, 2 \,  \max
\left( d_{11} , d_{22} \right) \,+\, \left[ \max \left( d_{11} , d_{12}
\right)  \rule[0mm]{0mm}{5mm} \right]^2 \right. \nonumber \\ && \left.
+\, \left[ \max \left( d_{21} , d_{22} \right) \rule[0mm]{0mm}{5mm}
\right]^2  \right\}^{1/2} \hspace{20mm} \forall i \, .
\label{reconnection2_2D} \end{eqnarray} where the quantities
\,$d_{jk}$\, are defined as: \begin{equation} d_{jk} \,=\, \rho \,
\left( \, \Delta x \, \left| \frac{\partial \beta^j}{\partial x^k}
\right| \,+ \, \frac{\left( \Delta x \right)^2}{2} \, \left|
\frac{\partial^2 \beta^j}{\partial x^1 \,  \partial x^2} \right| \:
\right) \,+\, \rho^2 \, \frac{\left( \Delta x \right)^2}{2} \, \left(
\frac{\partial \Gamma^j}{\partial x^k} \right) \end{equation} and the
maximum should be taken over all values of \,$x^i$\,.  As in the
one-dimensional case, the last condition is valid only to second order
in \,$\Delta x$\,.

\end{enumerate}

Conditions \ref{reconnection1_2D} and \ref{reconnection2_2D} are the
{\em causal reconnection conditions} in the two dimensional case.
They will guarantee that proper causal molecules can always be formed.

\subsection{Numerical examples.}\label{sec:4sub4}

To test the finite difference methods that we have developed, we will
consider two different situations:  a grid moving with a uniform speed,
and a grid rotating with constant angular velocity.

\subsubsection{Uniformly moving grid.}

We will first study the case of the grid moving with uniform speed, in
order to show the advantages of the time-symmetric scheme.  If the grid
is moving with velocity \mbox{\,$\vec{v} \,=\, \left( v^x,v^y
\right)$\,}, it is not difficult to see that: \begin{eqnarray}
g_{i\,j} &=& \delta_{i\,j} \\  \beta^i &=& v^i/c ,\quad\mbox{and} \\
\Gamma^x &=& \Gamma^y \,=\, 0 \, .  \end{eqnarray}

\begin{figure}
\psfig{file=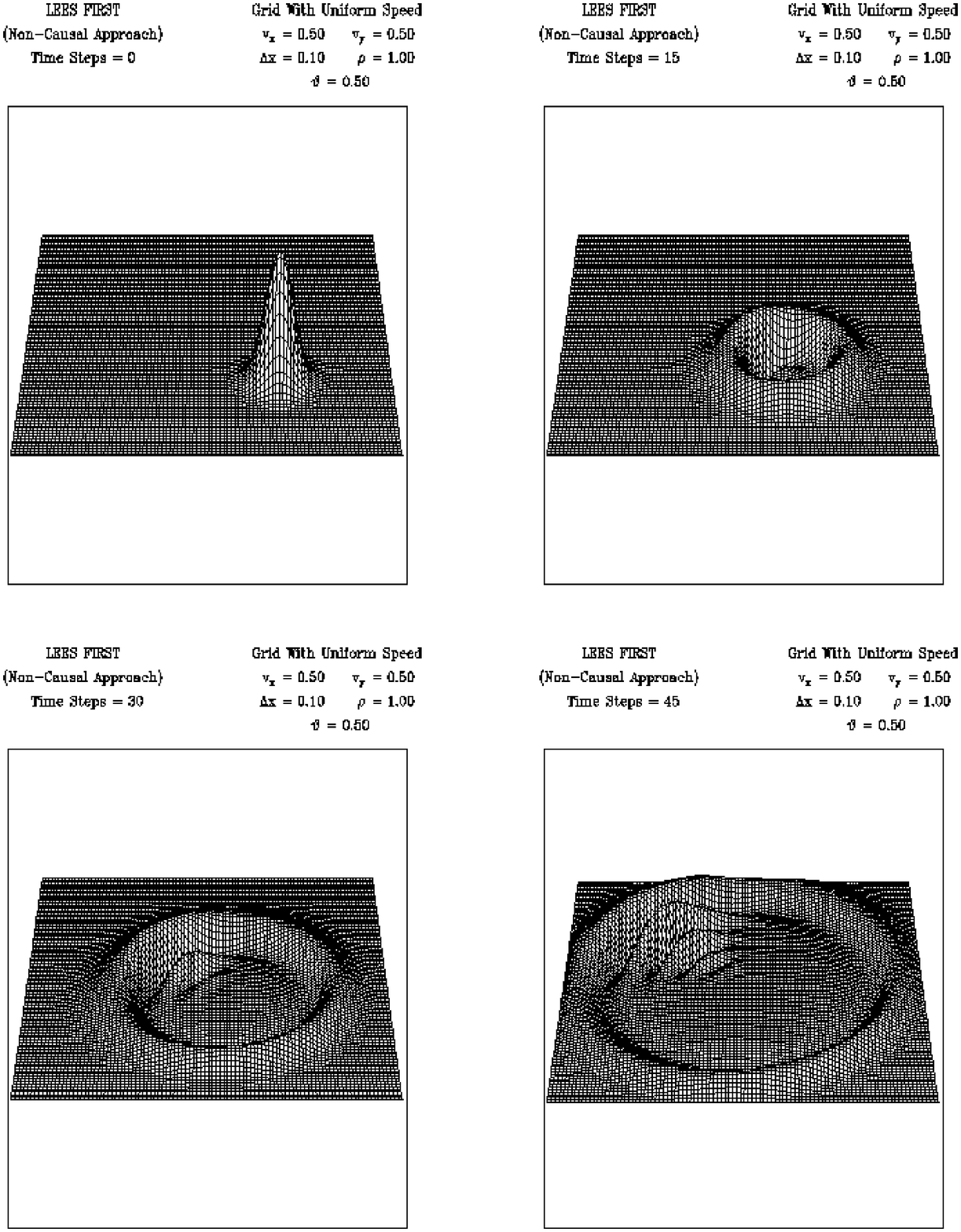,bbllx=-3cm,bblly=-3cm,bburx=21cm,bbury=28cm,height=20cm,width=16cm}
\caption{Uniform shift vector: Lees' first scheme.}
\label{fig:example:lees1}\end{figure}

Using these values for the coefficients, we have studied the numerical
solution to the wave equation for a number of examples, comparing the
three different ADI methods developed earlier.  The first set of graphs
(\Figref{fig:example:lees1}) show the result of one such calculation
for a scheme of Lees' first type.  In the graphs we show the grid
region \mbox{\,$[(0,10) \times (0,10)]$\,}, and we calculate the
evolution of a Gaussian wave packet originally at rest at the point
\,$(7,7)$\,.  For simplicity, we have imposed reflecting boundaries.
We have taken a time-step such that $\rho=1$, which means that we are
well beyond the Courant limit.\footnote{\, In a $n$ dimensional
problem, the Courant limit for the stability of an explicit scheme is
\mbox{\,$\rho \,=\, 1/\sqrt{n}$\,}.} The evolution is followed using a
grid with a speed given by: \[ \vec{v} \,=\,( \frac{1}{2} \, ,
\frac{1}{2} \, ), \qquad |\vec{v}| = 0.707 < 1,\] where we have taken
\mbox{\,$c \,=\, 1$\,}.

We see how an instability is beginning to grow even though the grid is
moving slower than the wave speed.  This is precisely in accordance
with the results of our local stability analysis.  This instability
grows slowly, as expected.  Nevertheless, it is clear that its presence
is unacceptable in a calculation of any  duration.  If we use a method
of Lee's second type, the instability takes longer to  develop, but
once it appears it grows very fast, much faster that with Lees' first
method.  The fact that the instabilities in general take longer to
appear with  Lees' second method can be traced to the particular wave
modes that are  involved.  As we can see in the graphs, the
instabilities in Lees' first method are  associated with relatively
long wavelengths (several grid points), and since these  modes are
already present in the initial data, they start growing right away.
In Lees' second method, however, the instabilities turn out to be
associated with  very short wavelengths (one or two grid points), which
do not contribute  significantly to the initial data.  These means
that, even though the instabilities  are more violent with this method,
it will take a long time for the unstable modes  to grow to the scale
of the real solution.

\begin{figure}
\psfig{file=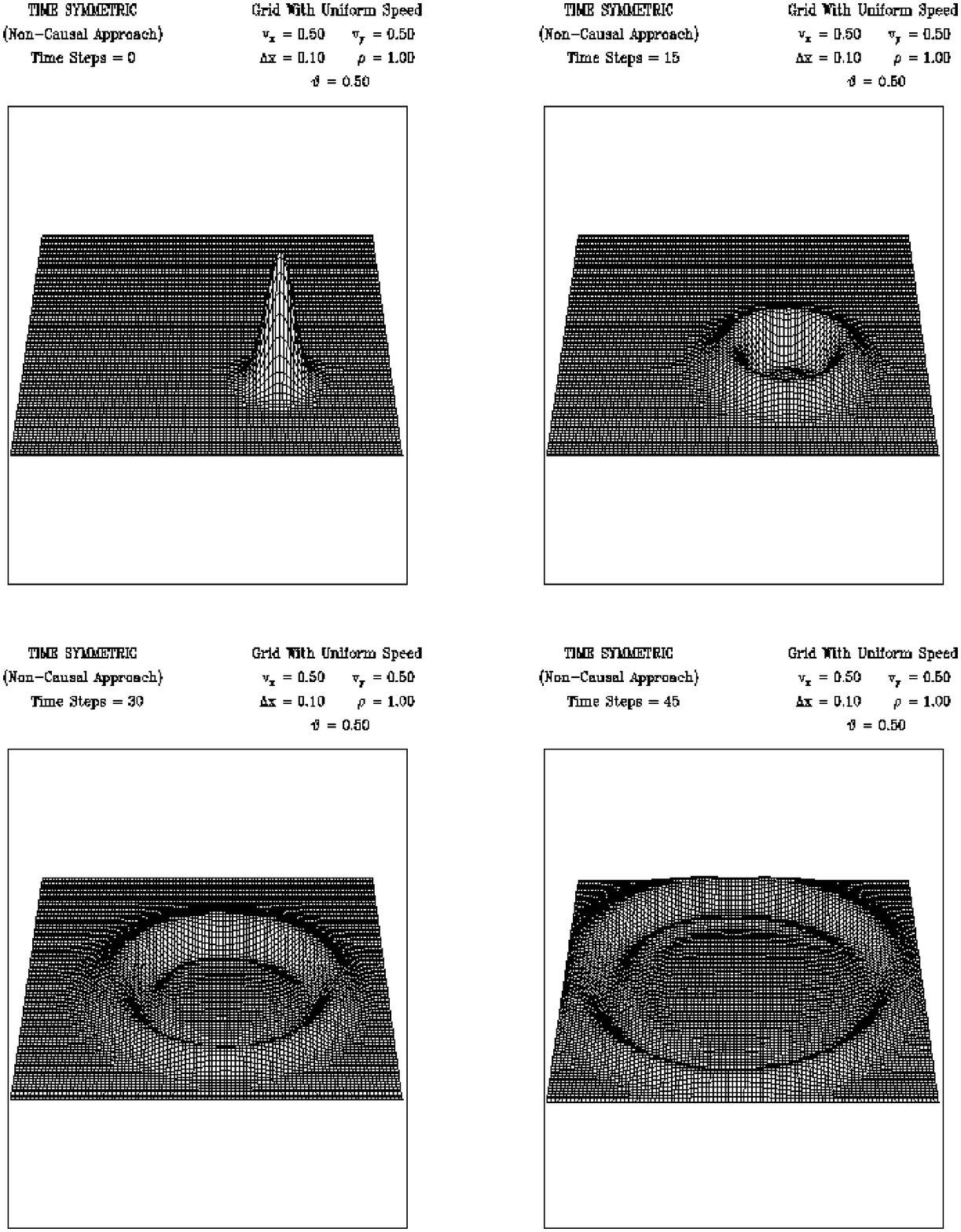,bbllx=-3cm,bblly=-3cm,bburx=21cm,bbury=28cm,height=20cm,width=16cm}
\caption{Uniform shift vector: time-symmetric scheme.}
\label{fig:example:timesymm} \end{figure}

In the next set of graphs (\Figref{fig:example:timesymm}) we have
applied the time-symmetric scheme to the same problem.   The
instability has completely disappeared.  This is again in agreement
with our previous conclusions, and shows the superiority of this
method.

We have performed similar calculations for many different values of the
grid speed and we have found the essentially similar results.  The
time-symmetric scheme remains stable as long as the grid moves slower
than the waves, while the other schemes present instabilities for quite
small grid velocities.

\subsubsection{Rotating grid.} 

In order to show the advantage of causal reconnection of the
computational molecules when the grid moves very fast, we will consider
now  an  example with a grid rotating with the constant angular
velocity  \,$\omega$\, .  It is not difficult to show that:
\begin{eqnarray}  g_{i\,j} &=& \delta_{i\,j} \\ \beta^x \,=\, -
\frac{\omega}{c}\, y \, , &\quad &  \beta^y \,=\,  \frac{\omega}{c} \,
x \, ,\\  \Gamma^x \,=\,- \frac{\omega^2}{c^2} \, x \, , &\quad &
\Gamma^y \,=\,  - \frac{\omega^2}{c^2}\, y \, .  \end{eqnarray}

To test for local stability when using  causal reconnection, we take
\[ \omega \,=\, 0.25 \, \]  in units in which $c=1$ and the grid
extends  over the range  \mbox{\,$[(-5,5) \times (-5,5)]$\,}.  This
means that the centers of the edges  of the grid will be moving faster
than the wave speed, with a  linear velocity  of $1.25$, while the
corners will be going even faster.

\begin{figure}
\begin{center}
\psfig{file=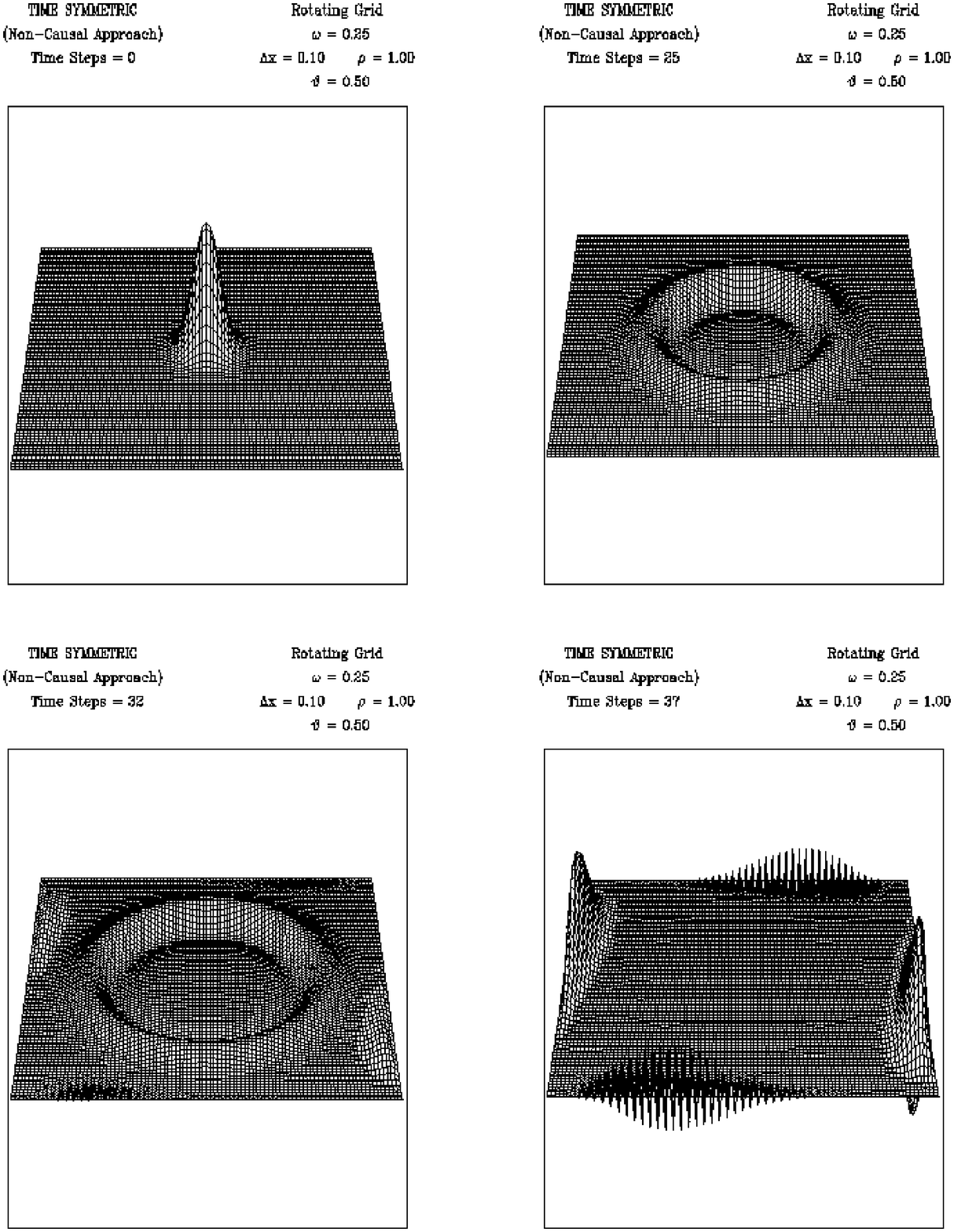,bbllx=-2cm,bblly=-5cm,bburx=22cm,bbury=26cm,height=20cm,width=16cm}
\end{center}
\caption{Rotating grid: non-causal approach.}
\label{fig:rotating:direct} \end{figure}

The next graphs show the results of a time-symmetric calculation, first
using a ``direct''  calculation (fixed computational molecule) and
second  using causal reconnection.  Again we use a time-step such that
$\rho=1$.  In \Figref{fig:rotating:direct}, we show the evolution of a
Gaussian wave packet originally at rest at the center of the grid,
using the direct approach.  We see how after \,$32$\, time-steps an
instability has appeared close to the boundaries.  Only  five
time-steps later, this instability has grown so large that the original
wave is no longer visible (the scale is automatically adjusted to
display the largest value of the function).

\begin{figure}
\psfig{file=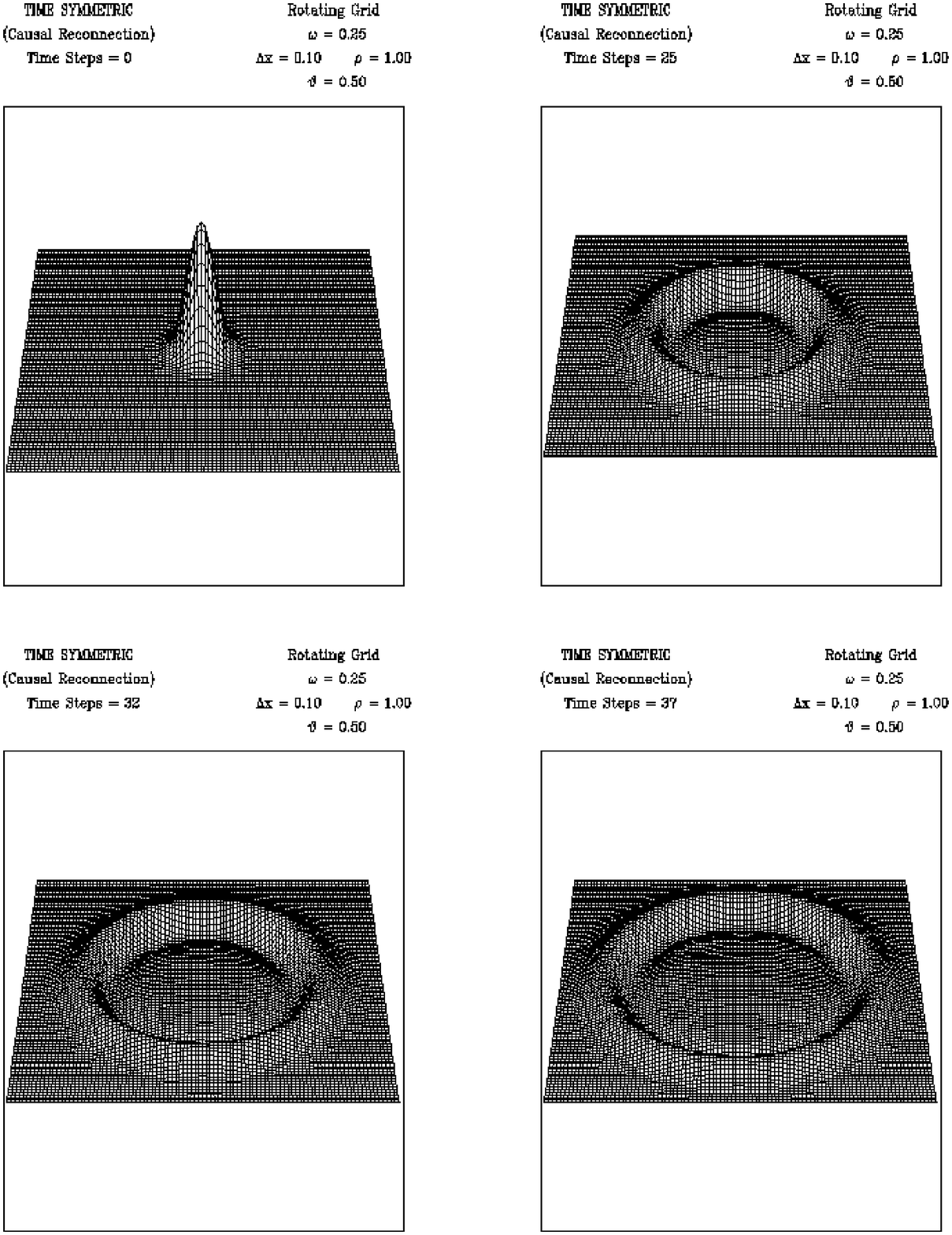,bbllx=-2cm,bblly=-5cm,bburx=22cm,bbury=26cm,height=20cm,width=16cm}
\caption{Rotating grid: causal reconnection.}
\label{fig:rotating:causal} \end{figure}

\Figref{fig:rotating:causal} shows the same calculation using causal
reconnection.  The instability is not present.  In fact, we have done
the same calculation with much larger values of the angular velocity
(up to \mbox{\,$\omega \,=\, 3.0$\,}, where the edge is travelling at
15 times the wave  speed), and the scheme remains locally stable.

These examples demonstrate dramatically that time-symmetric ADI can be
married with causal reconnection, and that together the two techniques
provide a robust difference approximation to the wave equation on a
moving grid. These methods are stable, offer all the computational
advantages of ADI schemes, and remain second order accurate in
\,$\Delta x$\, and \,$\Delta t$\,.

A comment on how to enforce causal reconnection at the boundaries seems
in order here.  In all our examples we have taken the practical
approach of setting the value of the wave function to zero whenever a
complete causal molecule cannot be formed.  This can happen not only at
the boundaries, but also at inner points close to the boundaries for
large enough grid speeds.  The philosophy behind this approach is
simple: If the causal molecule is incomplete, then we would need
information from outside the grid to evolve the wave function at that
place.  If we impose the condition that no information can come from
the outside, then we must take the value of the wave function as zero
at that point. This requirement can be relaxed somewhat be using an
outgoing wave boundary condition whenever we can still find a causally
related point in the previous time level, but not before that.  At
places where one can't even find a causally related point in the
previous time level, the only legitimate thing one can do is to set the
value of the wave function to zero.


\section{Conclusions.}\label{sec:V}

The wave equation we have studied here is a prototype for more complex
equations of mathematical physics, such as the Einstein field
equations.  In fact, many hyperbolic systems in mathematical physics
can be formulated in terms of the wave operator.  One would expect the
instabilities we have found here to be {\em generic:}  any numerical
approximation to a hyperbolic system on a shifting grid should exhibit
them.

Only experience will show us just how well our cures for these generic
instabilities transfer to more interesting equations.  However, the
instabilities we have described here are cured by the application of
two clear physical principles, causality and time-reflection
invariance. It seems clear that it would be asking for trouble {\em
not} to incorporate these principles into the design of algorithms for
the numerical integration of any fundamental physical equations.

We have, of course, studied in detail only one second-order
differential equation in one and two dimensions.  The restriction to
two dimensions is  not important.  The physical principles involved  do
not depend on the  number of dimensions, and the savings obtained by
using an ADI scheme  instead of a fully-implicit formulation increase
rapidly with the number of  dimensions. In many physical systems, it is
advantageous to formulate the  equations of the theory as first-order
differential equations.  This is true in  hydrodynamics and in many
studies of general relativity.  The general principles  of causality
and time-reflection invariance extend in a simple way to such
systems.   Time-symmetric ADI should prove relatively straightforward
to apply to more  complicated systems of equations, provided the
original differential equations  embody time symmetry.

Causality may be less straightforward in nonlinear equations, where the
structure  of the characteristic cone will depend on the solution, and
so the exact causal  relationships between time-levels cannot be
decided independently of solving  the equations.  However, causal
reconnection is implemented via an inequality:  one requires that grid
points should be {\em within} the characteristic cones of their
relatives at the previous time-step.  In most cases, one would hope
that the inequality  can be assured simply by extrapolation from the
behavior of the characteristic cones on the known time-steps.

An important area for the application of the techniques we have
developed here would be numerical fluid dynamics, where the study of
wave phenomena in supersonic flows is a natural place to expect
causality problems, and where the interest in three-dimensional
problems makes ADI essential in many cases.

In some restricted situations, it may be straightforward to apply these
techniques;  for example, a neutron star moving supersonically through
a grid in general relativity will, if treated in the standard way, use
acausal computational molecules.  Using  causal reconnection, adapted
to the characteristics of the fluid problem,  should prevent
instabilities of the type we have found here.

But the application of our techniques to more general problems in fluid
dynamics  will not be automatic.  Causal reconnection will have to be
generalized to deal  with hydrodynamic shocks.  At a shock, the regular
causal structure of the fluid  breaks down.  This does not mean that
causal reconnection cannot be implemented  there.  On the contrary, the
fact that a causal algorithm is constantly mapping the structure of the
characteristics means that it can be programmed  automatically to
locate and to identify shocks.

The idea of correctly representing the causal structure of the original
differential equation is not new, existing methods for handling  shocks
and related transport problems, such as upwind differencing
\cite{upwind}  and Godunov methods \cite{Godunov}, are already based on
the local structure of the characteristics of the fluid.  These ideas
have also been introduced in the numerical study of steady supersonic
flows, where the direction of flow behaves like a time coordinate and
the equations become hyperbolic.  Integration methods have been
developed that use retarded differences in the upstream direction to
maintain stability \cite{Murman(1971)}, \cite{Jameson(1974)}. All these
methods differ from causal reconnection in the fact that they keep
using only the nearest neighbours to build the computational
molecules.  We have recently become aware, however, of a paper by E.
Seidel and Wai-Mo Suen that introduces an idea they call ``causal
differencing'', that is very similar to our causal reconnection
\cite{Seidel(1992)}.

Fluid dynamics also presents special challenges to time-symmetric ADI.
The usual equations of inviscid gas dynamics are time-symmetric, but
the presence of viscosity or shocks introduces a fundamental
irreversibility into the problem.  We hope to treat the fluid dynamic
problem in a future paper.

We are confident, however, that the present techniques will generalize
easily to problems in numerical general relativity, such as that of the
motion of black holes  through fixed grids.  Causal reconnection should
allow the equations to remain stable  and causal.  Moreover,  the
computational advantages offered by ADI schemes, of  permitting stable
large time-steps (provided the physical situation allows such steps  to
remain accurate) while avoiding time-consuming sparse-matrix solutions,
can be  obtained without sacrificing accuracy or stability.  It is hard
now to imagine any  situation in numerical integrations of the vacuum
field equations of general relativity where one would use implicit
methods without employing time-symmetric ADI.

\renewcommand{\thesection}{}


\renewcommand{\theequation}{A.\arabic{equation}}

\setcounter{equation}{0}

\section{\hspace{-9mm} Appendix A.   Derivation of Wave Equation  on a
Shifting Grid.}\label{sec:appA}

In this appendix we will sketch the derivation of \Eqref{eq:wave2} by
making use of elegant tensorial techniques.  There are many alternative
approaches, of course, and a reader unfamiliar with tensors can obtain
the same result in a straightforward, but rather long, way simply by
making the following general change of variables in the original wave
equation from the physical (inertial) coordinates $\{\xi^\mu\}$ to the
computational coordinates $\{x^\alpha\}$:  \begin{equation}  x^i \,=\,
x^i(\xi^\mu) \, , \qquad  x^0 \,=\, \xi^0 \, , \end{equation} with the
associated change of derivatives \begin{equation}
\frac{\partial}{\partial \xi^i} \,=\, \frac{\partial x^k}{\partial
\xi^i} \, \frac{\partial} {\partial x^k} \, ,
\qquad\frac{\partial}{\partial \xi^0} \,=\, \frac{\partial x^k}
{\partial \xi^0} \, \frac{\partial}{\partial x^k} \,+\,
\frac{\partial}{\partial x^0} \, . \end{equation}

The functions that we have identified as the shift vector
\,$\beta^i$\,, the spatial metric \,$g_{\,i\,j}$\, and the
\,$\Gamma^i$\, coefficients in Equations~(\ref{eq:shift1}),
(\ref{eq:3metric}) and~(\ref{eq:gamma}) come out as part of the
algebra.    A reader who wants an introduction to the use of tensors in
mathematical  physics is invited to consult reference
\cite{geommeths}.

We will start from the expression of the wave equation in a general
coordinate system: \begin{equation}   \Box^2 \, \phi \,=\,
(\gamma^{\mu\nu} \,\phi_{;\,\mu})_{;\,\nu} \,=\, 0 \, ,  \end{equation}
where the semicolon stands for covariant derivative.  Using the
explicit expression for the covariant derivatives, the last equation
takes the form: \begin{equation} \gamma^{\mu\nu} \, \frac{\partial^2
\phi}{\partial x^\mu \,\partial x^\nu} - \Gamma^\lambda \,
\frac{\partial \phi}{\partial x^\lambda} =0 \, ,
\label{eq:wave1}\end{equation} where the coefficients
\,$\Gamma^\lambda$\, are defined in terms of the Christoffel symbols
as: \begin{equation} \Gamma^\lambda \,:=\, \gamma^{\mu\nu} \,
\Gamma^\lambda_{\mu\nu} \, .  \end{equation}

Our first task is, then, to find the inverse metric $\gamma^{\mu\nu}$.
The metric in the new coordinates is given by

\begin{equation}   \gamma_{\mu\nu} \,=\, \frac{\partial
\xi^\alpha}{\partial x^\mu} \frac{\partial \xi^\beta}{\partial x^\nu}
\: \: \eta_{\alpha \beta}  \, , \label{metric:def}\end{equation} with
$\eta_{\alpha\beta}$ the Minkowski metric tensor: \begin{equation}
\left.  \begin{array}{lll} \eta_{00} \:=\, -1 \, , \\  \eta_{\,i\,i}
\:=\, \:\: 1 \, \hspace{12mm} ( {\rm no \hspace{2mm} sum}) \, ,\\
\eta_{\mu\nu} \,=\, \:\: 0 \hspace{15mm} \mu \,\neq\, \nu \, .
\end{array}\right\}  \end{equation}

We now note that for a line of constant \,$\{\xi^i\}$\, we have  \[
0=\left.  \frac{d\xi^i}{dt} \right|_{\{\xi^i\}} =\, \frac{\partial
\xi^i}{\partial x^j} \left .\frac{d x^j}{d t} \right|_{\{\xi^i\}} +
\frac{\partial \xi^i}{\partial t} \, ,\]  which implies
\begin{equation}  \frac{\partial \xi^i}{\partial t} \,=\, -
\frac{\partial\xi^i}{\partial x^j} \left.  \frac{d x^j}{d t}
\right|_{\{\xi^i\}} \, . \end{equation} Using now the definition of the
shift vector (\Eqref{eq:shift1}) and writing \,$x^0 \,=\, ct$\, we
find:  \begin{equation}  \frac{\partial \xi^i}{\partial x^0} \,=\,
\beta^j \,\frac{\partial \xi^i}{\partial x^j}  \label{shift:2} \, .
\end{equation} This is an important relation, and we will use it to
rewrite the metric coefficients given by \Eqref{metric:def}.

For the mixed components in space and time of \,$\gamma_{\mu \nu}$\, we
find:  \begin{eqnarray*} \gamma_{\,0\,i} \,=\, \gamma_{\,i\,0} &=&
\sum_{l=1}^{n} \, \frac{\partial \xi^l}{\partial x^i} \, \frac{\partial
\xi^l}{\partial x^0} \\ &=&\, \beta^j \, \sum_{l=1}^{n}
\,\frac{\partial \xi^l}{\partial x^i} \, \frac{\partial \xi^l}{\partial
x^j} \, , \end{eqnarray*} and finally:  \begin{equation}  \gamma_{0\,i}
\,=\, \gamma_{\,i\,0} \,=\, \gamma_{\,i\,j} \,\beta^j \,=\, g_{\,i\,j}
\, \beta^j \, .  \end{equation}

In a similar way we can find the coefficient \,$\gamma_{00}$\,:
\begin{eqnarray*} \gamma_{00} \,&=&\, - \left( \frac{\partial
\xi^0}{\partial x^0} \right)^2  + \sum_{l=1}^{n} \, \left(
\frac{\partial \xi^l}{\partial x^0} \right)^2 \\ &=&\, -1 +
\sum_{l=1}^{n} \, \beta^j \, \frac{\partial \xi^l}{\partial x^j} \,
\beta^i \, \frac{\partial \xi^l}{\partial x^i} \, , \end{eqnarray*} and
from this we obtain: \begin{equation}  \gamma_{00} \,=\, -1 +
g_{\,i\,j} \, \beta^i \beta^j \, . \end{equation}

We will adopt the convention that the indices of the shift vector can
be raised and lowered by using only the spatial metric:
\begin{equation}  \beta_{\,i} \,=\, g_{\,i\,j} \, \beta^j \, ,
\hspace{10mm} \beta^i \,=\, g^{i\,j} \, \beta_{\,j} \, ,
\end{equation} where \,$g^{i\,j}$\, are the coefficients of the inverse
of the spatial metric matrix \,$g_{\,i\,j}$\,.

The coefficients of \,$\gamma_{\mu\nu}$\, can now be written as:
\begin{equation}  \gamma_{\mu\nu} \,=\,  \left(  \begin{array}{rr} (-1
+ \beta_i \beta^i) & \beta_k \\ \vspace{2mm} \\ \beta_j \hspace{9mm} &
g_{j\,k} \end{array} \right) \, .  \end{equation} Using the last
expression it is not difficult to see that the coefficients of the
inverse metric \,$\gamma^{\mu \nu}$\, will be given by:
\begin{equation}   \gamma^{\mu\nu} \,=\,  \left( \begin{array}{ll} -1 &
\hspace{9mm} \beta^k \\ \vspace{2mm} \\  \:\: \beta^j & (g^{j\,k} -
\beta^j \beta^k) \end{array} \right) \, .  \label{eq:4metric:inverse}
\end{equation} Having found \,$\gamma^{\mu\nu}$, we will now look for
an expression for the coefficients \,$\Gamma^i$\,.  Since the original
coordinates $\{\xi^\alpha\}$ define an inertial reference frame, the
Christoffel symbols can be expressed in terms of their transformation
to the general coordinates: \begin{equation}   \Gamma^\lambda_{\mu\nu}
\,=\, \frac{\partial x^\lambda}{\partial \xi^\alpha} \,
\frac{\partial^2 \xi^\alpha}{\partial x^\mu \, \partial x^\nu} \, .
\end{equation} From the last expression it is easy to see that:
\begin{equation}   \Gamma^0_{\mu\nu} \,=\, 0 \, ,  \end{equation}
which in turn means:  \begin{equation}   \Gamma^0 \,=\, 0 \, .
\end{equation} On the other hand, from the general expression for the
Christoffel symbols: \begin{equation}  \Gamma^\lambda_{\mu \nu} \,:=\,
\frac{1}{2} \, \gamma^{\lambda \alpha} \left[ \frac{\partial
\gamma_{\mu \alpha}}{\partial x^{\nu}} \,+\, \frac{\partial \gamma_{\nu
\alpha}}{\partial x^{\mu}} \,-\, \frac{\partial \gamma_{\mu
\nu}}{\partial x^{\alpha}} \right] \, ,  \end{equation} it is not
difficult to show that: \begin{equation}  \Gamma^i \,:=\,
\gamma^{\mu\nu} \, \Gamma^i_{\mu \nu} \,=\,
-\frac{1}{\sqrt{g}}\left\{\frac{\partial}{\partial t}
\left(\sqrt{g}\beta^i\right)+\frac{\partial}{\partial x^j}\left[
\sqrt{g}\left(g^{ij}-\beta^i\beta^j\right)\right]\right\}.
\end{equation}

Using the previous results, we can finally rewrite \Eqref{eq:wave1} in
the following way: \begin{equation}   (g^{i\,k} - \beta^i \beta^k) \,
\frac{\partial^2 \phi}{\partial x^i \, \partial x^k} + \frac{2
\beta^i}{c} \, \frac{\partial^2 \phi}{\partial x^i \, \partial t} -
\Gamma^i \, \frac{\partial \phi}{\partial x^i} - \frac{1}{c^2} \,
\frac{\partial^2 \phi}{\partial t^2} \,= \, 0 \, . \end{equation} This
is the final form of the wave equation in the coordinate system adapted
to the motion of the grid.


\renewcommand{\theequation}{B.\arabic{equation}}

\setcounter{equation}{0}

\section{\hspace{-9mm} Appendix B. Implementation of Causal
Reconnection.}\label{sec:appB}

In this appendix, we discuss one algorithm that  determines the
positions of the points that form the causal computational molecules.
We will consider the case of an arbitrary number of spatial dimensions
\,$n$\,.  The particular cases of one and two dimensions can then be
found in a straightforward way.

There are many different ways of finding the closest causally connected
grid points.  For example, if it is possible to find the transformation
of coordinates that takes us back to the original inertial reference
frame \begin{equation} \xi^i \,=\, \xi^i \left( x^j, t \right) \, ,
\label{inv.func} \end{equation} then we could use the fact that in that
reference frame the causal  structure is particularly simple: we would
simply select those points in the different time  levels that have the
closest values of \,$\left\{ \xi^i \right\}$\,.  This method, however,
will only be useful in a few special cases.  Indeed, in the general
case it may prove almost impossible to find the functional relation
(\ref{inv.func}).

With this in mind, we have developed a method that can be applied in
the general case.  Let us then assume that we are given a point in the
last time level \mbox{\,$\left\{ t_0 + \Delta t \right\}$\,} with
position \,$x^i_{t_0 + \Delta t}$\,.  Our aim is to find points
\,$x^i_{t_0}$\, and \,$x^i_{t_0 - \Delta t}$\, in the two previous time
levels in such a way as to  guarantee that a proper causal molecule
will be formed.  We have already seen  that this will happen if both
\,$x^i_{t_0 - \Delta t}$\, and \,$x^i_{t_0  + \Delta t}$\, are inside
the numerical light-cone of \,$x^i_{t_0}$\,.

Our algorithm to find the proper causal molecules linking the grids at
time-steps \,$t_0+\Delta t$, \,$t_0$, and \,$t_0-\Delta t$  assumes
that the causal reconnection condition holds.  (If it doesn't, then
remedial action, changing $\Delta t$ or $\Delta x^i$, is required.)
Our procedure  is the following.

\begin{enumerate}

\item Choose some point \,$x^i_{t_0 + \Delta t}$\,. The center of its
causal molecule will be that grid point \,$y^i_{t_0}$\,  for which the
following function reaches a minimum: \begin{equation} f_1 \left( y^i
\right) \,:=\, \sum_{i=1}^N \left\{ \left[ y^i \,-\, \beta^i \left(
y^j,t_0 \right) \Delta t \,+\, \frac{1}{2} \, \Gamma^i \left( y^j,t_0
\right) \left( \Delta t \right)^2 \right] - x^i_{t_0 + \Delta t} \,
\right\}^2 \, .  \end{equation} This minimum can easily be found by
standard multi-dimensional  search techniques.  Once the minimum is
found, we have our best  approximation to the exact inverse coordinate
transformation: the point $x^i$ is approximately at the same spatial
location as  $y^i$ in the original inertial frame.   \item Once we have
found the appropriate \,$y^i_{t_0}$\,, we look in the third time  level
\,$\left\{ t_0 - \Delta t \right\}$\, for the completion of the causal
molecule,  the point \,$z^i_{t_0 - \Delta t}$\, that minimizes the
function: \begin{equation} f_2 \left( z^i \right) \,:=\, \sum_{i=1}^N
\left\{ \left[ y^i_{t_0} \,+\, \beta^i \left( y^j_{t_0},t \right)
\Delta t \,+\, \frac{1}{2}  \, \Gamma^i \left( y^j_{t_0},t \right)
\left( \Delta t \right)^2 \right] - z^i \, \right\}^2 \, .
\end{equation} This is much easier than the previous step because we
already have the  point \,$y^i_{t_0}$\,, so we don't have to calculate
again the functions  \,$\beta$\, and \,$\Gamma$\,.  In fact, minimizing
\,$f_2$\, is equivalent to  finding the point \,$z^i$\, in the third
time level that is closest to the  axis of the light cone of
\,$y^i_{t_0}$\, in the original inertial frame.  We  do not have to go
through the  grid again to find this point.   \item So far we have
constructed only one molecule.  One needs to repeat the above steps for
all points at time-step $t_0+\Delta t$, but of course the best guess
for a causal molecule for any grid point is simply to translate the
molecule found for its neighbor.  This will occasionally fail, but only
by one grid point.  So  the minimization steps will require a computing
effort that is only proportional to the  number of grid points.

\end{enumerate}

It is not difficult to prove that, whenever the causal reconnection
conditions hold,  this algorithm does indeed produce a proper causal
molecule.  For reasons indicated above, the computational effort is
proportional to the number of grid points.  In complex problems, such
as general relativity, this is likely to be a very small overhead.


\section{\hspace{-9mm} Acknowledgements.}

We want to thank G.D. Allen for many useful discussions and comments.
One of the authors (M. Alcubierre) also thanks the `Universidad
Nacional Aut\'{o}noma de M\'{e}xico' for financial support.



\begin{thebibliography}{99}

\bibitem{MTW} C.W. Misner, K.S. Thorne and J.A. Wheeler, {\em
Gravitation}, W.H. Freeman and Co., U.S.A., 1973.

\bibitem{York} J.W. York, `Kinematics and Dynamics of General
Relativity' \, in:  {\em Sources of Gravitational Radiation}, ed. L.L.
Smarr, pp. 83-126, Cambridge University Press, U.S.A., 1979.

\bibitem{recepies} W.H. Press, B.P. Flannery, S.A. Teukolsky and W.T.
Vetterling,\hspace{1mm} {\em Numerical \linebreak recipes: The Art of
Scientific Computing}, Cambridge University Press, U.S.A., 1989.

\bibitem{Richtmyer} R.D. Richtmyer and K.W. Morton, \hspace{1mm} {\em
Difference Methods for Initial-Value \linebreak Problems}, 2nd. ed.,
Interscience, U.S.A, 1967.

\bibitem{Strang(1968)} G. Strang,  {\em SIAM J. Num. Anal.}, {\bf 5},
p. 506 (1968).

\bibitem{Marchuck(1982)} G.I. Marchuck, {\em Methods of Numerical
Mathematics}, 2nd. Ed., Springer-Verlag, U.S.A., 1982.

\bibitem{Lees} M. Lees, {\em J. Soc. ind. appl. Math.}, {\bf 10}, p.
610 (1962).

\bibitem{Fairweather} G. Fairweather and A.R. Mitchell, {\em J. Inst.
Maths.  Applics.}, {\bf 1}, p. 309 (1965).

\bibitem{upwind} G.E. Farsythe and W.R. Wasow, {\em Finite-Difference
Methods for Partial Differential Equations}, John Wiley and Sons,
U.S.A., 1967.

\bibitem{Godunov} M. Holt, {\em Numerical Methods in Fluid Dynamics},
Springer-Verlag, U.S.A., 1977.

\bibitem{Murman(1971)} E.M. Murman and J.D. Cole, {\em AIAA Journal},
{\bf 9}, p. 114, (1971).

\bibitem{Jameson(1974)} A. Jameson, {\em Comm. Pure Appl. Math.}, {\bf
27}, p. 283 (1974).

\bibitem{Seidel(1992)} E. Seidel and Wai-Mo Suen, {\em Phys. Rev.
Lett.}, {\bf 69} No. 13, p. 1845 (1992).

\bibitem{geommeths} B.F. Schutz, {\em Geometrical Methods of
Mathematical Physics}, Cambridge University Press, Cambridge, UK,
1980.

\end{thebibliography}
\end{document}